\documentclass[11pt]{article}
\usepackage{subcaption}
\usepackage{mystyle}
\usepackage{needspace}

\theoremstyle{plain}
\DeclareRobustCommand{\reptheoremheading}{\Cref{\reptheoremlabel}}
\newtheorem*{reptheorembody}{\reptheoremheading}
\newenvironment{reptheorem}[1]
  {\def\reptheoremlabel{#1}\begin{reptheorembody}}
  {\end{reptheorembody}}
  
\usepackage[sort]{natbib}
\usepackage{booktabs}

\usetikzlibrary{arrows.meta}
\usetikzlibrary{decorations.pathreplacing}

\definecolor{ink}{HTML}{17324D}
\definecolor{teal}{HTML}{007F86}
\definecolor{rust}{HTML}{BB472D}
\definecolor{muted}{HTML}{586573}

\tikzset{
  >={Latex[length=2.1mm]},
  site/.style={circle,draw=ink,fill=white,line width=.8pt,
    minimum size=7mm,inner sep=1pt},
  flow/.style={-{Latex},line width=1.1pt,draw=rust},
  every picture/.style={font=\small}
}

\DeclareMathOperator{\cost}{cost}

\renewcommand{\top}{\operatorname{top}}
\DeclareMathOperator{\distortion}{distortion}

\newcommand{\OPT}{\mathsf{OPT}}
\algnewcommand{\pois}{\text{Poisson}}
\newcommand{\Mopt}{M^*}

\newcommand{\dif}{\mathop{}\!\mathrm{d}}
\newcommand{\rsd}{\mathsf{RSD}}
\newcommand{\ps}{\mathsf{PS}}
\newcommand{\tps}{\mathsf{TruncPS}}

\crefname{construction}{construction}{constructions}
\Crefname{construction}{Construction}{Constructions}
\crefformat{construction}{#2construction~#1#3}
\Crefformat{construction}{#2Construction~#1#3}

\begin{document}
\title{Sublinear Metric Distortion for Matching}

\author{
Moses Charikar\\
Stanford University
\and
Alireza Haqi\\
Stanford University
\and
Jabari Hastings\\
Stanford University
\and
Yichen Huang\\
Harvard University
\and
Prasanna Ramakrishnan\thanks{Emails:
  \texttt{moses@cs.stanford.edu},
  \texttt{ahaqi@stanford.edu},
  \texttt{jabarih@stanford.edu},
  \texttt{yichenhuang@g.harvard.edu}, and
  \texttt{pras\_r@seas.harvard.edu}.}\\
Harvard University
}

\date{\vspace{15pt}\small{\today}}

\maketitle

\begin{abstract}

We study metric distortion in one-sided matching. Given $n$ agents and $n$ items in an unknown metric space, our goal is to match agents to items so that the total distance between matched pairs is small. However, instead of direct access to the distances in the metric, we are only given each agent's ranking of the items by distance. What is the smallest worst-case approximation factor (called the \emph{distortion}) that a mechanism can guarantee in expectation? Despite significant interest over the last decade, the best-known guarantee has been the $O(n)$ distortion of \emph{Random Serial Dictatorship} since the beginning. 

We give a matching mechanism with distortion $O(n^{\log_4 3})$, the first sublinear upper bound. The key idea is to use a \emph{partial} matching derived from a truncated version of either Random Serial Dictatorship or the Probabilistic Serial rule. We then complete this matching using a new, general-purpose \emph{cluster completion} algorithm that ensures unmatched agents receive nearby items. Along the way, we pin down the optimal distortion of Random Serial Dictatorship and Probabilistic Serial up to logarithmic and constant factors, respectively, and give improved lower bounds for \emph{truthful} mechanisms.  
\end{abstract}

\thispagestyle{empty}

\clearpage
\pagenumbering{roman}

\setcounter{tocdepth}{2}
\tableofcontents

\clearpage
\pagenumbering{arabic}

\section{Introduction}

Matching mechanisms are a hallmark of a well-oiled society. These mechanisms orchestrate how
doctors are assigned to hospitals, how students are placed in schools, how tenants receive housing units, how conference submissions are routed to reviewers, and so much more. In many of these settings, the resulting assignment can be highly consequential, with a lasting impact on everyone involved. One of the central goals of computational social choice is to understand which matching mechanisms give participants the best collective outcomes. 

We consider one-sided matching markets in which $n$ agents must be matched to $n$ items, and agents have ranked preferences over items.
One of the most prominent mechanisms is \emph{Random Serial Dictatorship (RSD)} \citep*{Abdulkadiroglu:1998aa}, also called \emph{Random Priority}, 
which processes the agents in a uniformly random order, and greedily matches each to her favorite available item. RSD is \emph{truthful}, meaning no agent can get a better outcome by misreporting her preferences. Another well-studied mechanism is the \emph{Probabilistic Serial rule (PS)} \citep*{Bogomolnaia:2001aa}, under which agents simultaneously ``eat'' their favorite available item at the same constant rate, moving on to their next favorite whenever an item runs out. The fraction of each item consumed by each agent can be viewed as a fractional matching, and we can sample an integral matching using the Birkhoff--von Neumann decomposition. While PS is not truthful, it is \emph{envy-free}, meaning no agent prefers another agent's allocation to her own in expectation. Though RSD and PS seem fundamentally different, they are closely related. Che and Kojima \cite{Che:2010aa} showed that the two are asymptotically equivalent in large markets,\footnote{Roughly speaking, if we take any matching instance and create $q$ copies of each agent and item, then the assignment probabilities under RSD and PS (aggregating copies) converge to one another as $q \to \infty$.} suggesting that the benefits of one may transfer to the other in practice.

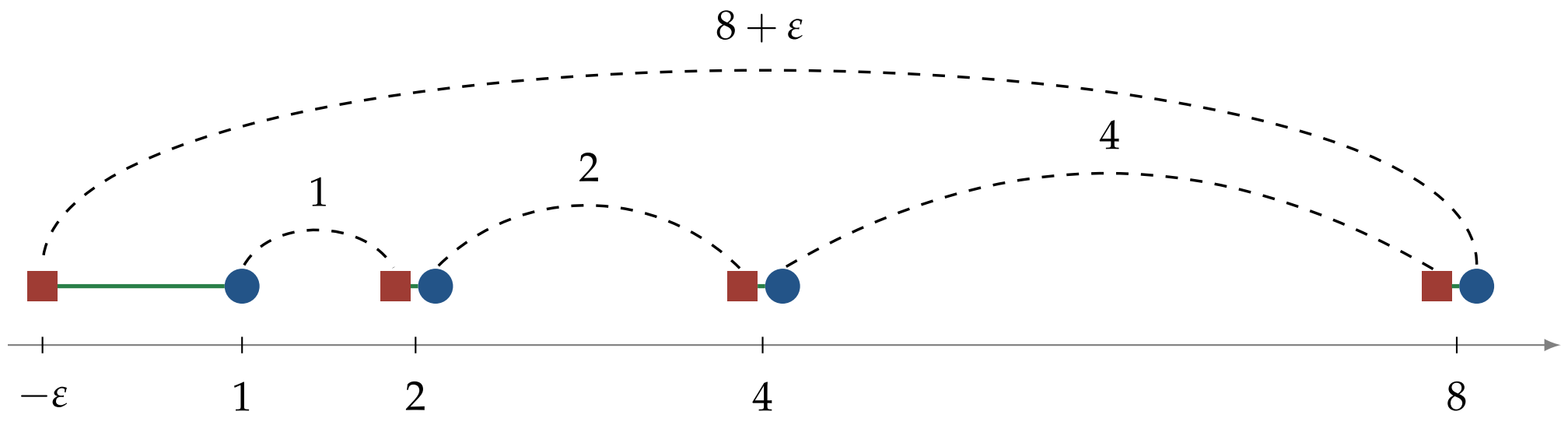
\begin{figure}[ht]
\centering
\pgfmathsetlengthmacro{\lineunit}{0.8*\textwidth/8.95}
\begin{tikzpicture}[x=\lineunit,y=1cm,>=latex,font=\small]
  \path[use as bounding box] (-.45,-.5) rectangle (8.75,2.65);
  \definecolor{agentblue}{RGB}{35,84,136}
  \definecolor{itemred}{RGB}{159,60,52}
  \definecolor{optimalgreen}{RGB}{42,128,73}
  \tikzset{
    agent/.style={circle,fill=agentblue,inner sep=0pt,
      outer sep=0pt,minimum size=8.4525pt},
    item/.style={rectangle,fill=itemred,inner sep=0pt,
      outer sep=0pt,minimum size=7.245pt},
    matching/.style={-,black,dashed,line width=.65pt,
      shorten <=1pt,shorten >=1pt},
    optimal/.style={-,optimalgreen,line width=1pt}
  }
  \def\epsdisplay{.15}
  \def\pairsep{.115}
  \draw[gray,->] (-.35,0) -- (8.6,0);
  \foreach \x in {1,2,4,8} {
    \draw (\x,-.07) -- (\x,.07);
    \node[below=5pt] at (\x,0) {$\x$};
  }
  \draw (-\epsdisplay,-.07) -- (-\epsdisplay,.07);
  \node[below=5pt] at (-\epsdisplay,0) {$-\varepsilon$};

  \node[item] (b0) at (-\epsdisplay,.5) {};
  \node[agent] (a0) at (1,.5) {};
  \foreach \x/\i in {2/1,4/2,8/3} {
    \node[item] (b\i) at (\x-\pairsep,.5) {};
    \node[agent] (a\i) at (\x+\pairsep,.5) {};
  }

  \foreach \i in {0,1,2,3} {
    \draw[optimal] (b\i.east) -- (a\i.west);
  }

  \draw[matching] (a0.north) to[out=60,in=120]
    node[midway,above=2pt,text=black] {$1$} (b1.north);
  \draw[matching] (a1.north) to[out=45,in=135]
    node[midway,above=2pt,text=black] {$2$} (b2.north);
  \draw[matching] (a2.north) to[out=30,in=150]
    node[midway,above=2pt,text=black] {$4$} (b3.north);
  \draw[matching] (a3.north) .. controls +(0,2.25) and +(0,2.25) ..
    node[midway,above=3pt,text=black] {$8+\varepsilon$} (b0.north);
\end{tikzpicture}
\caption{An illustrative example of metric distortion with $n  =4$ agents (blue dots) and items (red squares) on a line, with agents at $2^{0}, 2^{1}, \dots, 2^{n-1}$ and items at $-\eps, 2^{1}, \dots, 2^{n-1}$. The green matching is optimal, with cost $1 + \eps$. A Serial Dictatorship with agents choosing from left to right leads to a cascade of unfortunate choices, giving the dashed matching with cost $2^n-1+\varepsilon$.}
\label{fig:intro-example}
\end{figure}

These matching mechanisms are extensively studied\footnote{See \cite[Section~5]{Moulin:2019aa} and \cite[Sections~3.1--3.2]{Sonmez:2025aa} for work surveying RSD and PS in the Economics literature, and \Cref{sec:related} for further discussion.} and have attractive theoretical properties, but how close do they actually get to the best collective outcomes? The \emph{metric distortion} framework offers a compelling way to approach this question \citep*{Anshelevich:2015aa,Anshelevich:2018aa}. It imagines that agents and items lie in an unknown metric space and that an agent's cost for an item is the distance between them. Agents rank items in order of cost, and the \emph{social cost} of a matching is the sum of the agents' costs. Given only these rankings, how closely can a mechanism approximate the minimum social cost? Its \emph{distortion} is the worst-case ratio of its expected social cost to the optimum, over all instances and all metrics consistent with the rankings. (See \Cref{fig:intro-example} for an example.)

The metric distortion framework was originally introduced in the context of \emph{voting} (motivated by popular spatial models like the \emph{political spectrum} \cite{Hotelling:1929aa,Downs:1957aa}), and has grown into a vibrant research area (See \Cref{sec:related} for a detailed discussion). Part of the appeal is that this exploration deepens our understanding of voting rules by allowing us to probe their weaknesses and expose how their choices can lead to catastrophic failures. But another, more unexpected benefit is that the pursuit of \emph{optimal} distortion has motivated the development of simple and natural new voting rules, which have now closed the gap for deterministic rules \cite{Anshelevich:2015aa,Gkatzelis:2020aa} and nearly closed it for randomized rules \cite{Charikar:2022aa,Cai:2026ab}.

Although the metric viewpoint is also natural for matching (e.g., students may prefer nearby schools, reviewers may prefer papers close to their expertise), the optimal distortion has proved much harder to pin down.
Caragiannis, Filos-Ratsikas, Frederiksen, Hansen, and
Tan \cite{Caragiannis:2016aa,Caragiannis:2024ab} initiated this work, showing that a deterministic \emph{Serial Dictatorship} 
has worst-case distortion $2^{n} - 1$, while RSD has distortion $O(n)$ via a simple inductive argument. Anari, Charikar, and
Ramakrishnan \cite{Anari:2023aa} later devised a new deterministic matching mechanism with distortion $O(n^2)$ called \emph{RepMatch}, which forms clusters of nearby agents with a ``representative'' in each, and merges clusters when their representatives' favorite items overlap. They also established the first nontrivial lower bound of $\Omega(\log n)$ for all mechanisms, and posed the \emph{thin matching} conjecture which would imply that \emph{any} randomized mechanism can be made deterministic with little loss in distortion. Haqi and Oveis Gharan  \cite{Haqi:2026aa} recently resolved this conjecture up to polylogarithmic factors, giving a deterministic mechanism with distortion $\tilde{O}(n)$ using RSD as a black box. Despite these advances, no mechanism has been known to have sublinear distortion, leaving an exponential gap with the logarithmic lower bound. Closing this gap has been a stubborn challenge \cite[Open Problem~2]{Anshelevich:2021aa} that Filos-Ratsikas, Gkatzelis, Latifian, Rewinski, and Voudouris \cite{Filos-Ratsikas:2025aa} call ``quite possibly one of the most challenging open questions in the distortion literature.'' 

Curiously, one promising candidate for achieving sublinear 
distortion is RSD itself. \cite{Caragiannis:2016aa,Caragiannis:2024ab} give a lower bound of only
$\Omega(n^{0.29})$, and Ramakrishnan \cite{Ramakrishnan:2026aa} conjectured that the upper bound can be improved to $O(n^{1-\epsilon})$ for some constant $\epsilon > 0$. The metric distortion of RSD also has broader significance due to a connection to \emph{online metric matching}. In particular, this distortion is precisely the competitive ratio of the greedy algorithm when requests arrive in a uniformly random order. In this context,  Gairing and Klimm \cite{Gairing:2019aa} proved the same $\Omega(n^{0.29})$ and $O(n)$ bounds. Progress has been elusive in both settings, but any improvement in one would carry over to the other.

\subsection{Our contributions and technical overview}

Our contributions fall into four broad directions: understanding the distortion of RSD (\Cref{sec:rsd}), doing the same for PS (\Cref{sec:ps}), obtaining sublinear distortion by modifying these mechanisms (\Cref{sec:rsd-rep,sec:coordinated-merging}), and giving new lower bounds for truthful mechanisms (\Cref{sec:truth}). The corresponding sections are largely self-contained, and we invite the reader to take a nonlinear approach based on interest. See 
\Cref{tab:distortion-results} for a summary of the bounds.

\subsubsection{Distortion of RSD} 

We settle the distortion of RSD up to logarithmic factors, showing that the early upper bound of \cite{Caragiannis:2016aa,Caragiannis:2024ab} is essentially tight.

\begin{theorem}\label{thm:rsd-lowerbound}
There are matching instances in which RSD has distortion $\Omega(n/\sqrt{\log n})$.    
\end{theorem}

This result disproves the conjecture of Ramakrishnan \cite{Ramakrishnan:2026aa}, and also gives the following corollary for online metric matching.

\begin{corollary}\label{cor:intro-greedy}
For online metric matching in the random-order model, the greedy algorithm has competitive ratio $\tilde{\Theta}(n)$. 
\end{corollary}

The starting point is the cascade behind \Cref{fig:intro-example}, in which an agent steals an item that is best for another agent, forcing that agent to take an item farther away, and so on. Replacing the single agent and item at each position with exponentially growing groups of agents and items (e.g., $r^t$ agents and items at position $2^t$), we can make it more likely that the cascade reaches the rightmost group under a random order. Balancing this with the distances between positions, \cite{Caragiannis:2016aa,Caragiannis:2024ab} give a lower bound of $\Omega(n^{0.29})$. Our construction allows the cascade to follow more routes. We arrange locations in a grid of levels and columns, with distances growing exponentially with the level, much like the line. Each location can have several parents in the preceding level, and displaced agents from any of them can take its items and propagate the cascade.

Branching gives the cascade more opportunities to reach deep levels, but we also need to ensure that the total number of agents used in the construction is small. Making every group smaller would save agents, but would also make the cascade more vulnerable to fluctuations in the random order. We therefore prune grid locations that contribute little to reaching the deepest level, while choosing the sizes of the remaining groups to preserve the cascade. 
In our analysis of the resulting construction, we aim to show that the expected number of deepest-level agents matched to items outside that level is at least one. It is natural to view RSD as assigning each agent an independent, uniformly random arrival time in $[0,1]$, with each item disappearing when an arriving agent claims it. We build on this view by considering an auxiliary process in which items can also disappear at deadlines of our choosing, even if no agent has claimed them. We bound the expected number of these artificial removals and couple this process with RSD to establish the desired expectation guarantee. 
Matching an agent from the deepest level to an item outside that level costs $\Omega(n/\sqrt{\log n})$ times the optimal matching cost, so our lower bound on the expected number of these assignments gives the claimed distortion. 

\subsubsection{Distortion of PS}
Next, we consider the distortion of PS for the first time in the metric distortion literature. Given its equivalence with RSD in large markets, it may be reasonable to expect that PS has roughly linear distortion. Indeed, the example in \Cref{fig:intro-example} gives a straightforward $\Omega(n)$ lower bound.\footnote{To see why, notice that for every integer $t\ge 0$, the agent at $2^t$ eats the item at $-\eps$ in the last $2^{-(t+1)}$ fraction of time. Each agent incurs a cost at least $1/2$, and thus the total cost is at least $n/2$.}
However, we show that the distortion of PS is actually \emph{exponential} in the worst case, and give a matching upper bound up to a constant factor.

\begin{theorem}\label{thm:ps-tight-intro}
The worst-case distortion of PS is $\Theta(c^n)$ where $c = \sqrt{10}/3 \approx 1.054$.    
\end{theorem}
The lower bound (\Cref{thm:lb-ps}) again comes from a cascade, modified to make the rightmost agent start eating the leftmost item earlier. Unlike \Cref{fig:intro-example}, each intermediate layer now has two locations, each containing an agent and an item. The upper item is exhausted first, freeing its agent to start eating the next upper item. When the lower agent is released, she joins her and helps exhaust that item. On the basic line, each additional layer halves the fraction of the leftmost item that the rightmost agent eats; in the repeating part of our construction, these staggered arrivals multiply that fraction by $5/9$ instead. Since her distance to the leftmost item doubles with each additional layer, her expected cost grows by a factor of $10/9$ per layer. There are $n/2$ layers in our construction, giving distortion $\Omega((10/9)^{n/2})$. 
The upper bound (\Cref{thm:ub-ps}) analyzes how successive phases of eating can amplify cost. We first establish a rate inequality that bounds the instantaneous increase in cost in terms of the optimal matching and the cost already incurred. Writing $\OPT$ for the cost of the optimal matching, the inequality implies that the cost accumulated by time $t<1$ is at most $\frac{t}{1-t} \OPT$. We then turn to the late stages of the mechanism. We identify groups of agents that, if left alone, would keep eating their current items until time $1$. Agents outside these groups can exhaust their items early, releasing more agents who may then disturb other groups. We track this cascade by bounding how far each released agent may have to travel, weighted by how much time she has left to eat. The main technical step is a displacement inequality that controls the growth of this quantity in terms of the number of agents released, with the same exponential base as in the lower bound. 

At first glance, this result seems completely at odds with the expectation that PS should behave similarly to RSD. However, we show that a slight modification brings it back in line. The key is to leverage the observation that in the worst case, PS incurs almost all of its cost at the very end of the process. With this in mind, consider what happens if we stop PS at some prescribed time and return the partial fractional matching. We call the resulting mechanism \emph{TruncatedPS} (see \Cref{ssec:tps}). 
Assuming that each agent's rate of consumption is $1$ item per unit of time, it is not hard to see that we need to run past time $1 - \frac1n$, otherwise an item ranked last by all agents would not yet have been touched. We can show that if we stop at time $1 - \frac{1}{2n}$, then every item has been at least half-eaten, and the doubled weights contain a complete fractional matching. With doubling, the mechanism  has distortion $O(n)$, matching RSD.  

\begin{theorem}\label{thm:ub-ps-truncated}
TruncatedPS (with doubling) has distortion $O(n)$ in all matching instances.
\end{theorem}

However, in \Cref{thm:lower-bound-truncated-ps} we show that our branching construction for RSD gives a lower bound of $\Omega(n/\sqrt{\log n})$ for TruncatedPS (with doubling), so this modification alone gets stuck in the same place.

\subsubsection{Cluster completion and sublinear distortion} 
To find a matching mechanism with better distortion, we build on the observation that PS incurs most of its cost at the end of the process. In fact, the same is true for RSD --- \cite*{Anari:2023aa} showed that \emph{TruncatedRSD}, which stops before the last $k$ agents are matched, produces a partial matching with cost $O(n/k)$ times the optimal perfect matching. This raises a natural question: can we complete the partial matching without losing the cost advantage of the prefix? We show that this is possible, obtaining the following sublinear distortion bound.

\begin{theorem}\label{thm:rsd+repmatch}
There exists a randomized mechanism with distortion $O(n^{\log_4 3})$, where $\log_4 3 \approx 0.7925$.
\end{theorem}

This result establishes the first sublinear distortion bound for matching. The result of Haqi and Oveis Gharan \cite{Haqi:2026aa} immediately gives a deterministic mechanism with distortion $\widetilde{O}(n^{\log_4 3})$ as a corollary. 

Our completion procedure, which we call \emph{Cluster Completion}, uses a partition of the agents and items into balanced clusters: At a high level, imagine that we precompute a balanced partition of agents and items into clusters (with equally many agents and items in each cluster) such that every agent is close to all items in her cluster. We can think of the clusters as a kind of ``safety net'', and the goal is to complete the matching while guaranteeing that every agent either keeps her early RSD assignment, or is matched to an item in her cluster. 

The difficulty is that an unmatched agent's cluster may contain no unmatched item. We therefore allow a chain of reassignments: give her an occupied item from her cluster whose matching partner is from another cluster, rematch its previous matching partner to an item in her own cluster, and continue until reaching
an unused item. We show that such a chain can always be chosen
to visit each cluster at most once. Thus, if $w_j$ is the
largest agent--item distance within cluster $j$, completing one
agent adds cost at most $\sum_j w_j$. Writing $\OPT$ for the
optimal matching cost, a partition satisfying
$\sum_j w_j\le O(n^\theta)\OPT$ therefore gives
overall distortion $O(\frac{n}{k} + kn^\theta)$, which balances to $O(n^{(1 + \theta)/2})$ by choosing $k = n^{(1 - \theta)/2}$. The same approach can also be adapted to partial \emph{fractional} matchings, so the same bound applies to \emph{either} TruncatedRSD or TruncatedPS.

The remaining challenge is to construct such a partition with
$\theta<1$. In this regard, we observe that the
$w_j$ terms are precisely the kind of quantities tracked in
the analysis of \emph{RepMatch}~\cite{Anari:2023aa}. RepMatch
iteratively merges clusters whose representatives claim
overlapping sets of favorite items. Once these sets are disjoint,
the clusters define a balanced partition.\footnote{RepMatch then produces a matching by matching arbitrarily in each cluster, whereas we only use the cluster to guide the completion.} Hastings and Ramakrishnan~\cite{Hastings:2025aa} studied a
size-based variant that retains the representative of the larger
cluster at each merge. Inspired by their rule, our variant
chooses the retained representative in a \emph{randomized}
fashion, with probabilities determined by the two cluster sizes.
The resulting partition satisfies the desired cost bound in
expectation with $\theta=\log_2(3/2)$, yielding the claimed theorem.

The central role of RepMatch in cluster completion leads us to ask whether better matching mechanisms can build on its core idea of iteratively merging nearby clusters. In particular, the way that RepMatch only uses the representatives' preferences and then arbitrarily matches agents and items within clusters feels like an inefficient use of the  available information. With this in mind, we introduce another new matching mechanism called \emph{Coordinated Merging} which \emph{also} has sublinear distortion.

\begin{theorem}\label{thm:sing-layer-sublinear-distortion}
In all matching instances, the Coordinated Merging mechanism has distortion $O(n^{\alpha})$ where $\alpha \approx 0.95084$.
\end{theorem}

While this is a weaker guarantee than \Cref{thm:rsd+repmatch}, it gives a different route to sublinear distortion which is markedly different from anything resembling RSD. We also view this mechanism as a proof-of-concept that a more careful version of RepMatch can have better distortion, and we are optimistic that similar ideas may be useful in future work on the problem.
We give a high-level description of the approach below.

 RepMatch takes two clusters with a common item of interest among their top preferred items, with the number of preferred items equal to the respective cluster size, and  merges them. We can think of this process in terms of a \emph{conflict graph} with
one vertex for each cluster, and an edge between two clusters whenever their sets of preferred items overlap.
The new algorithm instead computes a maximal independent set, denoted $S$, aiming to capture a global property of the conflict graph.
The algorithm randomly chooses between the following two procedures, using probabilities that balance their expected cost against the reduction in the number of clusters.

\begin{itemize}[itemsep=0em,topsep=0em]
    \item We allocate to each cluster in $S$ its favorite items.\footnote{More precisely, a weighted lottery selects either $S$ or a single cluster outside it, with $S$ having the highest selection probability. Each selected cluster then receives its favorite items. See \Cref{alg:large-independent-set} for details.}
    \item We merge every cluster outside $S$ with one of its neighbors in $S$ (the details take some care, described in \Cref{subsubsection:mergingAnalysis}). 
\end{itemize}

In either case, we recursively solve the resulting instance, which has fewer clusters. When $S$ is large, assigning its clusters their favorite items removes many clusters from the instance. When $S$ is small, merging clusters into it leaves few representatives for the recursive call, while the expected merging cost is at most a constant times the optimum. The algorithm ultimately chooses between the two procedures at random, with probabilities determined by the size of $S$.

\subsubsection{Lower bounds for truthful mechanisms} 

Finally, we turn our attention to developing a stronger understanding of \emph{truthful mechanisms} in the metric distortion framework. Truthfulness is one of the central focuses of the broader mechanism design literature and is one of the core appeals of (Random) Serial Dictatorships. This property has also been central to the study of the metric distortion of matching. \cite{Caragiannis:2016aa,Caragiannis:2024ab} focus entirely on truthful mechanisms, and the distortion survey of \cite[Open Problem~2]{Anshelevich:2021aa} asks how the requirement of truthfulness impacts the optimal distortion for metric matching (alongside asking for the optimal distortion in general). Despite this focus on this question, the only general lower bound for truthful mechanisms is the $\Omega(\log n)$ bound that applies to all mechanisms.

We show, for the first time, how to explicitly leverage truthfulness to obtain a stronger lower bound, even for randomized mechanisms that need only be truthful in expectation.

\begin{theorem}\label{thm:tie-sqrtlog}
Every truthful-in-expectation matching mechanism has distortion at least $2^{\Omega(\sqrt{\log n})}$.
\end{theorem}

The core idea is to build on the lower bound of \cite{Caragiannis:2016aa,Caragiannis:2024ab} for RSD, which uses an instance on the line like \Cref{fig:intro-example} but with groups of agents and items that grow exponentially. We can show that if a mechanism gives a sufficiently large probability of matching an agent to the leftmost item, then the agent stands to benefit if she misreports her preferences by impersonating the group to the right. If the mechanism has low distortion, then agents in that group must be very likely to get matched to collocated items, so the impersonating agent increases her chances of getting a higher ranked item. This tension forces the distortion of a truthful mechanism to be high.\footnote{Since the kinds of misreports we use involve one agent impersonating another, the argument can be adapted to prove the same lower bound against \emph{envy-free} mechanisms. Intuitively, an agent would only ever want to impersonate an agent she envies.}

We then look to understand the distortion of \emph{deterministic} truthful mechanisms, and show that under Pareto optimality, all such mechanisms have distortion at least as bad as a deterministic Serial Dictatorship.

\begin{theorem}\label{thm:det-po}
Every deterministic truthful mechanism that is Pareto optimal has distortion at least $2^n - 1$.
\end{theorem}

Intuitively, Pareto optimality does not really constrain distortion, since any Pareto improvement necessarily improves the distortion as well. Thus, we can interpret this result as showing how truthfulness forces some kind of social inefficiency in deterministic mechanisms, either by violating Pareto optimality or incurring exponential distortion. 

The proof uses the instance from \Cref{fig:intro-example}, and argues that Pareto optimality and truthfulness force a deterministic mechanism to match the rightmost agent to the leftmost item for some relabeling of the agents. It helps to relabel the agents randomly, and show that the resulting randomized mechanism (which is truthful-in-expectation, Pareto optimal, and \emph{anonymous} as well) gives positive probability to this match of extremes. Pareto optimality forces agents to match to nearby items, either to the left or one item to the right. On the other hand, truthfulness implies that instances that differ in one agent's preference maintain the same probability of that agent matching to the shared prefix of the two preferences. We can use these two facts to consider a web of instances connected by one agent moving at a time, and argue step-by-step that some of the probability mass of matching to the leftmost item gets pushed towards the right.\\

\Cref{tab:distortion-results} below summarizes the main distortion bounds proved in the paper.

\begin{table}[!ht]
\centering
\small
\renewcommand{\arraystretch}{1.25}
\setlength{\tabcolsep}{3.8pt}
\begin{tabular}{@{}lcccl@{}}
\toprule
 & \multicolumn{2}{c}{Previous} & This paper & Pointer \\
\cmidrule(lr){2-3}
 & Lower bound & Upper bound & & \\
\midrule
RSD
 & $\Omega(n^{0.29})$ & $O(n)$
 & $\widetilde \Omega(n)$ & \Cref{thm:rsd-lowerbound} \\
PS
 & $\Omega(n)$ & ---
 & $\Theta((10/9)^{n/2})$ & \Cref{thm:ps-tight-intro} \\
TruncatedPS (with doubling)
 & --- & ---
 & $\widetilde \Theta(n)$ & \Cref{thm:ub-ps-truncated,thm:lower-bound-truncated-ps} \\
Best randomized
 & {$\Omega(\log n)$} & $O(n)$
 & $O(n^{\log_4 3})$ & \Cref{thm:rsd+repmatch} \\
Truthful in expectation
 & $\Omega(\log n)$ & {$O(n)$}
 & $2^{\Omega(\sqrt{\log n})}$ & \Cref{thm:tie-sqrtlog} \\
Deterministic, truthful, Pareto Optimal
 & --- & $2^n-1$
 & $\geq 2^n-1$ & \Cref{thm:det-po} \\
\bottomrule
\end{tabular}
\caption{Summary of our results. A dash indicates that no previous bound is listed. \cite{Caragiannis:2016aa,Caragiannis:2024ab} proved the prior $\Omega(n^{0.29})$ and $O(n)$ bounds for RSD, and the $2^{n} - 1$ bounds for Serial Dictatorship. The $\Omega(\log n)$ lower bounds are due to \cite{Anari:2023aa}. {The $\Omega(n)$ lower bound for PS is a straightforward consequence of \Cref{fig:intro-example}.}
}
\label{tab:distortion-results}
\end{table}

\subsection{Related work}\label{sec:related}

\paragraph{Metric Distortion.}
The metric distortion framework was introduced by \cite{Anshelevich:2015aa,Anshelevich:2018aa} in the context of voting, in which voters have preferences over candidates, and the goal is to choose a single winning candidate (or distribution over candidates). A long line of work \cite{Goel:2017aa,Munagala:2019aa,Kempe:2020aa,Gkatzelis:2020aa,Kizilkaya:2022aa,Kizilkaya:2023ab} culminated in showing that the optimal metric distortion for deterministic rules is exactly 3, and another recent flurry \cite{Charikar:2022aa,Charikar:2024aa,Frank:2026aa,Ye:2026aa,Shah:2026aa,Cai:2026ab} has put the optimal distortion for randomized rules between 2.1126 and 2.1246. Beyond the search for optimal distortion, researchers have also used the framework as a way of understanding the impact of more permissive or restrictive models. Richer models allow voters to deliberate in small groups~\cite{Fain:2017aa,Goel:2025aa,Munagala:2025aa}, or let voters report how strongly they prefer one candidate to another~\cite{Abramowitz:2019aa}. More restrictive models use only aggregate pairwise comparisons~\cite{Goel:2017aa,Munagala:2019aa,Charikar:2025aa}, truncated rankings~\cite{Kempe:2020ab,Anagnostides:2022aa}, answers to a few pairwise comparisons per voter~\cite{Ebadian:2024aa}, ballots from a small sample of voters~\cite{Gross:2017aa,Fain:2019aa}, or bound the randomness of the voting rule~\cite{Cai:2026aa}.

Over a similar time frame, the progress in the matching setting has been harder to come by. Apart from the
works mentioned earlier, a few papers have found success tackling special cases of the problem. On the line, \citep{Hastings:2026aa} explore how a mechanism's distortion changes with a prediction. For this learning-augmented setting, they   establish $1$-consistency and $3$-robustness, which is optimal.
\cite{Gkatzelis:2026aa} explore how the distortion is affected by the combination of ordinal preferences, approval information or knowledge of inter-item distances. 

Metric distortion has also been applied to maximum matchings with ordinal information. In contrast to our minimization setting, constant distortion bounds exist for both truthful and randomized mechanisms \citep{Anshelevich:2016aa,Anshelevich:2016ab}.

\paragraph{Utilitarian Distortion for Matching.} 

A sibling of metric distortion is \emph{utilitarian distortion}, in which agents have unknown values for items, and a mechanism must find a matching within a small factor of the one with maximum value, using only ordinal preferences. The utilitarian distortion setting was also introduced in the context of voting by Procaccia and Rosenschein \cite{Procaccia:2006aa} and studied in the matching setting by Filos-Ratsikas, Frederiksen, and Zhang  \cite{Filos-Ratsikas:2014aa}. With arbitrary utilities, distortion is unbounded for deterministic mechanisms, and the optimal randomized distortion is $n$ by taking a uniformly random matching\footnote{Not to be confused with RSD; a uniformly random matching makes no use of the preferences.} (see \cite{Amanatidis:2022aa,Amanatidis:2024aa}). With these dead ends, researchers primarily study  utilities that are \emph{normalized}, most commonly to be unit-sum for each agent. With this assumption RSD has the optimal randomized distortion of $\Theta(\sqrt{n})$ \cite{Filos-Ratsikas:2014aa} and any deterministic mechanism that always gives at least one agent her first choice has the optimal deterministic distortion of $\Theta(n^2)$ \cite{Amanatidis:2022aa}.

Another line of work supplements preference rankings with limited access to cardinal values. For unrestricted nonnegative valuations, \cite{Amanatidis:2022aa} give deterministic algorithms with distortion $O(n^{1/(\lambda+1)})$ using $O(\lambda\log n)$ value queries per agent. \cite{Amanatidis:2024aa} obtain $O(\sqrt{n})$ distortion with just two queries per agent. More recently, \cite{Ebadian:2025aa} establish the optimal trade-off for a constant query budget: for every fixed integer $\lambda\geq 1$, $\lambda$ queries per agent suffice for distortion $O(n^{1/\lambda})$, matching the lower bound for deterministic algorithms \citep{Amanatidis:2022aa}. Other elicitation models use threshold queries \citep{Ma:2021aa} or ask agents to classify items into utility ranges \citep{Latifian:2025aa}.

\cite{Filos-Ratsikas:2026aa} study utilitarian distortion in stable matching, measuring welfare relative to the maximum-welfare stable matching.
\cite{Caragiannis:2026aa} study $b$-matching, where agents may receive multiple items and each agent's item values are drawn i.i.d.\ from an unknown agent-specific distribution. They establish the optimal distortion of $e/(e-1)$ for expected welfare.

\paragraph{RSD and PS in Economics.} 
Since their introduction, RSD and PS have attracted sustained interest in the economics community. Much of this work studies the properties that make these mechanisms appealing for practical applications.

For RSD, several results provide strong axiomatic foundations. Bade \cite{Bade:2020aa} shows that uniformly randomizing the agents' roles in any deterministic, truthful, Pareto-optimal mechanism yields RSD, provided the mechanism is \emph{nonbossy}.\footnote{Nonbossiness means an agent cannot change others' assignments by changing her report without also changing her own assignment.} Pycia and Troyan \cite{Pycia:2026aa} uniquely characterize RSD through ex post Pareto optimality, anonymity, and an obviously strategy-proof implementation, in which agents can recognize the incentive to act truthfully without complicated reasoning about others' behavior. Basteck and Ehlers \cite{Basteck:2025aa} show that RSD cannot be uniformly Pareto-improved upon by another truthful mechanism satisfying an additional invariance condition.\footnote{Their bounded invariance condition requires that rearranging items below an item $b$ in one agent's ranking does not change anyone's probability of receiving $b$. Improvements are measured by stochastic dominance and must hold at every preference profile, with a strict improvement somewhere.} Although RSD is not envy-free, \cite{Connor:2026aa} show that it approximately satisfies this property: an agent's expected utility for another agent's allocation is at most $\sqrt{2}$ times her expected utility for her own, and this bound is tight.

For PS, \cite{Hashimoto:2014aa} give axiomatic characterizations based on efficiency and envy-freeness under stochastic dominance, together with a condition ensuring that an agent cannot change her chances of receiving higher-ranked items by reordering lower-ranked ones. Although PS is not truthful in general, Kojima and Manea \cite{Kojima:2010aa} show that, for fixed item types and a fixed agent utility function, truthful reporting becomes weakly dominant when there are sufficiently many copies of each type. \cite{Budish:2013aa} extend PS to settings with additional constraints, such as quotas in school choice, while retaining corresponding efficiency and fairness guarantees.

\paragraph{Greedy Algorithm for Online Metric Matching.}
Our problem is related to online metric matching, where servers are given in advance and requests arrive sequentially, each requiring an immediate and irrevocable assignment to an available server. The natural \emph{Greedy} algorithm matches each request to its nearest available server. Its choices coincide with Serial Dictatorship for a fixed arrival order, and with RSD under a uniformly random arrival order.
Kalyanasundaram and Pruhs~\cite{Kalyanasundaram:1993aa} established Greedy's tight competitive ratio of $2^n-1$ under adversarial arrivals, even on the line. For uniformly random arrival orders, Gairing and Klimm~\cite{Gairing:2019aa} proved an $O(n)$ upper bound and an $\Omega(n^{0.29})$ lower bound. The equivalence with RSD means that our lower bound also establishes a nearly linear competitive ratio for Greedy in this model.

Greedy has also been studied with resource augmentation, where the online algorithm receives additional capacity at each server location while the offline benchmark uses the original capacities~\cite{Arndt:2023aa,Kalyanasundaram:2000aa,Caragiannis:2024ab}, and under stochastic models for request and server locations~\cite{Balkanski:2023greedy,Yang:2026greedy}. Further results on online matching beyond the greedy algorithm include~\cite{Arndt:2026aa,Bansal:2014aa,Harada:2025aa,Kalyanasundaram:1993aa,Kalyanasundaram:2023aa,Khuller:1994aa,Meyerson:2006aa,Seth:2026aa}.

\section{Preliminaries}\label{sec:prelims}
A matching \emph{instance} is a tuple $\sigma = (A, B, \{ \succ_a \}_{a\in A} )$,
where $A$ is a set of agents, $B$ is a set of items, and for each agent $a \in A$, $\succ_a$ is a total linear order over the items. If an agent $a$ prefers an item $b_i$ over $b_j$, then we write $b_i \succ_a b_j$.
All instances will have $|A| = |B| = n$ for some $n \geq 1$, and our guarantees are always parametrized by $n$.

\paragraph{Metric spaces.} 
Distances are conveyed by an unknown \emph{pseudometric} \allowbreak $d:(A\cup B) \times (A\cup B) \to \mathbb{R}_{\geq 0}$. 
By definition, the function $d$ satisfies the following  properties.
\begin{enumerate}[label=(\roman*),noitemsep]
	\item \textit{Indiscernibility of Identicals:} $d(x, y) = 0$ if $x = y$,
	\item \textit{Symmetry:} $d(x, y) = d(y, x)$,
	\item \textit{Triangle inequality:} $d(x, y) \leq d(x, z) + d(z, y)$.
\end{enumerate}
Note that our definition allows collocation of distinct vertices.
Our choice is primarily for ease of exposition; any ties that appear in our
realized metrics can be broken with appropriate perturbations of $\eps$.
Thus, all our results also extend to metric spaces.

A metric $d$ is \emph{consistent} with an instance $\sigma$ if, for every agent $a\in A$ and all items $b_i,b_j\in B$, $b_i\succ_a b_j$ implies $d(a,b_i)\leq d(a,b_j)$.
We let $\rho(\sigma)$ denote the set of all metrics consistent with $\sigma$. 

\paragraph{Matchings.} A \emph{fractional matching} is any matrix indexed by $A$ and $B$ 
(e.g., $X = (X_{a,b})_{a \in A, b \in B}$) with nonnegative entries such that each row sum and each column sum is in $[0, 1]$.  
If all the row sums and column sums are equal to $1$, we say that the fractional matching is \emph{perfect};
otherwise, we say that it is \emph{partial}. When the perfect fractional matching is \emph{integral} (i.e., every entry is $0$ or $1$),
we denote it with a set $M$ of disjoint pairs $\{a,b\}$ with $a\in A$ and $b\in B$; whenever $\{a,b\}\in M$, we write
$M(a)=b$ and $M(b)=a$ for ease of exposition.

In this work, we consider mechanisms $\mathcal M(\sigma)$ that may use randomness
in conjunction with the rankings in $\sigma$ to output a matching. 
We identify each randomized matching mechanism  
with its induced distribution over perfect integral matchings.
We also write $M \sim \mathcal M(\sigma)$ when we refer to a perfect integral matching
sampled from this distribution.
Note that any perfect fractional
allocation can be implemented as a distribution over
perfect integral matchings, by the Birkhoff--von Neumann theorem
\citep{Birkhoff:1946aa,Von-Neumann:1953aa}.

\paragraph{Distortion.}
The \textit{cost} of an integral matching $M$ is the 
sum of assigned distances:
\begin{equation*}
\cost(M, d) \coloneqq \sum_{a \in A} d(a, M(a)).
\end{equation*}
The \textit{cost} of a fractional matching $X$ is given by the weighted sum $\cost(X, d) \coloneqq \sum_{a \in A} \sum_{b \in B} X_{a,b} \cdot d(a, b)$.
The \textit{distortion} of a perfect integral matching $M$ with respect to a metric $d$ is the ratio of that matching's cost compared to that of the optimal (integral) matching $M^*$:\footnote{Throughout, we interpret $0/0$ as $1$ and $c/0$ as $+\infty$ for $c>0$.}
\begin{equation*}
\distortion(M, d) \coloneqq \frac{\cost(M, d)}{\min_{M^*} \cost(M^*, d)}.
\end{equation*}
The worst-case distortion of a mechanism $\mathcal M$ on an instance $\sigma$ is given by 
\begin{equation*}
\distortion(\mathcal M, \sigma) 
\coloneqq \sup_{d \in \rho(\sigma)}
	\frac{   \E_{M \sim \mathcal M(\sigma)}
\left [  \cost(M, d) \right ] }{\min_{M^*} \cost(M^*, d)}.
\end{equation*}

\paragraph{Truthfulness.}
A deterministic mechanism is \emph{truthful} if no agent can ever get a better outcome by misreporting her preferences. In utilitarian settings, the standard way of extending truthfulness to randomized mechanisms is to demand that no agent can improve her utility (reduce her cost) in expectation by misreporting. In our metric distortion setting, this definition manifests as follows.

\begin{definition}
A mechanism $\mathcal M$ is \emph{truthful-in-expectation} if for every instance $\sigma$, agent $a\in A$, metric $d\in\rho(\sigma)$, and instance $\sigma'$ that differs from $\sigma$ only in $a$'s ranking,
\begin{equation}\label{eq:tie-property}
\E_{M\sim\mathcal M(\sigma)}[d(a,M(a))] \leq \E_{M\sim\mathcal M(\sigma')}[d(a,M(a))].
\end{equation}
\end{definition}

Another natural way to express truthfulness is through \emph{stochastic dominance}. Rather than comparing expected
distances, we can ask that misreporting does not increase an agent's probability of receiving one of her top $k$ items, for any $k$. 
\begin{definition}
A mechanism $\mathcal M$ is \emph{SD-strategyproof} if for every agent $a\in A$, instances $\sigma$ and $\sigma'$ that differ only in $a$'s ranking, and prefix $S$ of $a$'s ranking in $\sigma$,
\begin{equation}\label{eq:sd-strat-property}
\Pr_{M\sim\mathcal M(\sigma)}[M(a)\in S] \geq \Pr_{M\sim\mathcal M(\sigma')}[M(a)\in S].
\end{equation}
\end{definition}

Note that this definition has no reference to the underlying metric space, and purely operates on the ordinal preferences defined by $\sigma$. It is a well-known fact that SD-strategyproofness is equivalent to truthfulness-in-expectation when agents can have arbitrary utilities over items (see \citep[Section~4 and Definition~5]{Bogomolnaia:2001aa}). Essentially the same argument works under our metric constraints.

\begin{proposition}\label{prop:equiv}
A mechanism is truthful-in-expectation if and only if it is SD-strategyproof.
\end{proposition}

\begin{proof}
Suppose that $\mathcal M$ is truthful-in-expectation, and fix $\sigma$, $a$, $\sigma'$, and $S$ as in the definition of SD-strategyproofness. Consider a metric $d$ such that all distances between agents and items are $1$, except $d(a, b) = 2$ if $b \notin S$. Then we have that, $$\E_{M \sim \mathcal M(\sigma)}[d(a, M(a))] = 2 - \Pr_{M \sim \mathcal M(\sigma)}[M(a) \in S]$$
and the same holds for $\sigma'$. Thus, \eqref{eq:tie-property} implies \eqref{eq:sd-strat-property} and $\mathcal M$ is SD-strategyproof.

Now suppose that $\mathcal M$ is SD-strategyproof, and fix $\sigma, a, \sigma'$ and $d$ as in the definition of truthfulness-in-expectation. Let $S_t = \{b \in B : d(a,b) \leq t\}$. Since $d$ is consistent with $\sigma$, each $S_t$ is a prefix of $a$'s ranking (possibly empty). Writing the expectations as integrals over tails, \eqref{eq:tie-property} is equivalent to
$$\int_0^\infty \Pr_{M \sim \mathcal M(\sigma)}[M(a) \notin S_t] \dif t \leq \int_0^\infty \Pr_{M \sim \mathcal M(\sigma')}[M(a) \notin S_t] \dif t$$
which is clearly implied by \eqref{eq:sd-strat-property}. Thus, $\mathcal M$ is truthful-in-expectation.
\end{proof}

In \Cref{sec:truth}, we will stick to the truthful-in-expectation terminology, but we will make frequent use of the property \eqref{eq:sd-strat-property}. 

Another helpful fact is that a mechanism can be made \emph{anonymous} without worsening its distortion or losing truthfulness-in-expectation, simply by randomly permuting the labels of the agents. Here, anonymity means that relabeling the agents only relabels the output distribution. For example, swapping $a_i$ and $a_j$'s preferences swaps their distribution of items under the mechanism. In particular, agents with identical rankings get identical distributions over items, since swapping their preferences does not change the instance.

This trick is common in the literature.\footnote{See \cite[Lemma~3.6]{Fiat:2002aa} for the same argument used to make auctions symmetric.} For example, \cite[Lemma~1]{Filos-Ratsikas:2014aa} prove an analogous statement in the utilitarian distortion setting, and their proof applies equally to our metric setting. For completeness, we include a proof in our language below.

\begin{proposition}\label{prop:anonymization}
Given a truthful-in-expectation mechanism with distortion at most $D$ on every instance, we can construct an anonymous truthful-in-expectation mechanism with the same distortion bound.
\end{proposition}

\begin{proof}
Given a mechanism $\mathcal M$ and a permutation $\pi$ of the agents, consider the mechanism $\mathcal{M}_\pi$ which relabels the agents according to $\pi$, then runs $\mathcal M$, and finally restores the original labels in the output. Let $\widehat{\mathcal M}$ be the mechanism that runs  $\mathcal{M}_\pi$ with a uniformly random permutation $\pi$.

$\widehat{\mathcal M}$ is anonymous since composing a uniformly random permutation with any fixed relabeling gives another uniformly random permutation. Thus, relabeling the input only relabels the output distribution.

Moreover, since each $\mathcal M_\pi$ is equivalent to running $\mathcal M$ on a relabeled instance, it has distortion at most $D$ and is truthful-in-expectation. For fixed instances and metrics, the distortion bound and the truthfulness conditions are linear inequalities in the matching probabilities, so they are preserved by convex combinations with weights independent of the instance. Thus, $\widehat{\mathcal M}$ has distortion at most $D$ and is truthful-in-expectation.
\end{proof}

\paragraph{Pareto optimality.}
A perfect integral matching $M$ is
\emph{Pareto optimal} if there is no perfect integral matching $M'$
such that, for every agent $a$, either $M'(a)=M(a)$ or
$M'(a)\succ_a M(a)$, with $M'(a)\succ_a M(a)$ for at least one agent.
Thus, no reassignment can make some agent strictly better off
without making another agent worse off, according to the ordinal
rankings in $\sigma$.

A randomized mechanism
$\mathcal M$ is \emph{ex-post Pareto optimal} if, for every instance
$\sigma$, every matching in the support of $\mathcal M(\sigma)$ is
Pareto optimal. Here ``ex-post'' means that Pareto optimality holds
for each realized matching after the randomness is resolved.

\newcommand{\Partition}{\textsc{RandRepPartition}}
\newcommand{\Acal}{\mathcal A}
\newcommand{\ClusterCompletion}{\textsc{ClusterCompletion}}
\newcommand{\Match}{\textnormal{\textsc{OrdinalMatch}}}
\newcommand{\TopItems}{\operatorname{Top}}
\newcommand{\indeg}{\operatorname{indeg}}
\newcommand{\outdeg}{\operatorname{outdeg}}

\section{Sublinear Distortion via Cluster Completion}
\label{sec:rsd-rep}
In this section, we present a randomized mechanism with $O(n^{\log_4 3})$ distortion. This is the first sublinear metric distortion bound for matching. Combined with the derandomization
techniques of~\cite{Anari:2023aa,Haqi:2026aa}, this yields a
deterministic mechanism with $\widetilde O(n^{\log_4 3})$ distortion.

\begin{reptheorem}{thm:rsd+repmatch}
There is a randomized matching mechanism $\mathcal A$ such that, for
every instance $\sigma$ with $n$ agents,
\begin{equation*}
 \distortion(\Acal,\sigma)=O(n^{\log_4 3}).
\end{equation*}
\end{reptheorem}

The central observation behind this improvement is that sequential mechanisms usually incur most of their cost toward the end of their execution: an RSD prefix leaving $k$ agents unmatched has expected cost
only $O(n/k)$ times the optimum~\cite*[Theorem~4]{Anari:2023aa}. This is intuitive, since the last agents are left with very limited choice and are hence costly. In~\Cref{sec:flow-formulation}, we will see similar behavior for PS (\Cref{cor:prefix}). This observation naturally raises the question: can we complete a partial matching produced by an RSD or PS prefix at low additional cost?

Completing a partial matching can require changing earlier assignments: an item needed by an unmatched agent may already be assigned. To control the cost of these changes, we introduce the algorithm \ClusterCompletion. We partition the agents and items into clusters, each containing equally many agents and items, and bound
the sum of the largest agent--item distances within the clusters.
Augmenting paths then fill the missing mass using new pairs
within these clusters. Each path visits a cluster
at most once, so the cost bound charges each cluster once per unit
of completed mass. From a technical perspective, a partition with bounded total cluster cost serves as a coarse version of an approximately optimal matching: it controls the cost of arbitrary assignments within each cluster and guides the augmenting paths. We then construct a random partition inspired by the representative-and-overlap structure of \textsc{RepMatch}~\citep*{Anari:2023aa,Hastings:2025aa}.

\Cref{sec:completion} defines balanced partitions and proves the
completion bound. \Cref{sec:partition} constructs the partition
and bounds its expected cost. \Cref{sec:combination} applies the RSD
and PS prefix bounds and chooses the stopping point to prove
\Cref{thm:rsd+repmatch}.

Throughout the analysis, fix an instance $\sigma$, a consistent
metric $d\in\rho(\sigma)$, and an optimal matching $\Mopt$. Write
$\OPT=\cost(\Mopt,d)$.

\subsection{ClusterCompletion from a Balanced Partition}
\label{sec:completion}

We first show how to complete a partial fractional matching. Recall that a partial fractional matching is a nonnegative matrix $X$ indexed by $A\times B$ whose row and column sums are at most one. Let $k:=n-\sum_{a,b}X_{a,b}$ be the total \emph{unmatched mass} on each side, i.e., the fractional number of agents whose assignments remain to be completed. We complete $X$ using a balanced partition of the agents and items.

\begin{definition}[Balanced Partition]\label[definition]{def:partition}
A \emph{balanced partition} is a family
$\mathcal P=(A_j,B_j)_{j=1}^m$ such that $(A_j)_j$ partitions $A$,
$(B_j)_j$ partitions $B$, and $|A_j|=|B_j|$ for every $j$. For each $j$, we refer to $A_j \cup B_j$ as a \emph{cluster}.
For a fixed metric $d$, define the \emph{partition cost} of $\mathcal P$ by
\begin{equation*}
 W(\mathcal P,d)=\sum_{j=1}^m w_j,
 \qquad w_j=\max_{a\in A_j,\ b\in B_j}d(a,b).
\end{equation*}
\end{definition}

Let us provide some intuition for how to complete a partial matching with a balanced partition. For now, think of $X$ as an \emph{integral} partial matching. Our goal is to match an unmatched agent $a \in A_i$ to an item in $B_i$, within the same cluster. Of course, we might have to reassign matched pairs if all items in the cluster are occupied. In that case, we find an item $b$ in $B_i$ whose matching partner $a'$ belongs to another cluster $A_j$. We match $a$ to $b$ and reassign $a'$ to some item $b' \in B_j$. Thus, any reassigned agent is also matched within its cluster. We might need to repeat the above process if $B_j$ is also fully occupied, and the process continues until we reach an unused item. Importantly, we make sure that this alternating path visits each cluster only once, and hence the cost of matching one unmatched agent is $W(\mathcal P, d)$.

\Cref{alg:block-repair} makes the above idea concrete and works for general fractional partial matchings. When $X$ is not integral, we match an additional mass of $\delta$ at cost $\delta \cdot W(\mathcal P, d)$. To avoid visiting the same cluster more than once, we build an auxiliary graph and decompose the maximum flow on the graph into simple paths.

\begin{center}
\myalg{alg:block-repair}{$\ClusterCompletion(X,\mathcal P)$}{
\textbf{Input:} A partial fractional matching $X$ and a balanced
partition $\mathcal P=(A_j,B_j)_{j=1}^m$. \medskip \\
\textbf{Build the graph.} Initialize $M\gets X$, with entries
$m_{a,b}$. Use one vertex per cluster, a source $s$, and a sink $t$.
Add the following arcs, omitting those of weight zero.
\begin{itemize}[noitemsep]
\item For each pair $a\in A_i$, $b\in B_j$ with $i\ne j$, add an arc $j\to i$
  of weight $X_{a,b}$, labeled $\{a,b\}$.
\item For each agent $a\in A_i$, add an arc $s\to i$ of weight
  $u_a=1-\sum_bX_{a,b}$, labeled $a$.
\item For each item $b\in B_j$, add an arc $j\to t$ of weight
  $v_b=1-\sum_aX_{a,b}$, labeled $b$.
\end{itemize}
\textbf{Complete along paths.} While positive weight leaves $s$:
\begin{itemize}[noitemsep]
    \item Choose a simple positive-weight path $s,j_0,\ldots,j_r,t$. Call its endpoint labels $a_0,b_{r+1}$ and its old-pair labels $\{a_h,b_h\}$, $1\le h\le r$, in path order.
    \item Let $\delta$ be the smallest arc weight on this path.
    \item Remove $\delta$ from each old pair $\{a_h,b_h\}$ in $M$ and add
$\delta$ to each within-cluster pair $\{a_h,b_{h+1}\}$, for
$0\le h\le r$. 
    \item Decrease every arc weight on the path by $\delta$.
\end{itemize}
\textbf{Return} $M$.
}
\end{center}

An old pair $\{a,b\}$ with $a\in A_i$, $b\in B_j$, and $i\ne j$
gives an arc $j\to i$: assigning mass of item $b$ within cluster $j$
releases the same amount at agent $a$ in cluster $i$. Source arcs record
agent deficits, and sink arcs record item deficits. The arc
weights form a flow, from which the procedure extracts simple
source--sink paths.

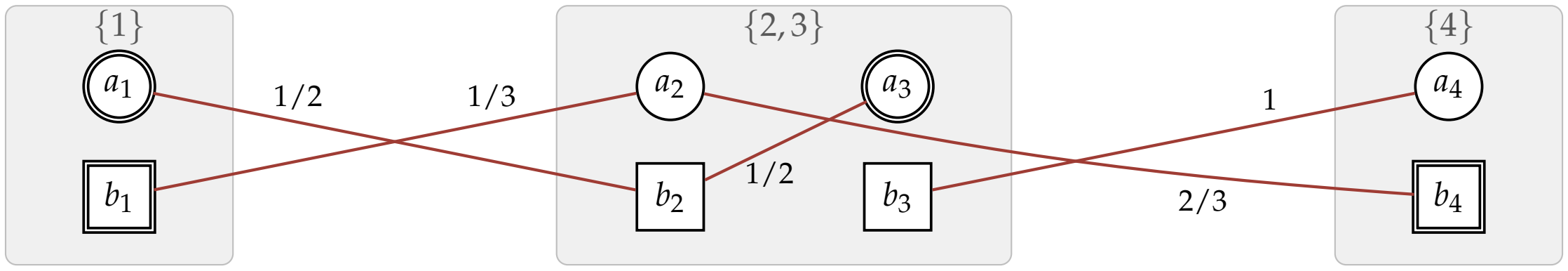
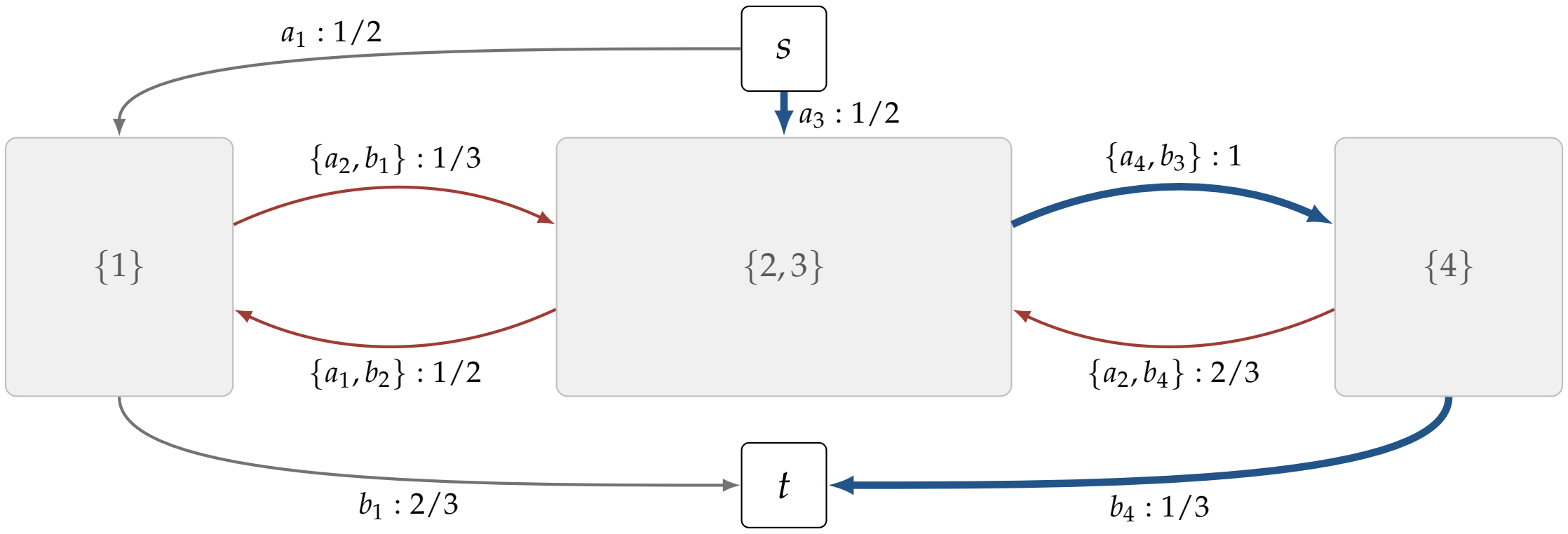
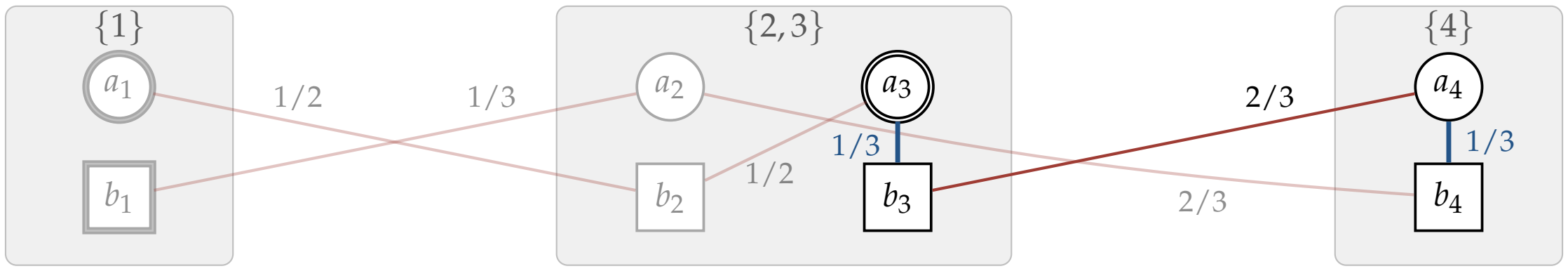
\begin{figure}[!htb]
\centering
\definecolor{ccred}{RGB}{159,60,52}
\definecolor{ccblue}{RGB}{35,84,136}
\captionsetup{skip=4pt}
\captionsetup[subfigure]{skip=2pt}
\tikzset{
  example agent/.style={circle,draw=black,fill=white,
    minimum size=6.3mm,inner sep=1pt,line width=.7pt},
  example item/.style={rectangle,draw=black,fill=white,
    minimum size=6.3mm,inner sep=1pt,line width=.7pt},
  example endpoint/.style={rounded corners=2pt,draw=black,
    fill=white,minimum size=8mm,inner sep=3pt,font=\normalsize},
  example link/.style={draw=ccred,line width=.8pt},
  example unchanged/.style={example link,draw opacity=.3,text opacity=.45},
  example inactive/.style={draw opacity=.3,text opacity=.45},
  example arc/.style={-{Latex[length=1.7mm]},draw=ccred,line width=.8pt},
  example deficit/.style={-{Latex[length=1.7mm]},draw=black!55,line width=.8pt},
  example selected/.style={-{Latex[length=2.5mm]},draw=ccblue,line width=2pt},
  example new/.style={draw=ccblue,line width=1.3pt},
  example label/.style={fill=none,draw=none,inner sep=1pt,font=\scriptsize}
}

\begin{subfigure}[t]{\textwidth}
\centering
\begin{tikzpicture}[x=1.65cm,y=.8cm,font=\footnotesize]
  \path[use as bounding box] (-.15,-1.65) rectangle (9.05,1.85);
  \draw[rounded corners=3pt,draw=black!25,fill=black!6]
    (0,-1.45) rectangle (1.3,1.6);
  \draw[rounded corners=3pt,draw=black!25,fill=black!6]
    (3.15,-1.45) rectangle (5.75,1.6);
  \draw[rounded corners=3pt,draw=black!25,fill=black!6]
    (7.6,-1.45) rectangle (8.9,1.6);
  \node[text=black!65] at (.65,1.35) {$\{1\}$};
  \node[text=black!65] at (4.45,1.35) {$\{2,3\}$};
  \node[text=black!65] at (8.25,1.35) {$\{4\}$};

  \node[example agent,double,double distance=.5pt]
    (befa1) at (.65,.65) {$a_1$};
  \node[example item,double,double distance=.5pt]
    (befb1) at (.65,-.65) {$b_1$};
  \node[example agent] (befa2) at (3.8,.65) {$a_2$};
  \node[example item] (befb2) at (3.8,-.65) {$b_2$};
  \node[example agent,double,double distance=.5pt]
    (befa3) at (5.1,.65) {$a_3$};
  \node[example item] (befb3) at (5.1,-.65) {$b_3$};
  \node[example agent] (befa4) at (8.25,.65) {$a_4$};
  \node[example item,double,double distance=.5pt]
    (befb4) at (8.25,-.65) {$b_4$};

  \draw[example link] (befa1) --
    node[example label,pos=.3,above=2pt] {$1/2$} (befb2);
  \draw[example link] (befa2) --
    node[example label,pos=.3,above=2pt] {$1/3$} (befb1);
  \draw[example link] (befa2) to[bend right=4]
    node[example label,pos=.72,below=2pt] {$2/3$} (befb4);
  \draw[example link] (befa3) --
    node[example label,pos=.60,below=2pt] {$1/2$} (befb2);
  \draw[example link] (befa4) --
    node[example label,pos=.30,above=2pt] {$1$} (befb3);
\end{tikzpicture}
\caption{The partition and initial matching $X$, with unmatched mass $k=1$.}
\label{fig:partition-example}
\end{subfigure}

\vspace{3pt}
\begin{subfigure}[t]{\textwidth}
\centering
\begin{tikzpicture}[x=1.65cm,y=.8cm,font=\footnotesize,
  flow block/.style={rectangle,rounded corners=3pt,
    draw=black!25,fill=black!6,text=black!65,
    minimum height=2.44cm,inner sep=0pt}]
  \path[use as bounding box] (-.15,-3.1) rectangle (9.05,3.25);

  \node[flow block,minimum width=2.145cm]
    (exj1) at (.65,.075) {$\{1\}$};
  \node[flow block,minimum width=4.29cm]
    (exj2) at (4.45,.075) {$\{2,3\}$};
  \node[flow block,minimum width=2.145cm]
    (exj3) at (8.25,.075) {$\{4\}$};
  \node[example endpoint] (exs) at (4.45,2.65) {$s$};
  \node[example endpoint] (ext) at (4.45,-2.5) {$t$};

  \draw[example deficit] (exs.west)
    .. controls (2.7,2.65) and (.65,2.65) ..
    node[example label,pos=.5,above=2pt] {$a_1:1/2$}
    (exj1.north);
  \draw[example selected] (exs.south) --
    node[example label,right=2pt] {$a_3:1/2$} (exj2.north);

  \draw[example arc] ([yshift=4mm]exj1.east)
    .. controls (1.9,1.15) and (2.55,1.15) ..
    node[example label,above=2pt] {$\{a_2,b_1\}:1/3$}
    ([yshift=4mm]exj2.west);
  \draw[example arc] ([yshift=-4mm]exj2.west)
    .. controls (2.55,-1) and (1.9,-1) ..
    node[example label,below=2pt] {$\{a_1,b_2\}:1/2$}
    ([yshift=-4mm]exj1.east);
  \draw[example selected] ([yshift=4mm]exj2.east)
    .. controls (6.35,1.15) and (7,1.15) ..
    node[example label,above=2pt] {$\{a_4,b_3\}:1$}
    ([yshift=4mm]exj3.west);
  \draw[example arc] ([yshift=-4mm]exj3.west)
    .. controls (7,-1) and (6.35,-1) ..
    node[example label,below=2pt] {$\{a_2,b_4\}:2/3$}
    ([yshift=-4mm]exj2.east);

  \draw[example deficit] (exj1.south)
    .. controls (.65,-2.5) and (2.7,-2.5) ..
    node[example label,pos=.6,below=2pt] {$b_1:2/3$} (ext.west);
  \draw[example selected] (exj3.south)
    .. controls (8.25,-2.5) and (6.2,-2.5) ..
    node[example label,pos=.6,below=2pt] {$b_4:1/3$} (ext.east);
\end{tikzpicture}
\caption{Cluster graph. Each shaded block is one flow vertex;
  thick blue arrows mark a selected simple path.}
\label{fig:block-graph-example}
\end{subfigure}

\vspace{3pt}
\begin{subfigure}[t]{\textwidth}
\centering
\begin{tikzpicture}[x=1.65cm,y=.8cm,font=\footnotesize]
  \path[use as bounding box] (-.15,-1.65) rectangle (9.05,1.85);
  \draw[rounded corners=3pt,draw=black!25,fill=black!6]
    (0,-1.45) rectangle (1.3,1.6);
  \draw[rounded corners=3pt,draw=black!25,fill=black!6]
    (3.15,-1.45) rectangle (5.75,1.6);
  \draw[rounded corners=3pt,draw=black!25,fill=black!6]
    (7.6,-1.45) rectangle (8.9,1.6);
  \node[text=black!65] at (.65,1.35) {$\{1\}$};
  \node[text=black!65] at (4.45,1.35) {$\{2,3\}$};
  \node[text=black!65] at (8.25,1.35) {$\{4\}$};

  \node[example agent,double,double distance=.5pt,example inactive]
    (afta1) at (.65,.65) {$a_1$};
  \node[example item,double,double distance=.5pt,example inactive]
    (aftb1) at (.65,-.65) {$b_1$};
  \node[example agent,example inactive] (afta2) at (3.8,.65) {$a_2$};
  \node[example item,example inactive] (aftb2) at (3.8,-.65) {$b_2$};
  \node[example agent,double,double distance=.5pt]
    (afta3) at (5.1,.65) {$a_3$};
  \node[example item] (aftb3) at (5.1,-.65) {$b_3$};
  \node[example agent] (afta4) at (8.25,.65) {$a_4$};
  \node[example item] (aftb4) at (8.25,-.65) {$b_4$};

  \draw[example unchanged] (afta1) --
    node[example label,pos=.3,above=2pt] {$1/2$} (aftb2);
  \draw[example unchanged] (afta2) --
    node[example label,pos=.3,above=2pt] {$1/3$} (aftb1);
  \draw[example unchanged] (afta2) to[bend right=4]
    node[example label,pos=.72,below=2pt] {$2/3$} (aftb4);
  \draw[example unchanged] (afta3) --
    node[example label,pos=.60,below=2pt] {$1/2$} (aftb2);
  \draw[example link] (afta4) --
    node[example label,pos=.30,above=2pt] {$2/3$} (aftb3);
  \draw[example new] (afta3.south) --
    node[example label,pos=.6,left=3pt,text=ccblue] {$1/3$}
    (aftb3.north);
  \draw[example new] (afta4.south) --
    node[example label,pos=.5,right=3pt,text=ccblue] {$1/3$}
    (aftb4.north);
\end{tikzpicture}
\caption{After one $1/3$-mass augmentation.
  Dim nodes and edges are unaffected by the augmentation.}
\label{fig:after-repair}
\end{subfigure}

\caption{A $1/3$-mass augmentation: the initial matching, its cluster
graph with the selected simple path, and the updated matching.
The exchange removes $1/3$ from $\{a_4,b_3\}$ and adds $1/3$
to each of $\{a_3,b_3\}$ and $\{a_4,b_4\}$.
Circles labeled $a_i$ are agents
and squares labeled $b_i$ are items; double outlines mark positive
unmatched mass.}
\label{fig:partition-graph}
\end{figure}

Each exchange assigns $\delta$ units of item mass to every agent on its
path. Every agent after the first gives up the same amount of
its old assignment, so its total assigned mass stays unchanged. 
The exchange fills $\delta$ of the deficits at the first agent
$a_0$ and the final item $b_{r+1}$. Furthermore, both endpoints of every newly assigned pair belong to the same cluster.
\Cref{fig:partition-graph} illustrates one such exchange.

\begin{lemma}[Completion from a Balanced Partition]\label[lemma]{lem:repair}
Given a partial fractional matching $X$ with unmatched mass $k$
and a balanced partition $\mathcal P$, \Cref{alg:block-repair}
uses only $X$ and the cluster memberships to return a perfect
fractional matching $M$ satisfying
\begin{equation}\label{eq:repair-cost}
 \cost(M,d)\le\cost(X,d)+\sum_j\min\{k,|A_j|\}w_j
          \le\cost(X,d)+kW(\mathcal P,d)
\end{equation}
for every $d\in\rho(\sigma)$.
\end{lemma}
\begin{proof}
We first prove that the procedure returns a perfect fractional
matching. Let $u_j=\sum_{a\in A_j}u_a$ and
$v_j=\sum_{b\in B_j}v_b$. Write $\indeg(j)$ and $\outdeg(j)$
for the initial total weights of old arcs entering and leaving
cluster $j$. Assignments within a cluster contribute equally to its
agent and item totals, so their contributions cancel. Since $|A_j|=|B_j|$, we obtain $u_j+\indeg(j)=v_j+\outdeg(j)$ for all $j$. Thus, the arc weights form a flow of value $k$.
Flow conservation ensures that a positive-weight source--sink
path exists whenever positive weight leaves the source.
Extracting a path preserves conservation in the residual graph
and sets at least one arc weight to zero. The procedure therefore
terminates with all source and sink arcs exhausted. The extracted path weights
sum to $k$.

Each exchange preserves the assigned mass of its internal agents
and items and fills $\delta$ of each endpoint's deficit. A path
with a single cluster assigns $\delta$ directly to its endpoint
pair. Residual arc weights ensure that the mass removed from any
old pair never exceeds its mass in $X$. Hence $M$ is nonnegative.
Exhausting the source and sink arcs makes every row and column
sum to one.

Each simple path adds mass at most once per cluster. Since the path
weights sum to $k$, the total new mass in cluster $j$ is at most $k$.
It is also at most $|A_j|$: the final matching has total mass
$|A_j|$ in its rows, and its retained old mass is nonnegative.
Each added unit in cluster $j$ costs at most $w_j$. Summing these
charges and discarding the nonnegative cost of removed old mass
proves \eqref{eq:repair-cost}.
\end{proof}

\subsection{Constructing a Partition via Randomized RepMatch}
\label{sec:partition}

We now construct a balanced partition with expected partition cost at most
$2n^\theta\OPT$, where $\theta=\log_2(3/2)$. We use the
representative-and-overlap framework of the \textsc{RepMatch}
algorithm~\cite{Anari:2023aa, Hastings:2025aa}, with randomized
representative selection.

Let $\TopItems(a,s)$ denote the first $s$ items in agent $a$'s
ranking in $\sigma$. Each current agent cluster $S\subseteq A$ has a representative $r_S\in S$
and claims $\TopItems(r_S,|S|)$. Start with singleton clusters. Whenever two
claimed prefixes overlap, merge their clusters and retain one
representative. If their sizes are $p\ge q$,
keep the smaller cluster's representative with probability
\begin{equation*}
 z(p,q)=\left(1+\frac qp\right)^\theta-1.
\end{equation*}

\begin{center}
\myalg{alg:partition}{$\Partition(\sigma)$}{
\textbf{Input:} An ordinal instance $\sigma$. \medskip \\
Start with singleton clusters $\{a\}$, each represented by
  $a$ and claiming its favorite item.
While two claimed prefixes overlap:
\begin{itemize}[noitemsep]
\item Choose such clusters
  $S,T$ by a fixed rule, with $p=|S|\ge q=|T|$. Merge them,
  keeping $r_T$ with probability $z(p,q)$ and $r_S$ otherwise.
\item The retained representative claims its first $p+q$ items.
\end{itemize}
When the claimed prefixes are disjoint, enumerate the final agent
  clusters as $A_1,\ldots,A_m$. Set
  $B_j=\TopItems(r_{A_j},|A_j|)$ for each $j$
  and return $\mathcal P=(A_j,B_j)_{j=1}^m$.
}
\end{center}

When $p=q$, the representative is chosen by a fair coin; when
$p>q$, the larger cluster's representative is retained with
probability greater than $1/2$. To see the role of $\theta$, suppose two clusters each have the same size
$p$, with all their agents and optimal items within distance $x$
of their representatives. A shared item places the representatives within distance $2x$,
so the merged cluster has radius at most $3x$ around either
representative. Thus, two clusters of size $p$ and radius $x$ are merged into one cluster of size $2p$ and radius $3x$. When $2^\theta=3/2$, we have $2x/p^\theta = 3x/(2p)^\theta$, so the total radius normalized by size to the power of $\theta$ remains the same. For unequal cluster sizes, the choice $z(p,q)$ makes the normalized sum nonincreasing in expectation.

\begin{lemma}[Partition Bound]\label[lemma]{lem:partition}
\Cref{alg:partition} returns a balanced partition $\mathcal P$ such that, for all $d\in \rho(\sigma)$,
\begin{equation*}
 \E[W(\mathcal P,d)]\le2n^\theta\OPT.
\end{equation*}
\end{lemma}
\begin{proof}
At termination, the claimed prefixes are disjoint and their lengths
sum to $n$, so they partition $B$ and the output is balanced.
For each current cluster $S$, let
$\Mopt(S)=\{\Mopt(a):a\in S\}$ and define the radius
\begin{equation*}
 D_S=\max_{x\in S\cup\Mopt(S)}d(r_S,x).
\end{equation*}
This ball contains the agents in $S$ and their $|S|$ distinct
optimal items. The representative's first $|S|$ items therefore
also lie within $D_S$. Since each agent is within $D_S$ of the
representative, every within-cluster pair has length at most $2D_S$.

To bound the sum of the final radii, track the normalized sum
\begin{equation*}
 \Phi=\sum_S D_S/|S|^\theta.
\end{equation*}
Condition on the history before the next merge of clusters $S,T$,
with $p=|S|\ge q=|T|$. A shared claimed item places their representatives
at distance at most $D_S+D_T$. If $r_S$ is retained, every point
of $T\cup\Mopt(T)$ is within $D_T+(D_S+D_T)$ of $r_S$, while
the points of $S\cup\Mopt(S)$ remain within $D_S$.
The new radius is therefore at most $D_S+2D_T$.
Symmetrically, retaining $r_T$ gives a radius of at most $2D_S+D_T$.
Writing $z=z(p,q)$, the conditional expected radius after the merge is at most
\begin{equation*}
 (1-z)(D_S+2D_T)+z(2D_S+D_T)
 =(1+z)D_S+(2-z)D_T.
\end{equation*}

The definition of $z$ gives
$(1+z)=(p+q)^\theta/p^\theta$.
To bound the other coefficient, convexity of $x\mapsto x^{-\theta}$
gives
\begin{equation*}
 p^{-\theta}+q^{-\theta}
 \ge2\left(\frac{p+q}{2}\right)^{-\theta}
 =\frac{3}{(p+q)^\theta}.
\end{equation*}
Multiplying by $(p+q)^\theta$ and rearranging yields
\begin{equation*}
 2-z=3-\frac{(p+q)^\theta}{p^\theta}
 \le\frac{(p+q)^\theta}{q^\theta}.
\end{equation*}

Thus
\begin{equation*}
 \E\!\left[
 \frac{D_{S\cup T}}{(p+q)^\theta}\,\middle|\,\text{history}\right]
 \le\frac{(1+z)D_S+(2-z)D_T}{(p+q)^\theta}
 \le\frac{D_S}{p^\theta}+\frac{D_T}{q^\theta},
\end{equation*}
and $\Phi$ is a supermartingale. Let $\Phi_0$ and
$\Phi_{\mathrm{final}}$ denote the initial and final potentials.
Each singleton has radius $D_{\{a\}}=d(a,\Mopt(a))$. Iterating the
conditional bound over the at most $n-1$ merges gives
\begin{equation*}
 \E[\Phi_{\mathrm{final}}]\le\Phi_0=\sum_a d(a,\Mopt(a))=\OPT.
\end{equation*}
Since $w_j\le2D_{A_j}$ and $|A_j|\le n$, we have
\begin{equation*}
 W(\mathcal P,d)\le2\sum_jD_{A_j}
 \le2n^\theta\sum_j\frac{D_{A_j}}{|A_j|^\theta}.
\end{equation*}
Taking expectations proves the claim.
\end{proof}

\subsection{Prefixes from RSD and PS}
\label{sec:combination}

It remains to choose a prefix rule and its stopping point. By \Cref{lem:repair,lem:partition},
completing a prefix with unmatched mass $k$ adds expected cost at
most $2kn^\theta\OPT$. We combine this bound with the RSD and PS prefix costs, then choose $k$ to balance the two terms. We first construct $\mathcal P\gets\Partition(\sigma)$.

\paragraph{RSD Prefix.}
Random serial dictatorship (RSD) processes agents in a uniformly
random order and assigns each agent her favorite available item. Its
prefix leaving $k$ agents unmatched satisfies the following bound.

\begin{lemma}[Truncated RSD; {\cite[Theorem~4]{Anari:2023aa}}]
\label[lemma]{lem:rsd}
For every integer $1\le k\le n$, the RSD prefix leaving $k$ agents
and $k$ items unmatched has expected cost at most $(n-k)\OPT/(k+1)$.
\end{lemma}

For an integer $1\le k\le n$, run truncated RSD for $n-k$ agents to
obtain $X$, and return $M=\ClusterCompletion(X,\mathcal P)$.
Combining the completion bound with \Cref{lem:rsd,lem:partition}
gives
\begin{equation}\label{eq:reduction}
 \E[\cost(M,d)]
 \le\left(\frac{n-k}{k+1}+2kn^\theta\right)\OPT.
\end{equation}

\paragraph{PS Prefix.}
Under probabilistic serial (PS), every agent consumes her favorite
available item at unit rate. Let $X(t)$ denote the partial fractional matching obtained by running PS for $t$ units of time. Every row of $X(t)$ sums to $t$. Thus, for any real $0<k\le n$,
stopping at time $1-k/n$ leaves unmatched mass $k$.
We will prove in \Cref{cor:prefix} that this prefix satisfies
\begin{equation*}
 \cost(X(1-k/n),d)\le\frac{n-k}{k}\OPT.
\end{equation*}
Complete this
prefix using $M=\ClusterCompletion(X(1-k/n),\mathcal P)$. By
\Cref{lem:repair,lem:partition}, the resulting perfect fractional
matching satisfies
\begin{equation}\label{eq:ps-reduction}
 \E[\cost(M,d)]
 \le\left(\frac{n-k}{k}+2kn^\theta\right)\OPT.
\end{equation}

\paragraph{Choosing the Prefix Length.}
In \eqref{eq:reduction} and \eqref{eq:ps-reduction}, stopping earlier
decreases the bound on the prefix cost and increases the bound on the completion cost.
Choose $k$ of order $n^{(1-\theta)/2}$ to balance the terms $n/k$
and $kn^\theta$.

\begin{proof}[Proof of \Cref{thm:rsd+repmatch}]
Write $\gamma=(1-\theta)/2$. For the RSD construction, take
$k=\lceil n^\gamma\rceil$. Then
$n^\gamma\le k\le2n^\gamma$, and \eqref{eq:reduction} gives
\begin{equation*}
 \E[\cost(M,d)]
 \le\left(\frac nk+2kn^\theta\right)\OPT
 \le5n^{(1+\theta)/2}\OPT.
\end{equation*}
For the PS construction, take $k=n^\gamma$ in
\eqref{eq:ps-reduction}. The completed PS matching satisfies
\begin{equation*}
 \E[\cost(M,d)]
 \le\left(\frac nk+2kn^\theta\right)\OPT
 =3n^{(1+\theta)/2}\OPT.
\end{equation*}
Note that $(1+\theta)/2 = (1 + \log_2{\frac 3 2})/2 = \log_4 3$.
Both constructions use only the ordinal instance $\sigma$.
Since $d\in\rho(\sigma)$ was arbitrary, each defines a randomized
matching mechanism with $O(n^{\log_4 3})$ distortion.
\end{proof}

\section{Sublinear Distortion by Coordinated Merging}

\label{sec:coordinated-merging}

In this section, we provide a randomized mechanism with distortion $O(n^\alpha)$, where $\alpha \approx 0.95$, with a weaker guarantee than that of \Cref{thm:rsd+repmatch}. However, our proposed algorithm uses a different set of ideas that {we believe} are interesting on their own.

To motivate our algorithm, we first recall the deterministic \emph{RepMatch} mechanism of Anari, Charikar, and Ramakrishnan \cite{Anari:2023aa}, using the cluster notation adopted throughout this section. At each iteration, the input to the algorithm is as follows:

\begin{itemize}
    \item $k$ clusters of agents with sizes $\ell_1, \ell_2, \dots, \ell_k$, as in \cite{Anari:2023aa}. We use the terms \emph{size} and \emph{demand} interchangeably for clusters throughout this section.
    \item For each cluster $1 \leq i \leq k$, a preference list $B_i$ of size $\ell_i$ consisting of the top $\ell_i$ items ranked by its \emph{representative}, a designated agent in cluster $i$. All preferences and distances for the cluster refer to this representative.
\end{itemize}

At each iteration, \emph{RepMatch} selects an arbitrary pair of clusters whose preferred item sets overlap and merges the smaller cluster into the larger one. If no such pair exists, the algorithm assigns each cluster its preferred items and terminates.

Our algorithm combines the merging idea of \emph{RepMatch} with a few other ideas. It uses the global structure of conflicts to coordinate several merges at once. We represent these conflicts by a graph and use the sizes of its maximal independent sets to guide the algorithm.

\begin{definition}[Conflicting preference graph]\label{def:conflicting-preference-graph}
Consider a current instance with $k$ clusters with sizes $\ell_1,\dots,\ell_k$. For each cluster $i$, let $B_i$ be the set of its $\ell_i$ favorite available items. The \emph{conflicting preference graph} is the simple undirected graph $\mathcal G=(V,E)$ with
\[
V=\{1,\dots,k\},\qquad
E=\bigl\{\{i,j\}:1\leq i<j\leq k,\ B_i\cap B_j\neq\varnothing\bigr\}.
\]
\end{definition}

We will analyze the algorithm by induction on the number of clusters. The size of a maximal independent set in $\mathcal G$\footnote{More precisely, after the tentative allocation, the algorithm computes an MIS in the subgraph of $\mathcal G$ induced by clusters with positive residual demand and adds the isolated vertices of $\mathcal G$ to obtain $S$. This set need not be maximal in the full graph. We refer to $S$ as an MIS for convenience; the analysis does not require maximality in $\mathcal G$.} suggests two ways to reduce this number while controlling the cost:

\begin{itemize}
    \item $\mathcal{G}$ has a large maximal independent set (MIS): We can assign the clusters in the MIS their favorite items at low cost, removing many clusters and leaving a smaller residual instance.
    
    \item $\mathcal{G}$ has a small MIS: We can merge all the clusters into the MIS at low cost, leaving an instance with few clusters.
\end{itemize}

These two cases give the intuition behind the algorithm. The full procedure combines the two operations using carefully chosen probabilities and additional steps to make the induction work.

Before presenting the algorithm, we establish a simple bound on the optimal cost of the residual instance after a partial allocation to the clusters.

\begin{lemma} \label{lemma:residual-optimal-value}
    Consider a current instance with $k$ clusters with demands $\ell_1, \ell_2, \dots, \ell_k$ with optimal value $\OPT$. For each cluster $i$, let $P_i$ be a partial assignment to cluster $i$ such that $P_i \cap P_j = \emptyset$ for $1 \leq i < j \leq k$. Then the optimal value of the residual instance $\OPT_{\text{res}}$ after associating the set $P_i$ with cluster $i$ is at most:
    \[
    \OPT + \sum_{i = 1}^k \sum_{b \in P_i} d(i, b).
    \]
\end{lemma}
\begin{proof}
Apply the flow-decomposition argument in the proof of \Cref{lem:transport}, replacing each cluster $i$ with $\ell_i$ co-located agents and using $(P_i)$ as the partial assignment. Integrality of the flow gives a residual allocation of cost at most $\OPT+\sum_i\sum_{b\in P_i}d(i,b)$.
\end{proof}

\subsection*{Sublinear Distortion via Coordinated Merging} \label{section:single-layer-algorithm}

\begin{reptheorem}{thm:sing-layer-sublinear-distortion}
    There exists an algorithm $\mathcal{A}$ such that for any ordinal preference instance $\sigma$,
        $$
            \distortion(\mathcal A, \sigma) \leq 2n^{\alpha} - 1
        $$
    where $\alpha \approx 0.95084$. The algorithm strengthens the merging approach of \emph{RepMatch} by coordinating merges through independent sets.
\end{reptheorem}

With a suitable choice of parameters $\alpha$ and $\beta$, \Cref{alg:single-layered-matching} achieves the distortion bound in \Cref{thm:sing-layer-sublinear-distortion}. The algorithm begins by running \Cref{alg:trial-weighted-pivoting} to obtain:
\begin{itemize}
    \item A maximal independent set (MIS) $S$.
    \item A tentative allocation $X_i$ for each cluster $i$. \footnote{The reason we refer to $X_i$ as a tentative allocation is that it is going to be used only in the merging sub-procedure.}
    \item A merging map $\operatorname{ctr}: \text{cluster} \rightarrow S$.
\end{itemize}
The algorithm then flips a coin, with probabilities determined by the size of $S$, to choose between the following two procedures:

\begin{itemize}
    \item \emph{Serve}: Choose $J$ from $\{S\}\cup\{\{i\}:i\in V\setminus S\}$ according to a weighted lottery (a distribution). Assign each cluster in $J$ its favorite items (see \Cref{def:serving-independent-set}), then recursively solve the residual instance.
    \item \emph{Merge}: Give each cluster $i$ its tentative allocation, then merge the clusters into $S$ according to $\operatorname{ctr}$. Recursively solve the merged instance and distribute the representatives' items back to the original clusters.
\end{itemize}

We leave $\alpha$ and $\beta$ unspecified until the very end (see \Cref{def:single-layer-critical-parameters} for exact values).
The rest of this section is organized as follows:
\begin{itemize}
    \item In \Cref{subsection:MIS-and-tentative-allocation}, we present the construction of the MIS, the tentative allocations, and the merging map $\operatorname{ctr}$.
    \item In \Cref{susbsubsection:serving-analysis}, we describe the weighted lottery for choosing either $S$ or a single cluster outside $S$. We analyze the cost of assigning the selected clusters their favorite items and show how a large independent set improves the bound.
    \item In \Cref{subsubsection:mergingAnalysis}, we present the procedure for merging clusters into $S$ using $\operatorname{ctr}$ and analyze its cost.
\end{itemize}

We now introduce the definitions and notation used in the analysis.

\begin{definition}[Item multiplicity]\label{def:prefix-multiplicity}
For each currently available item $b$, let
\[
h_b\coloneqq\bigl|\{i\in V:b\in B_i\}\bigr|
\]
be the number of clusters that include $b$ among their preferred items.
\end{definition}

\begin{definition}[Private items]\label{def:private-items}
An item $b$ is \emph{private} if $h_b=1$. The private items of cluster $i$ form the set
\[
\text{Private}_i\coloneqq\{b\in B_i:h_b=1\}.
\]
\end{definition}

\begin{definition}[Shared items]\label{def:shared-items}
An item $b$ is \emph{shared} if $h_b\geq2$. The shared items of cluster $i$ form the set
\[
C_i\coloneqq\{b\in B_i:h_b\geq2\}=B_i\setminus\text{Private}_i.
\]
\end{definition}
\begin{notation}[Costs]\label{not:prefix-costs}
Let $\OPT$ denote the optimum allocation cost of the current instance, and let $F(t)=2t^\alpha$ for $\alpha\in(0,1)$ to be determined later.
Let $A$ be the total private cost, defined more concretely as follows:
\begin{align*}
    A&\coloneqq\sum_{i\in V}\sum_{b\in \text{Private}_i}d(i,b).
\end{align*}
Let $G$ be the total shared cost, and for any set of clusters $J\subseteq V$, let $g(J)$ be the cost of their preferred items:
\begin{align*}
G&\coloneqq\sum_{i\in V}\sum_{b\in C_i}d(i,b),
& g(J) \coloneqq\sum_{i\in J}\sum_{b\in B_i}d(i,b).
\end{align*}
Then $g(V)=A+G$.
\end{notation}

\subsection{Maximal Independent Set and Tentative Allocations} \label{subsection:MIS-and-tentative-allocation}

\Cref{alg:trial-weighted-pivoting} constructs a maximal independent set and a merging map for the two procedures. The main algorithm uses a coin flip to choose between merging and assigning preferred items, with probabilities determined by the independent set's size and the algorithm's parameters.

The algorithm initializes the tentative allocation $X_i$ of each cluster $i$ to $\text{Private}_i$, the set of its private preferred items. It then tentatively assigns each shared item $b$ to one of the $h_b$ clusters that prefer it, choosing each with probability $\frac{1}{h_b}$.

Each cluster $i$ now has residual weight $v_i = \ell_i - |X_i|$, and these weights determine the distribution over maximal independent sets. When all $v_i$'s are equal, the procedure reduces to randomized greedy MIS: it samples a uniformly random permutation of the vertices and greedily adds vertices in that order while preserving independence. In general, it implements a weighted version of this procedure. Variants of this approach have been used for different purposes in \emph{Random PIVOT} \cite{Fischer:2025aa}, \emph{KWIKCLUSTER} \cite{Ailon:2008aa} and \emph{Max} \cite{Boppana:2020aa}.

The MIS stage of \Cref{alg:trial-weighted-pivoting} uses $v_i$ as the weight of each vertex (cluster) $i$. At each step, it chooses a vertex $j$ from the active set $R$ with probability $\frac{v_j}{\sum_{r \in R} v_r}$. It adds $j$ to the MIS and removes $j$ and its active neighbors from $R$, repeating until $R$ is empty. The merging map sends each selected vertex to itself and each removed neighbor to the selected vertex that removed it.

\begin{center}
\myalg{alg:trial-weighted-pivoting}{Uniform item association and MIS generation}{
\textbf{Input:} Clusters $V=\{1,\dots,k\}$ with demands $(\ell_i)_{i\in V}$, preferred item sets $(B_i)_{i\in V}$, conflicting preference graph $\mathcal G=(V,E)$, and sets $(\text{Private}_i,C_i)_{i\in V}$. \medskip \\
\textbf{Tentative allocation.}
\begin{itemize}[noitemsep]
\item Initialize $X_i=\text{Private}_i$ for every cluster $i$.
\item Independently assign each shared item $b\in\bigcup_{i\in V}C_i$ to one of the clusters in $\{a\in V:b\in B_a\}$, choosing each with probability $1/h_b$.
Add $b$ to that cluster's tentative allocation.
\item Set $v_i=\ell_i-|X_i|$ for each $i$.
\item Initialize $R=\{i\in V:v_i>0\}$ and $I=\varnothing$.
\end{itemize}
\textbf{Choose representatives.} While $R$ is nonempty:
\begin{itemize}[noitemsep]
\item Choose a cluster $j\in R$ with probability $v_j/\sum_{a\in R}v_a$.
\item Add $j$ to $I$.
\item Let $N_j=\{j\}\cup\{i\in R:\{i,j\}\in E\}$.
\item Set $\operatorname{ctr}(i)=j$ for every $i\in N_j$.
\item Remove $N_j$ from $R$.
\end{itemize}
Set $S=I\cup\{i\in V:C_i=\varnothing\}$ and \textbf{return} $(X,v,I,\operatorname{ctr},S)$.
}
\end{center}

The following weighted symmetry property of the merging map $\operatorname{ctr}$ will be useful in the analysis.\footnote{This property strongly resembles the reversibility of a Markov chain in terms of the equality of flows between vertices.}

\begin{lemma}[Symmetry of the merging map]\label{lem:merging-map-symmetry}
Condition on the tentative allocation $X$ in \Cref{alg:trial-weighted-pivoting}, so that the residual demands $(v_i)$ are fixed. For clusters $i,j$ with $v_i,v_j>0$, let
\[
\pi_{i,j}\coloneqq\Pr[\operatorname{ctr}(i)=j\mid X],
\]
where the probability is over the subsequent choices of clusters in the while loop. Then
\begin{equation}\label{eq:merging-map-symmetry}
v_i\pi_{i,j}=v_j\pi_{j,i}.
\end{equation}
\end{lemma}

\begin{proof}
The case $i=j$ is immediate, and both probabilities vanish if $i,j$ are not adjacent. Otherwise, consider the first selected cluster whose removal step removes either $i$ or $j$; both clusters remain active until this step. Conditional on the preceding history and on the next selected cluster removing at least one of them, the probabilities that this cluster is $j$ and $i$ are $v_j/D$ and $v_i/D$, respectively, where $D$ is the total residual demand of active clusters equal or adjacent to $i$ or $j$. These choices give $\operatorname{ctr}(i)=j$ and $\operatorname{ctr}(j)=i$, respectively, while selecting any other cluster at this step makes both assignments impossible. Thus the two conditional probabilities, multiplied by $v_i$ and $v_j$, are equal. Averaging over the preceding history proves the identity.
\end{proof}

\subsection{Algorithm for Serving an MIS}\label{susbsubsection:serving-analysis}

We now describe how to benefit from a large MIS produced by \Cref{alg:trial-weighted-pivoting}. We begin by defining what it means to serve an independent set.
 
\begin{definition}[Serving of an Independent Set]\label{def:serving-independent-set}
Let $J\subseteq V$ be a nonempty independent set of $\mathcal G$. \emph{Serving $J$} means assigning every cluster $i\in J$ its set of preferred items $B_i$, removing the clusters in $J$ and the items in $\bigcup_{i\in J}B_i$, and restricting the remaining preference lists to the available items. The demands of the remaining clusters are unchanged. The resulting instance, denoted by $\sigma_J$, has $k-|J|$ clusters. Let $M_J$ also denote the partial assignment made by serving the set $J$.
\end{definition}

Assigning the clusters in a nonempty proper independent set $J$ their preferred items leaves $k-|J|$ clusters. To balance the recursive factors $F(k-|J|)$ in the analysis, we choose among the possible sets $J$ with probabilities proportional to $(1-|J|/k)^{-\alpha}$. Indeed, the function $w(z)=(1-z)^{-\alpha}$ satisfies
\[
w(|J|/k)F(k-|J|)=F(k).
\]
Thus, multiplying the recursive factor by the selection probability gives the same quantity for every choice of $J$.

\begin{center}
\myalg{alg:large-independent-set}{Serving an independent set by a weighted lottery ($\mathcal A_{\mathrm{serve}}$)}{
\textbf{Input:} A current instance $\sigma$ with $k\geq2$ clusters, an exponent $\alpha\in(0,1)$, and an independent set $\emptyset \neq S\subsetneq V$. \medskip \\
Set $w(z)=(1-z)^{-\alpha}$ and $\mathcal J=\{S\}\cup\bigl\{\{i\}:i\in V\setminus S\bigr\}$.\\
Choose $J\in\mathcal J$ according to the weighted lottery
\[
\pi(J)=\frac{w(|J|/k)}{w(|S|/k)+(k-|S|)w(1/k)}.
\]
Assign each cluster $i\in J$ its preferred items $B_i$.\\
Remove these clusters and items to obtain $\sigma_J$ as in \Cref{def:serving-independent-set}.\\
\textbf{Return} $(\sigma_J,M_J)$, where $M_J$ is this partial assignment.
}
\end{center}

\begin{lemma}[Expense of executing \Cref{alg:large-independent-set}]\label{lem:large-independent-set-expense}
Consider an instance with $k\geq2$ clusters and a nonempty proper independent set $S$. Use the lottery of \Cref{alg:large-independent-set} with $w(z)=(1-z)^{-\alpha}$, where $0<\alpha<1$, and write
\[
s=|S|,\qquad x=s/k,\qquad
W=w(x)+(k-s)w(1/k),\qquad
L_k(x)\coloneqq\frac{k-s+2}{W}.
\]
Let $F(t)=2t^\alpha$ and let $\OPT$ be the optimum cost of the current instance in a fixed consistent metric. Suppose that, for every instance $\sigma'$ with $1\leq t<k$ clusters, algorithm $\mathcal{A}$ has distortion of at most $F(t)-1$.
If $\cost(\mathcal A_{\mathrm{serve}})$ denotes the total cost of running $\mathcal A_{\mathrm{serve}}$ (\Cref{alg:large-independent-set}) and recursively solving the residual instance with $\mathcal{A}$ (\Cref{alg:single-layered-matching}), then
\begin{align*} 
\E[\cost(\mathcal A_{\mathrm{serve}})\mid S]+\OPT
&\leq\frac{F(k)}{W}\left((k-s+1)\OPT+g(S)+\sum_{i\in V\setminus S}g(\{i\})\right)\\
&\leq F(k)\OPT L_k(x).
\end{align*}
In particular, if $w(x)+\alpha(1-x)\geq2+\eta$ for some $\eta\in[0,1]$, then
\[
\E[\cost(\mathcal A_{\mathrm{serve}})\mid S]+\OPT
\leq\left(1-\frac{\eta}{k+2}\right)F(k)\OPT.
\]
\end{lemma}

\begin{proof}
The lottery chooses from $\mathcal J=\{S\}\cup\{\{i\}:i\in V\setminus S\}$. Serving $J\in\mathcal J$ costs $g(J)$ and leaves an optimum $\OPT_{\text{res},J}\leq \OPT+g(J)$ by \Cref{lemma:residual-optimal-value}.

Every $J\in\mathcal J$ is nonempty and proper, so the recursive hypothesis gives
\[
\E[\cost(\mathcal A_{\mathrm{serve}})\mid J]+\OPT
\leq g(J)+(F(k-|J|)-1)\OPT_{\text{res},J}+\OPT
\leq F(k-|J|)(\OPT+g(J)).
\]
The exact weights satisfy $w(|J|/k)F(k-|J|)=F(k)$. Averaging over the lottery therefore yields
\begin{align*}
\E[\cost(\mathcal A_{\mathrm{serve}})\mid S]+\OPT
&\leq\sum_{J\in\mathcal J}\frac{w(|J|/k)}{W}F(k-|J|)(\OPT+g(J))\\
&=\frac{F(k)}{W}\left((k-s+1)\OPT+g(S)+\sum_{i\in V\setminus S}g(\{i\})\right)\\
&\leq F(k)\OPT L_k(x).
\end{align*}
For the last inequality, the sets in $\mathcal J$ partition the clusters, and each cluster's favorite items cost no more than the items it receives in an optimum allocation. Thus $\sum_{J\in\mathcal J}g(J)=g(V)\leq \OPT$.

Finally, $w(1/k)\geq1+\alpha/k$. If $w(x)+\alpha(1-x)\geq2+\eta$, then $W\geq k-s+2+\eta$, and hence
\[
L_k(x)\leq\frac{k-s+2}{k-s+2+\eta}
\leq1-\frac{\eta}{k+2},
\]
where we used $s\geq1$ and $0\leq\eta\leq1$.
\end{proof}

\begin{remark}
    The bound in \Cref{lem:large-independent-set-expense} depends only on the size of the maximal independent set $S$ and holds regardless of the tentative allocations. The tentative allocations will, however, be used in the merging procedure, along with the properties of the merging map established in \Cref{lem:merging-map-symmetry}.
\end{remark}

\subsection{Algorithm for Merging into an MIS} \label{subsubsection:mergingAnalysis}

We now present the merging procedure and show how a small MIS improves its cost bound. The procedure first gives each cluster its tentative allocation, then merges the clusters using the map produced by \Cref{alg:trial-weighted-pivoting}. We bound the following two quantities:
\begin{itemize}
    \item The expected cost of the tentative allocations.
    \item The optimal cost of the residual merged instance.
\end{itemize}
Before deriving these bounds, we introduce the following notation:
\begin{notation}[Allocation and merging costs]\label{not:allocation-merging-costs}
For the tentative allocation $X$ produced by \Cref{alg:trial-weighted-pivoting}, write
\[
\widehat g\coloneqq\sum_{i\in V}\sum_{b\in X_i\cap C_i}d(i,b),
\qquad
g_X\coloneqq\sum_{i\in V}\sum_{b\in X_i}d(i,b)=A+\widehat g
\]
for the cost of the assigned shared items and the total allocation cost, respectively.
\end{notation}
\begin{notation}\label{not:merging-cost}
    Given a tentative allocation $X$ produced by \Cref{alg:trial-weighted-pivoting} with residual demands $v_i=\ell_i-|X_i|$ and merging map $\operatorname{ctr}$,\footnote{For clusters $i$ with $v_i=0$, for which the merging map is otherwise undefined, we set $\operatorname{ctr}(i)$ to an arbitrary element of $S$. This convention has no effect on the allocation or merging cost, since these clusters have no residual demand.}
    define the cost of merging as follows:
\[
\cost(\operatorname{ctr})\coloneqq\sum_{i\in V}v_i\,d\bigl(i,\operatorname{ctr}(i)\bigr).
\]
\end{notation}

Since \Cref{alg:trial-weighted-pivoting} assigns each shared item uniformly among the clusters that prefer it, the expected tentative allocation cost satisfies
\begin{equation} \label{eq:upper-bounding-tentative-allocation}
    \E[g_X] = \E[\widehat g  + A] = A + \sum_{i \in V} \sum_{b \in C_i} \frac{d(i, b)}{h_b} \leq A + \sum_{i \in V} \sum_{b \in C_i} \frac{d(i, b)}{2} \leq \frac{\OPT + A}2.
\end{equation}

By \Cref{lemma:residual-optimal-value}, making the tentative allocations leaves a residual instance with optimum cost at most $\OPT+g_X$. Merging the residual demands $v_i$ according to $\operatorname{ctr}$ increases this optimum by at most $\operatorname{cost}(\operatorname{ctr})$. Indeed, in an optimal residual allocation, reassigning cluster $i$'s $v_i$ items to $\operatorname{ctr}(i)$ increases their total cost by at most $v_i d(i,\operatorname{ctr}(i))$, by the triangle inequality. Summing over clusters gives a total increase of at most $\sum_{i\in V}v_i d(i,\operatorname{ctr}(i))=\operatorname{cost}(\operatorname{ctr})$. Therefore,
\begin{equation}
    \OPT_{\text{res}} \leq g_X + \operatorname{cost}(\operatorname{ctr}) + \OPT.
\end{equation}
In \Cref{lem:allocation-merging-cost-bound}, we prove that $\E[g_X + \operatorname{cost}(\operatorname{ctr}) + \OPT]  \leq \frac92 \OPT$. These bounds on the tentative allocation cost and the residual optimum are used in \Cref{lem:merging-branch-expense} to bound the expected distortion of the merging procedure.

\begin{center}
\myalg{alg:merge}{Merging algorithm $\mathcal A_{\mathrm{merge}}$}{
\textbf{Input:} A current instance $\sigma$ whose graph $\mathcal G$ has an edge, and $(X,v,I,\operatorname{ctr})$ from \Cref{alg:trial-weighted-pivoting}. \medskip \\
Give each cluster $i$ its tentative allocation $X_i$.\\
Record $M_X=\{(i,b):i\in V,\ b\in X_i\}$ and remove the allocated items.\\
Merge each positive residual demand $v_i$ into its representative $\operatorname{ctr}(i)$.\\
This gives an instance $\sigma'$ on representatives $j\in I$ with demands
\[
\ell'_j\coloneqq\sum_{i:\,v_i>0,\ \operatorname{ctr}(i)=j}v_i.
\]
Restrict each representative's preference list to the remaining items.\\
\textbf{Return} $(\sigma',M_X)$.
}
\end{center}

\begin{lemma}[Expense of executing \Cref{alg:merge}]\label{lem:merging-branch-expense}
Consider a current instance $\sigma$ with $k\geq2$ clusters and conflicting preference graph $\mathcal G=(V,E)$. Assume $E\neq\varnothing$; otherwise, assigning every cluster its set $B_i$ is optimal. Fix $0<\alpha<1$ and $0<\beta\leq1$. Suppose that, for every instance $\sigma'$ with $1\leq t<k$ clusters and optimum cost $\OPT_{\text{res}}$, the recursive algorithm $\mathcal A$ has distortion at most $F(t) -1$, where $F(t)=2t^\alpha$.

Run \Cref{alg:trial-weighted-pivoting} once to obtain $(X,v,I,\operatorname{ctr},S)$, and set $x=|S|/k$ and $p=\beta/(kx^\alpha)$ as in \Cref{alg:single-layered-matching}. Let $\cost(\mathcal A_{\mathrm{merge}})$ denote the total cost of running \Cref{alg:merge} on $\sigma$ with $(X,v,I,\operatorname{ctr})$, recursively solving the merged instance with $\mathcal A$, and distributing the representatives' items back to the original clusters. Then
\[
\E\bigl[p(\cost(\mathcal A_{\mathrm{merge}})+\OPT)\bigr]
\leq F(k)\OPT\frac{9\beta}{2k}.
\]
Note that the expectation is taken over the random choices in \Cref{alg:trial-weighted-pivoting} and in the recursive execution of $\mathcal A$.
\end{lemma}

\begin{proof}

Using $g_X$ and $\cost(\operatorname{ctr})$ from Notation~\ref{not:allocation-merging-costs} and Notation~\ref{not:merging-cost}, we first show that
\[
\E[\cost(\mathcal A_{\mathrm{merge}})\mid X,\operatorname{ctr}]+\OPT
\leq F(k)x^\alpha\bigl(\OPT+g_X+\cost(\operatorname{ctr})\bigr).
\]
Since $p x^\alpha=\beta/k$, using \Cref{lem:allocation-merging-cost-bound} we have
\begin{align*}
\E\bigl[p(\cost(\mathcal A_{\mathrm{merge}})+\OPT)\bigr]
&\leq F(k)\E\bigl[p x^\alpha(\OPT+g_X+\cost(\operatorname{ctr}))\bigr]\\
&=F(k)\frac{\beta}{k}\E[\OPT+g_X+\cost(\operatorname{ctr})]\\
&\leq F(k)\OPT\frac{9\beta}{2k}.
\end{align*}

By \Cref{lemma:residual-optimal-value}, allocating $X$ leaves a residual optimum of at most $\OPT+g_X$. Now condition on $X$ and $\operatorname{ctr}$. The optimum cost $\OPT_{\text{res}}$ of the merged instance $\sigma'$ is at most $\OPT+g_X+\cost(\operatorname{ctr})$: after allocating $X$, moving the $v_i$ residual items of cluster $i$ to $\operatorname{ctr}(i)$ increases their total cost by at most $\cost(\operatorname{ctr})$. Conversely, splitting the recursive allocation among the original clusters increases its cost by at most the same quantity. The induction hypothesis therefore gives
\begin{align*}
\E[\cost(\mathcal A_{\mathrm{merge}})\mid X,\operatorname{ctr}]+\OPT
&\leq g_X+\cost(\operatorname{ctr})+(F(|S|)-1)\OPT_{\text{res}}+\OPT\\
&\leq F(|S|)\bigl(\OPT+g_X+\cost(\operatorname{ctr})\bigr)\\
&\leq F(k)x^\alpha\bigl(\OPT+g_X+\cost(\operatorname{ctr})\bigr),
\end{align*}
as claimed.\footnote{The procedure merges the residual demands into the independent set $I$. Since $I\subseteq S$ and $F$ is increasing, $F(|I|)\leq F(|S|)$, which justifies using $|S|$ in the bound.}
\end{proof}

Now, to finish the cost analysis of the merging procedure, it remains to prove:
\[
\E\bigl[\OPT+g_X+\cost(\operatorname{ctr})\bigr]
\leq\frac92\OPT-\frac52A\leq\frac92\OPT.
\]
Before presenting the proof, we set up additional notation that helps us to upper bound the term $\operatorname{cost}(\operatorname{ctr})$.
\begin{notation}[Shared demand and radius]\label{not:shared-radius}
For each cluster $i$, let $u_i\coloneqq|C_i|$ be the number of shared items among its preferred items, and define its shared radius by
\[
\rho_i\coloneqq
\begin{cases}
\max_{b\in C_i}d(i,b),&C_i\neq\varnothing,\\
0,&C_i=\varnothing.
\end{cases}
\]
\end{notation}

\begin{lemma}[Expected allocation and merging cost]\label{lem:allocation-merging-cost-bound}
Consider a current instance $\sigma$ and fix a metric consistent with its preference lists. Let $\OPT$ be its optimum cost and $A$ its total private cost. Run \Cref{alg:trial-weighted-pivoting} once to obtain $(X,v,I,\operatorname{ctr},S)$. Then
\[
\E\bigl[\OPT+g_X+\cost(\operatorname{ctr})\bigr]
\leq\frac92\OPT-\frac52A\leq\frac92\OPT,
\]
where the expectation is over the random choices in \Cref{alg:trial-weighted-pivoting}.
\end{lemma}

\begin{proof}
Define
\[
\mu_i\coloneqq\sum_{b\in C_i}\frac1{h_b},\qquad
\bar g\coloneqq\sum_{i\in V}\sum_{b\in C_i}\frac{d(i,b)}{h_b}.
\]
Since each cluster's favorite items cost no more than its optimum allocation, $A+G=g(V)\leq \OPT$. The uniform item choices give 
\begin{equation}
    \E[\widehat g] =\bar g = \E[g_X] - A \quad \text{and} \quad \bar g \leq \frac{\OPT - A}{2}
\end{equation}
by an argument similar to that for \eqref{eq:upper-bounding-tentative-allocation}.
Also, we have $\E[v_i]=u_i-\mu_i$.

The only term we need to upper bound is the cost of merging clusters into the MIS. We prove
$$
    \E[\cost(\operatorname{ctr})] \leq 2\sum_i \E[v_i \rho_i].
$$
Consider a fractional partial assignment that allocates only shared items: each shared item $b$ is split equally among the $h_b$ clusters that prefer it, following the uniform association rule in \Cref{alg:trial-weighted-pivoting}. Private items remain unassigned and available in the residual instance. Then the optimal value of the residual instance is at most $\OPT + \bar g$ due to \Cref{lemma:residual-optimal-value} and its fractional variation.\footnote{For the fractional case, see the flow-decomposition argument in the proof of \Cref{lem:transport} in the PS analysis (\Cref{sec:ps}).} For a lower bound, minimize each cluster's residual cost separately, ignoring competition for items. Cluster $i$ has residual demand $|\text{Private}_i|+u_i-\mu_i$. A cheapest allocation in this relaxation gives it all its private items and fills the remaining $u_i-\mu_i$ units with items outside $B_i$, each costing at least $\rho_i$. Summing these per-cluster lower bounds gives
\begin{equation}\label{eq:single-layer-radius-charge}
\sum_{i\in V}(u_i-\mu_i)\rho_i + A\leq \OPT+\bar g.
\end{equation}

If $i,j$ are adjacent, a common item in $B_i\cap B_j$ gives $d(i,j)\leq\rho_i+\rho_j$. Conditional on $X$, \Cref{lem:merging-map-symmetry} therefore yields
\begin{align*}
\E[\cost(\operatorname{ctr})\mid X]
&\leq\sum_{i,j:\,v_i,v_j>0}v_i\pi_{i,j}(\rho_i+\rho_j)
=2\sum_{i\in V}v_i\rho_i.
\end{align*}
Here the equality follows by exchanging $i,j$ in the second sum and using $v_i\pi_{i,j}=v_j\pi_{j,i}$. Averaging and applying \eqref{eq:single-layer-radius-charge} and \eqref{eq:upper-bounding-tentative-allocation}, we obtain
\begin{align}
\E\bigl[\OPT+g_X+\cost(\operatorname{ctr})\bigr]
&\leq \OPT+A+\bar g+2(\OPT-A+\bar g)\notag\\
&\leq\frac92\OPT-\frac52A
\leq\frac92\OPT.\label{eq:single-layer-merging-charge}
\end{align}
\end{proof}

\subsection{The Full Algorithm and Its Analysis}
We now present the full algorithm and analyze its distortion using the bounds established for the two procedures.
\begin{center}
\myalg{alg:single-layered-matching}{Matching via Coordinated Merging $\mathcal A$}{
\textbf{Input:} An instance $\sigma$ with $k\geq1$ clusters $V$, demands $(\ell_i)_{i\in V}$, and preference lists; parameters $\alpha$ and $\beta$. \medskip \\
Compute $(B_i,\text{Private}_i,C_i)_{i\in V}$, $(h_b)$, and the graph $\mathcal G=(V,E)$.\\
If there are no conflicts, assign each cluster $i$ its preferred items $B_i$ and \textbf{return} the resulting allocation.\\
Otherwise, run \Cref{alg:trial-weighted-pivoting} once to obtain $(X,v,I,\operatorname{ctr},S)$.\\
Set $x=|S|/k$ and $p=\beta/(kx^\alpha)$.\\
Independently draw $H\sim\operatorname{Bernoulli}(p)$.\medskip \\
\textbf{If $H=0$:}
\begin{itemize}[noitemsep]
\item Discard $X$.
\item Run $\mathcal A_{\mathrm{serve}}(\sigma;\alpha,S)$ (\Cref{alg:large-independent-set}), obtaining $(\sigma_J,M_J)$.
\item Recursively solve $\sigma_J$ and \textbf{return} $M_J\cup\mathcal A(\sigma_J)$.
\end{itemize}
\textbf{Otherwise:}
\begin{itemize}[noitemsep]
\item Run $\mathcal A_{\mathrm{merge}}(\sigma;X,v,I,\operatorname{ctr})$ (\Cref{alg:merge}), obtaining $(\sigma',M_X)$.
\item Recursively compute $M'=\mathcal A(\sigma')$.
\item Split each representative $j$'s items in $M'$ arbitrarily among clusters $i$ with $v_i>0$ and $\operatorname{ctr}(i)=j$.\\
Give each cluster $i$ exactly $v_i$ items.
\item Let $M_{\mathrm{res}}$ be the resulting assignment and \textbf{return} $M_X\cup M_{\mathrm{res}}$.
\end{itemize}
}
\end{center}

\begin{definition}[Critical parameters]\label{def:single-layer-critical-parameters}
Let $\alpha_\star$ be the unique solution in $(0,1)$ of
\begin{equation}\label{eq:single-layer-critical-exponent}
\left(1-\left(\frac29\right)^{1/\alpha_\star}\right)^{-\alpha_\star}
+\alpha_\star\left(1-\left(\frac29\right)^{1/\alpha_\star}\right)=2,
\end{equation}
and define
\[
x_\star\coloneqq\left(\frac29\right)^{1/\alpha_\star},\qquad
\beta_\star\coloneqq
\frac{x_\star^{\alpha_\star+1}}{1-x_\star}
\left((1-x_\star)^{-\alpha_\star-1}-1\right).
\]

\end{definition}

\begin{restatable}{lemma}{criticalparameterlemma} \label{lemma:inequality-of-single-layer-critical-exponent}
Let $w_\star(z)=(1-z)^{-\alpha_\star}$. For every integer $k\geq2$ and every real $0<x<1$,
\begin{equation}\label{eq:single-layer-parameter-inequality}
\frac{9\beta_\star}{2k}
+\left(1-\frac{\beta_\star}{kx^{\alpha_\star}}\right)
\frac{k(1-x)+2}{w_\star(x)+k(1-x)w_\star(1/k)}
\leq1.
\end{equation}
Here $\alpha_\star, \beta_\star$ are defined in \Cref{def:single-layer-critical-parameters}.
\end{restatable}

Due to the purely algebraic nature of the proof, we present it in Appendix~\ref{app:single-layer-parameter-inequality}.

\begin{proof}[Proof of \Cref{thm:sing-layer-sublinear-distortion}] 
Use $\mathcal A$ from \Cref{alg:single-layered-matching} with parameters $\alpha=\alpha_\star$ and $\beta=\beta_\star$ from \Cref{def:single-layer-critical-parameters}, and write $w(z)=(1-z)^{-\alpha}$ and $F(t)=2t^\alpha$. For a current instance $\sigma$ with $k$ clusters and conflicting preference graph $\mathcal G=(V,E)$, let $\OPT$ be its optimum cost and $\cost(\mathcal A)$ the total cost returned by $\mathcal A$. We prove by strong induction on $k$ that
\begin{equation}\label{eq:single-layer-augmented-induction}
\E[\cost(\mathcal A)]+\OPT\leq F(k)\OPT.
\end{equation}
If there is no conflict between clusters, then assigning every cluster its set $B_i$ is feasible and optimal, so the left-hand side is $2\OPT\leq F(k)\OPT$.

Suppose $E\neq\varnothing$ and the claim holds for every instance with fewer than $k$ clusters. Let $(X,v,I,\operatorname{ctr},S)$ be the output of \Cref{alg:trial-weighted-pivoting}, and set $s=|S|$, $x=s/k$, and $p=\beta/(kx^\alpha)$. Since $1\leq|I|\leq s\leq k-1$, both recursive calls have fewer clusters. By \Cref{lem:merging-branch-expense,lem:large-independent-set-expense},
\begin{align*}
\E[\cost(\mathcal A)]+\OPT
&\leq F(k)\OPT\E\left[\frac{9\beta}{2k}+(1-p)L_k(x)\right]\\
&\leq F(k)\OPT.
\end{align*}
Here $L_k(x)$ is defined in \Cref{lem:large-independent-set-expense}, and the last inequality follows from \Cref{lemma:inequality-of-single-layer-critical-exponent} since we have the exact equality
$$
L_k(x) = \frac{k(1 - x) + 2}{(1 - x)^{-\alpha} + k(1 - x)(1 - 1/k)^{-\alpha}}.
$$ This closes the induction. Initially $k=n$, so \eqref{eq:single-layer-augmented-induction} gives distortion at most $2n^{\alpha_\star}-1$.
\end{proof}

\section{Random Serial Dictatorship}\label{sec:rsd}

\begin{reptheorem}{thm:rsd-lowerbound}
For arbitrarily large $n$, there exists an instance $\sigma$ with $n$ agents
such that
\[
  \distortion(\rsd,\sigma)
  = \Omega\!\left(\frac{n}{\sqrt{\log n}}\right).
\]
\end{reptheorem}

Our approach is a significant strengthening of an argument that,
to our knowledge, first appeared in the preliminary version of
\citep*{Caragiannis:2024ab}; a similar argument is presented in \citep{Gairing:2019aa}.
The basic idea is to construct an instance
in which a cascade of costly assignments arises as 
a result of indifference. Although agents have nearly the
same cost for two items, they will inadvertently select the item
with major downstream effects. When these consequential
decisions happen in a coordinated sequence, the last agent 
will incur a large cost because of the accumulation.

\cite{Caragiannis:2016aa,Caragiannis:2024ab} create a cascade using a line 
instance that consists of columns $ h = 0, 1, 2, \dots, H$.
The distance between consecutive columns grows exponentially;
column $0$ is located at $0$ on the real line and all other columns 
$h \neq 0$ are located at $2^{h- 1}$ on the real line.
At location $0$ there is only a single item, and at location $1$ there is only 
a single agent; all other locations have an equal number of agents and items.
Agents break ties in favor of items on the right. 

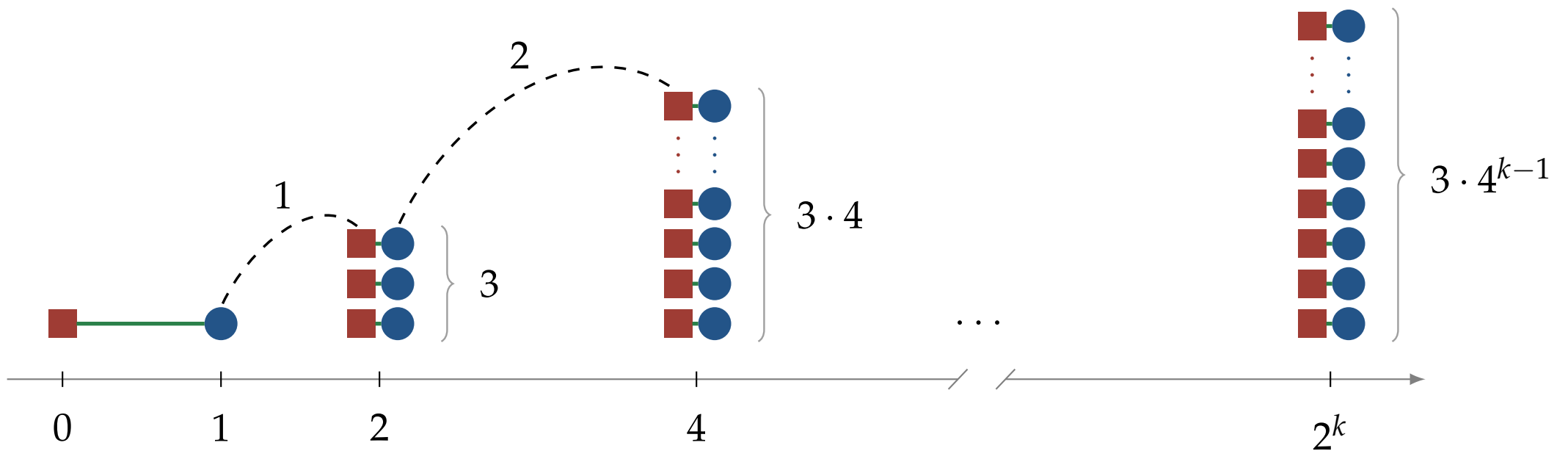
\begin{figure}[ht]
\centering
\pgfmathsetlengthmacro{\lineunit}{0.9*\textwidth/10.4}
\begin{tikzpicture}[x=\lineunit,y=1cm,>=latex,font=\small]
  \definecolor{agentblue}{RGB}{35,84,136}
  \definecolor{itemred}{RGB}{159,60,52}
  \definecolor{optimalgreen}{RGB}{42,128,73}
  \tikzset{
    agent/.style={circle,fill=agentblue,inner sep=0pt,
      outer sep=0pt,minimum size=8.4525pt},
    item/.style={rectangle,fill=itemred,inner sep=0pt,
      outer sep=0pt,minimum size=7.245pt},
    matching/.style={black,dashed,line width=.65pt,
      shorten <=1pt,shorten >=1pt},
    optimal/.style={optimalgreen,line width=1pt},
    countbrace/.style={decorate,
      decoration={brace,amplitude=3pt},
      draw=gray!75,line width=.45pt}
  }
  \def\pairsep{.115}

  \draw[gray] (-.35,0) -- (5.65,0);
  \draw[gray,->] (5.95,0) -- (8.6,0);
  \draw[gray] (5.59,-.09) -- (5.71,.09)
              (5.89,-.09) -- (6.01,.09);
  \foreach \x/\lab in {0/0,1/1,2/2,4/4,8/{2^k}} {
    \draw (\x,-.07) -- (\x,.07);
    \node[below=5pt] at (\x,0) {$\lab$};
  }

  \node[item] (b0) at (0,.5) {};
  \node[agent] (a0) at (1,.5) {};
  \draw[optimal] (b0.east) -- (a0.west);

  \foreach \x/\rows in {
      2/{.5,.86},
      4/{.5,.86,1.22,1.58},
      8/{.5,.86,1.22,1.58,1.94,2.30}} {
    \foreach \y in \rows {
      \node[item] (b) at (\x-\pairsep,\y) {};
      \node[agent] (a) at (\x+\pairsep,\y) {};
      \draw[optimal] (b.east) -- (a.west);
    }
  }

  \foreach \x/\rows in {
      4/{1.87,2.02,2.17},
      8/{2.59,2.74,2.89}} {
    \foreach \y in \rows {
      \fill[itemred] (\x-\pairsep,\y) circle[radius=.45pt];
      \fill[agentblue] (\x+\pairsep,\y) circle[radius=.45pt];
    }
  }

  \foreach \x/\pop/\top in {
      2/3/1.22,
      4/{3\cdot4}/2.46,
      8/{3\cdot4^{k-1}}/3.18} {
    \node[item] (b\x) at (\x-\pairsep,\top) {};
    \node[agent] (a\x) at (\x+\pairsep,\top) {};
    \draw[optimal] (b\x.east) -- (a\x.west);
    \draw[countbrace] (\x+.39,\top+.16) -- (\x+.39,.34)
      node[midway,right=6pt,text=black] {$\pop$};
  }

  \node at (5.8,.5) {$\cdots$};

  \draw[matching] (a0.north) to[out=65,in=140]
    node[midway,above=2pt] {$1$} (b2.north);
  \draw[matching] (a2.north) to[out=65,in=145]
    node[midway,above=2pt] {$2$} (b4.north);
\end{tikzpicture}
\caption{The RSD lower-bound construction with group sizes
$1,3,3\cdot4,\ldots,3\cdot4^{k-1}$ at positions
$1,2,4,\ldots,2^k$. Blue dots are agents and red squares are items. Green edges indicate the optimal
matching, and dashed edges illustrate the first cascade steps,
with ties broken to the right. }
\label{fig:rsd-line-lb}
\end{figure}

Glossing over some details, we discuss the crux of the argument.
The line instance is designed with a specific event 
in mind. Let $\mathcal E_0(M) $ denote the event that some
matching $M$ assigns an agent in the rightmost column the item at $0$.
Such a matching would have $\cost(M, d) \geq 2^{H - 1}$.
The masses of agents are placed so that for each column $h \geq 2$,
the ratio of the number of agents at column $h$ to the total number of agents
at all previous columns is $\lambda = 3$. Such a design makes
$n = 4^{H -1}$ 
and $\Pr_{M \sim \rsd(\sigma_n)} \left [ \mathcal E_0 (M) \right ] \geq
 \left ( 3/4\right )^{H - 1}$; the probability bound follows from counting 
the permutations that make some agent at column $h$ 
select an item after every agent at column $h - 1$, for every $h \geq 1$.
Putting it all together, one finds that the expected distortion is
at least 
\[
\Pr_{M \sim \rsd(\sigma_n)} \left [ \mathcal E_0 (M) \right ] \cdot 2^{H- 1} 
= 
\left ( \frac{3}{2} \right )^{H - 1} = n^{\log_4(3/2)} \approx n^{0.292}.
\]

In order to prove a stronger lower bound, we want to create a cascade, but with a greater degree of parallelism. Signposting what is to come,
we construct a branching instance with levels $h = 0, 1, 2, \dots, D$ and columns $j = 0, 1, 2, \dots, N$ where equal numbers of agents and items are co-located, except at location $(0, 0)$.
A small number $q$ of ``bad'' items will be located at distance $1$ from the vertex at $(0, 0)$, and we will be interested in the quantity $\mathcal C \mathcal E(M)$ that counts the number of agents in the matching $M$
in the final level $D$ that are assigned to items outside their location, each incurring a cost of at least $2^{D - 1}$.
The instance is constructed so that the expected value of $\mathcal C \mathcal{E}(M)$ is at least one;
we will show that $\E_{M \sim \rsd(\sigma_n)} \left [ \mathcal C \mathcal E(M) \right ] \geq 1$.
The optimal matching
will have total cost $q$. By tuning the parameters appropriately, we obtain a lower bound of $$\frac{2^{D - 1}}{q} = \Omega \left (\frac{n }{\sqrt{\log n}} \right).$$

\paragraph{Branching Instance.}
We define an instance indexed by pairs $(h, j) \in \left ( \mathbb  Z_{\geq 0} \times \mathbb Z_{\geq 0} \right ) \cup \{(-1, -1) \}$. The coordinates $h$ and $j$ denote the level and column of the instance, respectively.
Agents and items will be located at each of these vertices.
Let $m_{h,j}$ denote the (initial) fractional mass at vertex $(h,j)$, normalized so that $m_{0,0}=1$. We will later select a replication parameter $q$ and will set populations to $m^*_{h,j}=\lceil qm_{h,j}\rceil$.
Here $q$ is a positive integer, so $m^*_{0,0}=q$, making the optimal assignment cost for the instance $q$. \Cref{alg:rsd-integer-construction} computes initial masses for vertices and applies two pruning steps using only the unrounded masses of non-pruned parents, then assigns integer populations after all pruning decisions are final.
With the exception of vertices $(0, 0)$ and $(-1, -1)$, all vertices will
have an equal number of agents and items. We also refer to them as co-located items and agents. 
The metric for our branching instance is defined by the following weighted connected graph:
\begin{enumerate}[noitemsep]
\item There is a single edge of length $1$ between vertices $(-1, -1)$ and $(0, 0)$.
\item For each $h\geq 1$ and vertex $(h - 1, i)$, there is an edge of length $2^{h-1}$ to each vertex $(h, j)$ with $i < j$.
\end{enumerate}
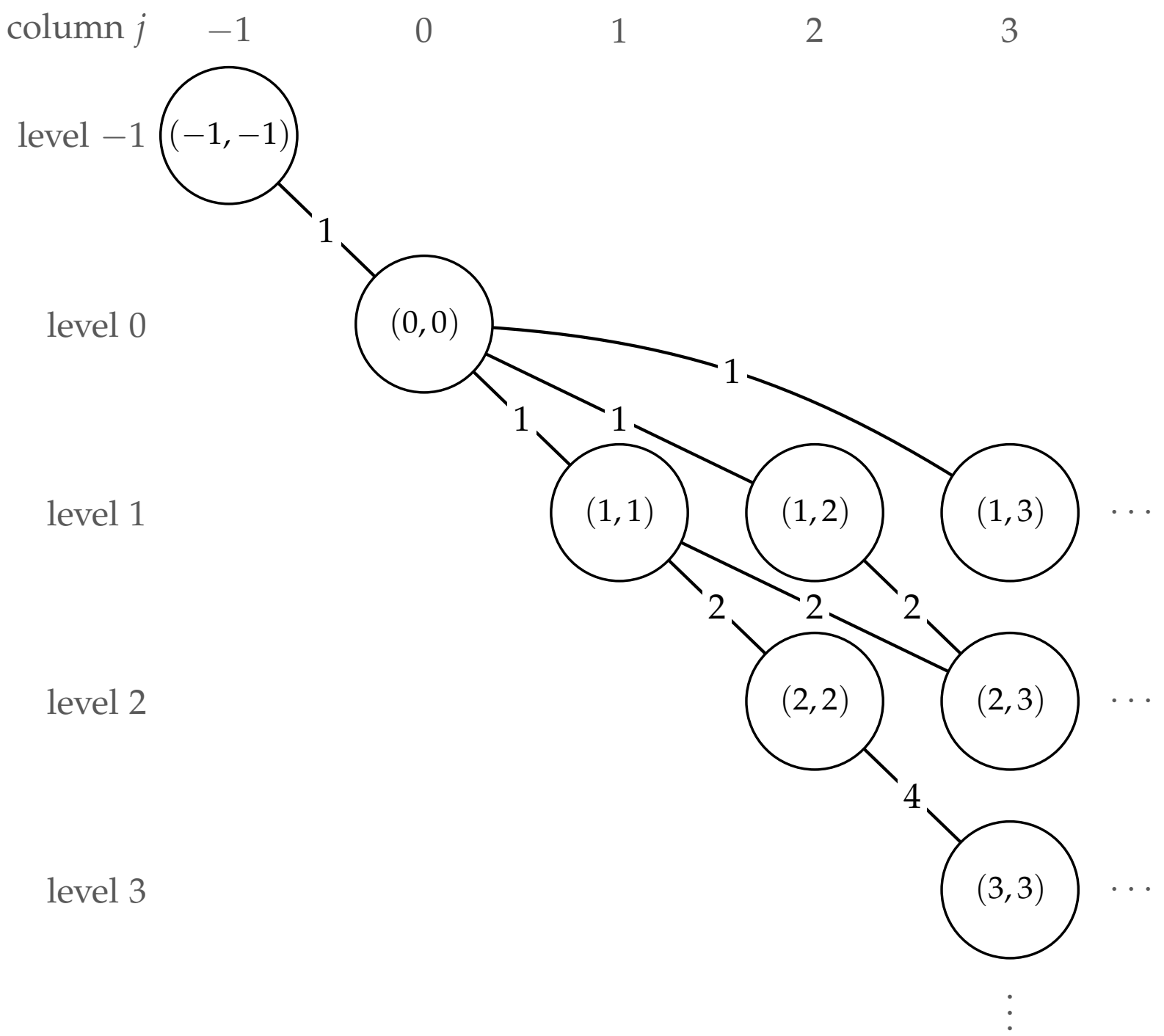
\begin{figure}[H]
\centering
\begingroup
\scriptsize
\pgfmathsetlengthmacro{\branchlabelwidth}{width("$(-1,-1)$")}
\pgfmathsetlengthmacro{\branchlabelheight}{height("$(-1,-1)$")}
\pgfmathsetlengthmacro{\branchlabeldepth}{depth("$(-1,-1)$")}

\begin{tikzpicture}[
  x=.8cm,y=1cm,font=\footnotesize,
  construction site/.style={
    circle,draw=black,fill=white,line width=.65pt,
    minimum size=10mm,inner sep=1pt,outer sep=0pt,
    font=\scriptsize,align=center,
    text width=\branchlabelwidth,
    text height=\branchlabelheight,
    text depth=\branchlabeldepth
  },
  construction edge/.style={draw=black,line width=.8pt},
  edge length/.style={
    midway,fill=white,inner sep=1.5pt,text=black
  }
]
  \node[anchor=east,text=black!65] at (-3.0,2.65) {column $j$};
  \foreach \j in {-1,0,1,2,3}
    \node[text=black!65] at (2.2*\j,2.65) {$\j$};
  \foreach \h in {-1,0,1,2,3}
    \node[anchor=east,text=black!65]
      at (-3.0,-1.7*\h) {level $\h$};

  \draw[construction edge] (-2.2,1.7)--(0,0)
    node[edge length] {$1$};

  \foreach \j in {1,2}
    \draw[construction edge] (0,0)--(2.2*\j,-1.7)
      node[edge length] {$1$};

  \draw[construction edge] (0,0) to[bend left=16]
    node[edge length] {$1$} (6.6,-1.7);

  \draw[construction edge] (2.2,-1.7)--(4.4,-3.4)
    node[edge length] {$2$};
  \draw[construction edge] (2.2,-1.7)--(6.6,-3.4)
    node[edge length] {$2$};
  \draw[construction edge] (4.4,-1.7)--(6.6,-3.4)
    node[edge length] {$2$};
  \draw[construction edge] (4.4,-3.4)--(6.6,-5.1)
    node[edge length] {$4$};

  \node[construction site] at (-2.2,1.7) {$(-1,-1)$};
  \node[construction site] at (0,0) {$(0,0)$};

  \foreach \h in {1,2,3} {
    \foreach \j in {\h,...,3}
      \node[construction site]
        at (2.2*\j,-1.7*\h) {$(\h,\j)$};
    \node[text=black!65] at (8.0,-1.7*\h) {$\cdots$};
  }
  \node[text=black!65] at (6.6,-6.1) {$\vdots$};
\end{tikzpicture}
\endgroup
\caption{The initial levels of the branching construction rooted at $(0,0)$, together with the special vertex $(-1,-1)$. Only vertices reachable from $(0,0)$ through increasing levels are shown, restricted to levels and columns $0$ through $3$. Every edge between the displayed vertices is drawn. For $h\geq 1$, an edge from $(h-1,i)$ to $(h,j)$, with $i<j$, has length $2^{h-1}$: the first three level transitions have lengths $1$, $2$, and $4$. The drawing is schematic and not to scale.}
\label{fig:branching-construction}
\end{figure}

The remaining distances are computed from shortest paths in the underlying weighted connected graph. It is also not hard to see that the optimal value of the instance we described is exactly $m^*_{0, 0}$.
We also introduce some terminology for a vertex $(h,j)$. For $h\geq 1$, its neighbors $(h-1,i)$ with $i<j$ are its \textit{parents}. An \textit{ancestor} is a vertex at a nonnegative level reached by following one or more successive parent edges. Its \textit{children} are its neighbors $(h+1,k)$ with $k>j$. We treat the special vertex $(-1,-1)$ separately.

\paragraph{Preferences.}
An agent at $v=(h,j)$ in the branching instance ranks the item locations in the following order:
\begin{equation}
\label{eq:preference}
\begin{aligned}
&\text{co-located items}
\ \succ\ \text{ancestors (decreasing level)}
\ \succ\ \text{children (increasing column)}
\\
&\qquad\succ\ \text{items at }(-1,-1)
\ \succ\ \text{remaining items (arbitrary tie-breaking by distance)}.
\end{aligned}
\end{equation}
At a fixed ancestor level, locations are ordered by increasing column. 
Ties within a location and among the remaining equal-distance items are broken arbitrarily.

It is not hard to see that this order is consistent with the shortest-path metric. The co-located items have distance $0$. An ancestor at level $a<h$ has distance $2^h-2^a$. Every child and the location $(-1,-1)$ have distance $2^h$. Any other path has length at least $2^h$.

\subsection*{Analyzing Distortion}

For the specific branching instance, we would like to prove that $\rsd$, in expectation, matches one of the agents at the deepest level to some item that is not located at the same location. This will make the distortion of the instance at least $2^{D - 1}/m^*_{0, 0}$ if $D$ denotes the last level of a vertex in the instance.
To show this, we first explain how we would like to view the $\rsd$ process. In the $\rsd$ process, we take a uniformly random permutation of the agents and, based on that order, give each agent its favorite item among the remaining ones. For our analysis, instead of generating a uniform permutation and assigning discrete positions, we assign each agent an independent $\text{Unif}[0, 1]$ arrival time, and $\rsd$ processes the agents in time order from zero to one.

To prove that at least one of the agents at level $D$ is matched outside its co-located items, we will show that there exists a time $t^* \in (0 , 1)$ such that:
\begin{itemize}
    \item Let $L$ denote the number of agents in the last level $D$ arriving after $t^*$.
    \item Let $R$ denote the number of items remaining in the last level $D$ at the time $t^*$.
\end{itemize}
Then, over the random choice of $\text{Unif}[0, 1]$ arrival times for the agents, we prove
\begin{equation}
    \E[(L - R)_+] \geq \E[L] - \E[R] \geq 1 . \label{eq:successful-cascade}
\end{equation}

Note that \eqref{eq:successful-cascade} immediately implies that the distortion is at least $\frac{2^{D - 1}}{m^*_{0, 0}}$, since after time $t^*$, at least $\E[(L - R)_+]$ agents at level $D$ are matched by $\rsd$, in expectation, to items outside level $D$.

Now we explain how we will prove \eqref{eq:successful-cascade}.
Computing $\E[L]$ is rather straightforward: it is exactly $(1 - t^*)\sum_j m^*_{D, j}$, which we will ensure is at least $2$ throughout the construction. However, computing $R$ turns out to be significantly challenging, and the remainder of the proof and the parameter choices focus on showing an upper bound of $1$ for $\E[R]$.

For the rest of this section, we assume that our instance is contained in a rectangle with corners $(0, 0)$ and $(D, N)$. We will clarify the dependence of $D$ and $N$ on $n$, the number of agents and items, later.

\subsection*{Artificial $\rsd$ and Deadlines}

To provide an upper bound for $\E[R]$ in \eqref{eq:successful-cascade}, instead of analyzing the $\rsd$ process, we consider an artificial $\rsd$ process with increasing deadlines $\{1 - u^*_j\}_{j = 1}^N$. At the time (deadline) $1 - u^*_j$, we remove all the remaining items in column $j$.
The intuition behind the artificial $\rsd$ process is that we expect all the items in column $j$ to have typically been consumed (selected) by the $\rsd$ process by the deadline $1 - u^*_j$. This assumption will help us show that, by the deadline $1 - u^*_{j + 1}$, all the items in column $j + 1$ have also typically been consumed. Then, at time $1 - u_N^* = t^*$, all the items have been consumed, implying that $R$ is zero most of the time.

However, these behaviors at the deadlines hold in the typical case, and we might need to delete some items at the deadlines.
We will give a coupling argument that handles these additional deletions in the $\rsd$ process.
More precisely, if we delete some items in column $j$ at deadlines $1 - u_j^*$, then we show that the expected number of items remaining at level $D$ in the actual $\rsd$ process is bounded above by the expected number of items deleted in the artificial process; see \Cref{subsection:Coupling-between-artificial-and-rsd}.
Here is the rough scheme of how we show that the number of item deletions in the artificial process is small, which motivates the construction of our instance.

Let's focus on an individual vertex $(h, j)$. At time $1 - u_{j - 1}^*$, every item in columns $1, 2, \dots, j - 1$ has been removed, and we expect every remaining item at $(h, j)$ to be removed by $1 - u_j^*$.
The expected number of agents from $(h, j)$ arriving before $1 - u_j^*$ is $m_{h, j}^* (1 - u_j^*)$. The expected number of agents from parents of $(h, j)$ arriving in the interval $[1 - u_{j - 1}^*, 1 - u_j^*]$ equals $(u_{j - 1}^* - u_j^*) \sum_{i < j} m_{h - 1, i}^*$.

\begin{figure}[H]
\centering
\begin{tikzpicture}[x=1cm,y=1cm]
  \node[anchor=east,align=right,text=muted] at (2.2,3.5)
    {Parents:\\level $h-1$};
  \foreach \x/\name in {3/p1,5/p2,7.4/pk}
    \node[site,minimum size=6mm] (\name) at (\x,3.5) {};
  \node[text=muted] at (6.2,3.5) {$\cdots$};
  \node[site,minimum size=6mm] (v) at (9.3,2.4) {$v$};
  \foreach \name in {p1,p2,pk}
    \draw[flow,draw=teal] (\name.south)--(v.west);
  \node[anchor=west,align=left] at (9.8,2.4)
    {$v=(h,j)$\\level $h$, column $j$};

  \draw[->,thick,ink] (0,0)--(12.1,0);
  \fill[teal!14] (3.0,.05) rectangle (7.2,.72);
  \fill[rust!10] (7.2,.05) rectangle (11.3,.72);
  \foreach \x/\lab in {0/0,3.0/{1 - u_{j-1}^*},7.2/{1 - u_j^*},11.3/1} {
    \draw[thick] (\x,-.12)--(\x,.15);
    \node[below=3pt] at (\x,0) {$\lab$};
  }
  \node[align=center,text=teal] at (5.1,.4)
    {$U_v$: parent arrivals};
  \node[align=center,text=rust] at (9.25,.4)
    {$V_v$: local late arrivals};
  \node[above=9mm,align=center,text width=3.8cm] at (3.0,0)
    {Earlier column items\\have been removed.};
  \node[above=9mm,align=center,text width=3.3cm] at (7.2,0)
    {Remove any items\\still at $v$.};
\end{tikzpicture}
\caption{At column $j$'s deadline $1-u_j^*$, artificial $\rsd$ deletes all items still at $v=(h,j)$.}
\label{fig:rsd-artificial-deadline}
\end{figure}
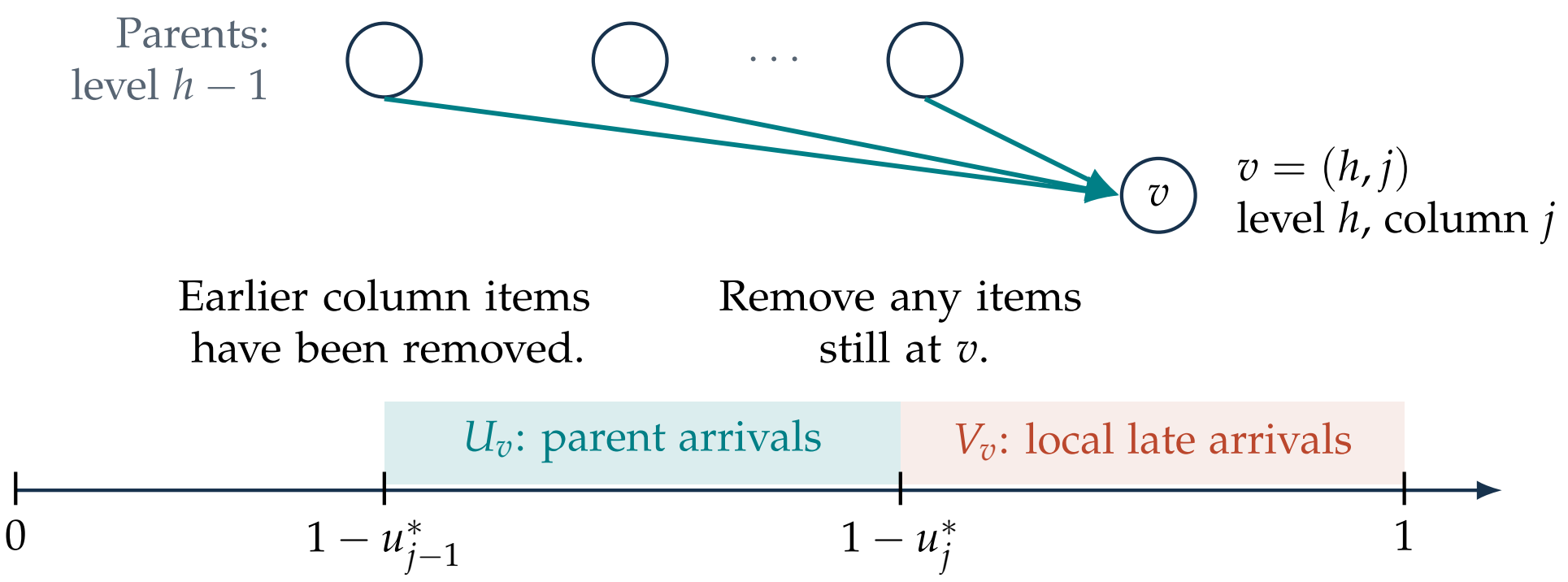

To ensure that the items at $(h, j)$ are removed by the deadline $1 - u_j^*$, we need to show roughly that enough parent agents of $(h, j)$ arrive in the interval $[1- u_{j - 1}^*, 1 - u_j^*]$ to be matched with whatever is left at $(h, j)$ by the deadline $1 - u_j^*$.

\begin{lemma}[Deletion bound at a location]
\label{lem:individual-node-deletion}

Fix a retained location $v=(h,j)$ with $h\geq1$ in the artificial $\rsd$ process. Let $U_v$ count agents at parents $(h-1,i)$, $i<j$, arriving in $[1-u_{j-1}^*,1-u_j^*]$, and let $V_v$ count co-located agents at $v$ arriving after $1-u_j^*$. Then the expected number of artificially removed items at $v$ at its deadline $1-u_j^*$ satisfies
\begin{equation}
 \E[\text{artificially removed items at }v]\leq\E[(V_v-U_v)_+].
 \label{eq:individual-node-deletion-upperbound}
\end{equation}
\end{lemma}

\begin{proof}
If no items are artificially removed at $v$, the bound is immediate. 
Otherwise, items at $v$ remain available until its deadline. Each of the $m^*_{h,j}-V_v$ local agents arriving before the deadline therefore takes an item at $v$. All item locations in columns $0,\ldots,j-1$ are empty by time $1-u_{j-1}^*$. By the preference order in \eqref{eq:preference}, every parent agent counted by $U_v$ also takes an item at $v$: its local and ancestor items and all children in earlier columns are unavailable. These are distinct agents, so the number of artificially removed items at $v$ is at most
\[
 m^*_{h,j}-(m^*_{h,j}-V_v)-U_v=V_v-U_v.
\]
Thus the number of artificially removed items at $v$ is at most $(V_v-U_v)_+$ almost surely. Taking expectations proves the claim.
\end{proof}

To ensure that all items at $(h, j)$ are removed by time $1 - u_j^*$, we will set parameters so that there is a sufficient surplus of parent agents in the interval $[1 - u_{j - 1}^*, 1 - u_j^*]$ to remove any remaining items at $(h, j)$. Thus, we roughly require:
\begin{equation}
\E[U_v] = (u_{j - 1}^* - u_j^*) \sum_{i < j} m_{h - 1, i}^* > (1 + \varepsilon_j)m_{h, j}^*u_j^* = \E[V_v] (1 + \varepsilon_j)
\label{eq:deletion-surplus-expectation}
\end{equation}
for the surplus parameters $\varepsilon_j$ specified in \eqref{eq:mass-pruning-thresholds}.
Note that $\E[(V_v - U_v)_+]$ will be small if $\varepsilon_j\E[V_v]$ is significantly larger than the standard deviation of $V_v$.
Both variables have binomial distributions, and if the difference between their expectations is as large as the standard deviation, then $\E[(V_v - U_v)_+]$ would be small.

Our eventual goal is to prove the following upper bound for $\E[R]$, as in \eqref{eq:successful-cascade}:
\begin{equation}
\E[R] \leq \sum_{(h, j)} \E[(V_{h, j} - U_{h, j})_+] \leq 1 \label{eq:bounding-removal-imply-bounding-last-level-item}
\end{equation}
Later we will see a procedure called \emph{mass pruning} (see \Cref{subsection:pruning-procedures}) that will ensure that every location $(h,j)$ with non-zero population has a sufficient surplus of parent agents compared to the fluctuations in the remaining items at $(h, j)$.
Later we will introduce another pruning procedure besides \emph{mass pruning}, called \emph{band pruning}, which restricts possible locations with non-zero mass.

\subsection{Detailed Explanation of the Branching Instance}

We first describe the construction of the lower-bound instance for $\rsd$. Besides the integers $D,N$ and the replication parameter $q$, the inputs are:
\begin{itemize}
    \item A sequence of positive numbers $\lambda_1, \lambda_2, \dots, \lambda_N$ such that $\sum_i \lambda_i = D$.
    \item Candidate locations $\mathcal{B}$ (induced by \textit{band pruning}).
    \item Positive cutoffs $\tau_j$ for fractional masses (used by \textit{mass pruning}).
\end{itemize}

At each candidate location $(h,j) \in \mathcal B$, compute $m_{h,j}=(\lambda_j/2)\sum_{i<j}m_{h-1,i}$ from retained parent masses. Keep this mass if $qm_{h,j}\geq\tau_j$; otherwise set $m_{h,j}=0$. After processing all columns, assign $m^*_{h,j}=\lceil qm_{h,j}\rceil$ agents and the same number of items at each retained candidate. Thus rounding is performed only after mass pruning.

\begingroup
\floatname{algorithm}{Construction}
\algtext*{EndIf}
\algtext*{EndFor}
\begin{algorithm}[H]
\caption{Construction of the branching instance from fractional masses}
\label[construction]{alg:rsd-integer-construction}
\small
\begin{algorithmic}[1]
\Require Integers $1\leq D\leq N$ and $q\geq1$; positive $\lambda_1,\ldots,\lambda_N$ with $\sum_j\lambda_j=D$.
\Require Candidate set $\mathcal B\subseteq\{(h,j):1\leq h\leq D,\ h\leq j\leq N\}$ and positive cutoffs $\tau_1,\ldots,\tau_N$.
\Ensure A metric preference instance $\sigma_n$ with retained masses $m_{h,j}$ and integer populations $m^*_{h,j}=\lceil qm_{h,j}\rceil$.
\State Set $m_{h,j}\gets0$ for $0\leq h\leq D$ and $0\leq j\leq N$.
\State Set $m_{0,0}\gets1$ and $m_{-1,-1}\gets0$.
\For{$j=1,\ldots,N$}
  \For{$h=1,\ldots,\min\{D,j\}$}
    \If{$(h,j)\in\mathcal B$}
      \State $m_{h,j}\gets\dfrac{\lambda_j}{2}\sum_{i=0}^{j-1}m_{h-1,i}$.
      \If{$qm_{h,j}<\tau_j$}
        \State $m_{h,j}\gets0$.
      \EndIf
    \EndIf
  \EndFor
\EndFor
\State Set $m^*_{h,j}\gets\lceil qm_{h,j}\rceil$ for all initialized locations.
\State $V\gets\{(-1,-1),(0,0)\}\cup\{(h,j)\in\mathcal B:m_{h,j}>0\}$.
\State Place $q$ agents and no items at $(0,0)$; place $q$ items and no agents at $(-1,-1)$.
\ForAll{$(h,j)\in V\setminus\{(-1,-1),(0,0)\}$}
  \State Place $m^*_{h,j}$ agents and $m^*_{h,j}$ co-located items at $(h,j)$.
\EndFor
\State Assign each agent the preference order as in \eqref{eq:preference}.
\State $n\gets q+\sum_{h=1}^D\sum_{j=1}^N m^*_{h,j}$.
\State \Return the resulting instance $\sigma_n$.
\end{algorithmic}
\end{algorithm}
\endgroup

\subsection{Motivation}

In this subsection, we first motivate the branching instance and the values of the fractional masses using a Poisson binomial distribution.

\paragraph{Poisson Binomial Distribution.} Take parameters $\lambda_1, \lambda_2, \dots, \lambda_N$. Consider independent Bernoulli random variables $r_i \sim \operatorname{Bernoulli}(\frac{\lambda_i}{\lambda_i + 2})$. Then $r_1, r_2, \dots, r_N$ form a zero-one sequence, or word, that can be interpreted as a path from the root to level $\sum_i r_i$, having one node at each level from one to $\sum_i r_i$.

For a subset (path) $S$, define its weight 
\[
w(S) := \prod_{i \in S} \frac{\lambda_i}{2}.
\]
Before either pruning step, the unrounded mass at $(h,j)$ equals the sum of the weights of all paths from the root to that location:
\begin{equation}
    m'_{h,j}=\sum_{\substack{S\subseteq[j]:\ |S|=h,\ j\in S}}w(S).
\end{equation}
Later, we define two pruning steps. After these steps, we will keep the same path interpretation; however, we allow paths only within specific subsets of the
$N \times D$ grid. We will give more details on pruning later in \Cref{subsection:pruning-procedures}.

In our branching instance construction, we will condition on the event that $\sum_i r_i = D$. Note that this event is very unlikely since $\E[\sum_i r_i] \approx D/2 + O(1)$\footnote{The approximation holds as long as $\sum_i \lambda_i^2 < O(1)$.}.

Let 
\[
B_0 := \sum_{S \subseteq [N]: |S| = D} w(S) \qquad Z = \prod_{i=1}^N \left(1 + \frac{\lambda_i}{2}\right).
\]
For the random occupied set $S:=\{i:r_i=1\}$ produced by independent Bernoulli random variables $r_i$, we have $\Pr(|S|=D)=B_0/Z$.

\begin{lemma}[Probability of level $D$ paths]
\label{lem:baseline-path-probability}
Suppose $\lambda_1,\ldots,\lambda_N>0$ satisfy
\begin{equation}    
 \sum_{i=1}^N\lambda_i=D,
 \qquad
 \sum_{i=1}^N\lambda_i^2\leq 1. \label{eq:total-and-conditional-mass}
\end{equation}
Then, for $B_0$ and $Z$ defined above,
\begin{equation}
 \frac{B_0}{Z}
 =\Theta\!\left(\frac{(\sqrt e/2)^D}{\sqrt D}\right).
 \label{eq:baseline-path-probability-asymptotic}
\end{equation}
\end{lemma}

\begin{proof}
Draw $D$ independent column indices $I_1,\ldots,I_D$ with $\Pr(I_\ell=i)=\lambda_i/D$. Each $D$-element subset occurs in $D!$ orders, so
\[
 B_0=\frac{(D/2)^D}{D!}\,
 \Pr(I_1,\ldots,I_D\text{ are all distinct}).
\]
By a union bound over pairs of indices,
\[
 \Pr(\text{some collision})
 \leq\binom D2\sum_i\left(\frac{\lambda_i}{D}\right)^2
 =\frac{D-1}{2D}\sum_i \lambda_i^2 \leq\frac12.
\]
To bound $Z$, apply $0\leq x-\log(1+x)\leq x^2/2$ with $x=\lambda_i/2$ and sum:
\[
 \begin{aligned}
 0\leq\frac D2-\log Z
 &=\sum_{i=1}^N\left[\frac{\lambda_i}{2}
     -\log\!\left(1+\frac{\lambda_i}{2}\right)\right]\\
 &\leq\frac18\sum_{i=1}^N\lambda_i^2 \leq \frac18.
 \end{aligned}
\]
Since $\sum_i \lambda_i^2\leq1$, we obtain $Z=\Theta(e^{D/2})$. The probability of no collision is at least $1/2$, so
\[
 \frac{B_0}{Z}
 =\Theta\!\left(\frac{e^{-D/2}(D/2)^D}{D!}\right).
\]
Finally, Stirling's formula gives
\[
 \frac{e^{-D/2}(D/2)^D}{D!}
 =\Theta\!\left(\frac{(\sqrt e/2)^D}{\sqrt D}\right),
\]
which proves \eqref{eq:baseline-path-probability-asymptotic}.
\end{proof}

\paragraph{Reference remaining times.}
For the chosen sequence $\lambda_1,\ldots,\lambda_N$, define
\begin{equation}
 u_0:=1,
 \qquad
 u_i:=\frac{u_{i-1}}{1+\lambda_i/2}
     =\prod_{k=1}^i\left(1+\frac{\lambda_k}{2}\right)^{-1},
 \qquad 1\leq i\leq N.
 \label{eq:reference-remaining-times}
\end{equation}
Since each $\lambda_i>0$, we have $1=u_0>u_1>\cdots>u_N>0$, so the reference times $1-u_i$ increase with the column index. An agent with a uniform arrival time arrives after $1-u_i$ with probability $u_i$, and in the interval $[1-u_{i-1},1-u_i]$ with probability $u_{i-1}-u_i$.

Applying $x-x^2/2\leq\log(1+x)\leq x$ to each factor in \eqref{eq:reference-remaining-times} yields
\[
 \exp\!\left(-\frac12\sum_{k=1}^i\lambda_k\right)
 \leq u_i
 \leq\exp\!\left(-\frac12\sum_{k=1}^i\lambda_k
                  +\frac18\sum_{k=1}^i\lambda_k^2\right).
\]
Under the assumption $\sum_k\lambda_k^2\leq1$, these bounds hold within absolute constant factors, uniformly in $i$. In particular, since $\sum_k\lambda_k=D$,
\begin{equation}
 u_N=\frac1Z=\Theta(e^{-D/2}),
 \qquad
 u_N B_0=\frac{B_0}{Z}
 =\Theta\!\left(\frac{(\sqrt e/2)^D}{\sqrt D}\right).
 \label{eq:reference-last-level-weight}
\end{equation}
The last estimate follows from \Cref{lem:baseline-path-probability}.

The interesting identity satisfied by consecutive reference times is:
$$
    u_{i - 1} - u_{i} = \frac{\lambda_i}{2}u_i.
$$

The raw masses $m'_{h, j}$ satisfy the following recursion, which strongly resembles \eqref{eq:deletion-surplus-expectation}:
$$
    (u_{i - 1} - u_{i})\sum_{j < i} m'_{h - 1, j} = u_i m_{h, i}'
$$
However, there is no surplus in the current reference times $u_i$. We will modify the deadlines in the next part to create the required surplus in \eqref{eq:deletion-surplus-expectation}.

\paragraph{Adjusting the deadlines to create surplus.}
We use a sequence $0<\varepsilon_j\leq1$, to be specified later (see \eqref{eq:mass-pruning-thresholds}), to adjust the reference schedule and obtain the surplus in \eqref{eq:deletion-surplus-expectation}. Define
\begin{equation}
 u_0^*:=1,
 \qquad
 u_j^*:=\frac{u_{j-1}^*}{1+(1+2\varepsilon_j)\lambda_j/2}
 =\prod_{i=1}^j\left(1+(1+2\varepsilon_i)\frac{\lambda_i}{2}\right)^{-1}.
 \label{eq:adjusted-remaining-times}
\end{equation}
The artificial deadline for column $j$ is $1-u_j^*$, and the final time is $t^*:=1-u_N^*$. These new deadlines satisfy the following surplus:
\begin{equation}
 u_{j-1}^*-u_j^*=(1+2\varepsilon_j)\frac{\lambda_j}{2}u_j^*.
 \label{eq:adjusted-time-balance}
\end{equation}
For a vertex $v=(h,j)$, the fractional mass recursion and the final rounding give
\[
 \frac{\lambda_j}{2}\sum_{i<j}m^*_{h-1,i}
 \geq q\frac{\lambda_j}{2}\sum_{i<j}m_{h-1,i}
 =qm_{h,j}>m^*_{h,j}-1.
\]
Mass pruning ensures $qm_{h,j}\geq\tau_j$ and hence $m^*_{h,j}\geq\tau_j$. Consequently, choosing $\tau_j\geq\frac1{\varepsilon_j}+2$ implies
\begin{equation}
 \E[U_v]
 >(1+2\varepsilon_j)(m^*_{h,j}-1)u_j^*
 \geq(1+\varepsilon_j)m^*_{h,j}u_j^*
 =(1+\varepsilon_j)\E[V_v].
 \label{eq:adjusted-deadline-surplus}
\end{equation}
This is precisely the surplus required in \eqref{eq:deletion-surplus-expectation}.

We will choose the sequences $\varepsilon_j$ and $\lambda_j$ so that $\sum_j\varepsilon_j\lambda_j$ is bounded by an absolute constant. The following lemma shows that this keeps the adjusted remaining times within a constant factor of the reference values.

\begin{lemma}[Bounded deadline adjustment]
\label{lem:bounded-deadline-adjustment}
Let $u_j$ and $u_j^*$ be defined by \eqref{eq:reference-remaining-times} and \eqref{eq:adjusted-remaining-times}, with $\lambda_j>0$ and $0<\varepsilon_j\leq1$. Then
\begin{equation}
 e^{-\sum_{j = 1}^i \varepsilon_j\lambda_j}\leq\frac{u_i^*}{u_i}\leq1.
 \label{eq:adjusted-time-comparison}
\end{equation}
In particular, $u_N^*/u_N=\Theta(1)$ whenever $\sum_{j = 1}^N \varepsilon_j \lambda_j = O(1)$.
\end{lemma}

\begin{proof}
Dividing the product formulas for $u_i^*$ and $u_i$ gives
\[
 \frac{u_i^*}{u_i}
 =\prod_{j=1}^i
   \frac{1+\lambda_j/2}{1+(1+2\varepsilon_j)\lambda_j/2}
 =\prod_{j=1}^i
   \left(1+\frac{2\varepsilon_j\lambda_j}{2+\lambda_j}\right)^{-1}.
\]
Every factor is at most one, proving the upper bound. For the lower bound, $\log(1+x)\leq x$ for $x\geq0$ yields
\[
 0\leq\log\frac{u_i}{u_i^*}
 =\sum_{j=1}^i\log\!\left(1+\frac{2\varepsilon_j\lambda_j}{2+\lambda_j}\right)
 \leq\sum_{j=1}^i\frac{2\varepsilon_j\lambda_j}{2+\lambda_j}
 \leq\sum_{j=1}^i\varepsilon_j\lambda_j,
\]
which concludes the proof of \eqref{eq:adjusted-time-comparison}.
\end{proof}

\begin{remark}
Recall that $\E[L]=u_N^*\sum_jm^*_{D,j}$. To obtain $\E[L]=\Theta(1)$, we choose
\[
 q=\Theta\!\left(\frac{1}{B_0u_N^*}\right)
  =\Theta\!\left(\sqrt D\,(2/\sqrt e)^D\right),
\]
using $B_0=\Theta((e/2)^D/\sqrt D)$ and the intended bound $u_N^*=\Theta(e^{-D/2})$.
If pruning preserves $\Omega(B_0)$ mass at level $D$ and bounds the total mass by $O(B_0)$, with rounding changing populations by only a constant factor, then
\[
 \begin{aligned}
 n&=\Theta(qB_0)=\Theta(e^{D/2}),\\
 \frac{2^{D-1}}q
  &=\Theta\!\left(\frac{n}{\sqrt D}\right)
   =\Theta\!\left(\frac{n}{\sqrt{\log n}}\right).
 \end{aligned}
\]
The constant factor in $q$ will ensure $\E[L]\geq2$. Band pruning controls the total population, while mass pruning ensures $\E[R]\leq1$; both preserve a constant fraction of the last-level mass.
\end{remark}

\subsection{Pruning Procedures} \label{subsection:pruning-procedures}

Recall that $B_0$ is the total weight of paths of length $D$. The first pruning step, called \emph{band pruning}, restricts paths to a set of locations $\mathcal B$. We choose $\mathcal B$ so that the total mass at all retained locations is $O(B_0)$, while preserving a constant fraction of the mass at level $D$.

The second step, called \emph{mass pruning}, removes locations in each column $j$ whose mass is below a prescribed threshold. This ensures that retained locations have enough mass to control the expected number of deletions in \eqref{eq:individual-node-deletion-upperbound}. We show that level $D$ still retains $\Omega(B_0)$ mass after this step.

For both pruning procedures, fix $\theta\geq64$ and set
\begin{equation}
 \lambda_i:=\frac{\theta}{\theta^3+N-i},\quad 1\leq i\leq N,
 \qquad D:=\sum_{i=1}^N\lambda_i.
\label{eq:harmonic-schedule}
\end{equation}
Through the rest of this section assume $D\in\mathbb Z$ for simplicity and carry out the computation given that assumption.
However, the assumption $D \in \mathbb{Z}$ can be satisfied by slightly perturbing $\lambda_1$ so that the total weight is integral. The perturbed construction has the same distortion guarantee; the perturbation and the verification of the required estimates are given in Appendix~\ref{app:rsd-integral-weight}.
An integral comparison bounds the sum of the squared weights:
\[
 \sum_i\lambda_i^2\leq\frac{\theta^2}{\theta^3-1}
 \leq\frac2\theta\leq\frac18.
\]

\subsubsection*{Band Pruning}

We define the band of candidate locations using two parameters, $0<\beta<1$ and $W>0$. Restricting paths to this band preserves a constant fraction of the total weight of paths ending at level $D$, while bounding the total retained mass by a constant times the mass at that level.

\begin{definition}[Candidate set]
\label{def:candidate-set}
Let $x_0:=0$, $x_j:=\sum_{i=1}^j\lambda_i$, and $k_j:=D-x_j=\sum_{i=j+1}^N\lambda_i$, so $x_N=D$.
The candidate set $\mathcal{B}_{\beta, W}$ with parameters $\beta$ and $W$ is
\begin{equation}
 \mathcal B_{\beta, W}:=\bigl\{(h,j):1\leq h\leq D,\ h\leq j\leq N,
           \ |h-x_j|\leq\beta k_j+W\bigr\}.
 \label{eq:candidate-band}
\end{equation}
The root $(0,0)$ is retained separately. Before mass pruning, set $m^{\mathcal B}_{0,0}:=1$, set all other locations' masses to zero, and for $1\leq h\leq D$ define
\[
 m^{\mathcal B}_{h,j}:=
 \begin{cases}
  \displaystyle\frac{\lambda_j}{2}\sum_{i=0}^{j-1}m^{\mathcal B}_{h-1,i},
       &(h,j)\in\mathcal B,\\
  0,&\text{otherwise},
 \end{cases}
\]
Write $B_{\mathrm{band}(\beta, W)}:=\sum_jm^{\mathcal {B}_{\beta, W}}_{D,j}$ for their last-level weight. Whenever $\beta$ and $W$ are clear from the context, we use $B_{\mathrm{band}}$ instead. By definition of the candidate set, each column $j$ contains at most $2\beta k_j+2W+1$ candidate locations.
\end{definition}

\begin{lemma}[Weight control of candidate set] \label{lemma:weight-of-candidate-set}
Suppose $\sum_i\lambda_i=D$ and $\sum_i\lambda_i^2\leq1/8$, and let $\mathcal{B}_{\beta, W}$ be the set in \Cref{def:candidate-set} with parameters $\beta$ and $W$ satisfying:
\begin{equation}    
0 < \beta \leq 1/100, \qquad W\geq\max\{6,\,8(128+64\beta^{-4})\}. \label{eq:lower-bound-parameter-W}
\end{equation}
Then
\begin{equation}
 \frac12B_0\leq B_{\mathrm{band}}\leq B_0,
 \qquad
 1+\sum_{(h,j)\in\mathcal B}m^{\mathcal{B}_{\beta, W}}_{h,j}\leq O_{\beta, W}(1)B_0.
 \label{eq:candidate-weight-control}
\end{equation}
More precisely, for $1\leq h\leq D$,
\begin{equation}
 \sum_jm^{\mathcal{B}_{\beta, W}}_{h,j}\leq O_{\beta, W}(1)B_0\rho^{D-h},
 \qquad \rho:=2e^{-1/(1+\beta)}<1.
 \label{eq:candidate-level-weight}
\end{equation}
\end{lemma}

\begin{proof}
Fix parameters $\beta$ and $W$ satisfying \eqref{eq:lower-bound-parameter-W}. We drop the subscripts in $\mathcal{B}_{\beta, W}$ and use $\mathcal{B}$ instead.
First we show that the band preserves last-level weight. Draw $D$ independent uniform points in $[0,D]$, partitioned into column intervals $(x_{j-1},x_j]$, and let $F(s)$ count points in $[0,s]$. Let $\mathcal A$ be the event that no two points occupy the same column. As in the proof of \Cref{lem:baseline-path-probability},
\[
 \Pr(\mathcal A^c)\leq\binom D2\sum_j(\lambda_j/D)^2\leq1/4,
 \qquad
 B_0=\frac{(D/2)^D}{D!}\Pr(\mathcal A).
\]
By \Cref{lemma:path-concentration}, we know that, with probability at least $3/4$,
\begin{equation}
 |F(s)-s|\leq\beta\min\{s,D-s\}+W
 \qquad\text{for every }s\in[0,D].
 \label{eq:uniform-counting-band}
\end{equation}

The no-collision and band events hold together with probability at least $1-1/4-1/4= 1/2$. A particular set of $D$ occupied columns has probability $D!D^{-D}\prod_{j\in S}\lambda_j$, so
\[
 B_{\mathrm{band}}\geq\frac{(D/2)^D}{D!}\frac{1}{2}
 \geq\frac12B_0.
\]
The upper bound $B_{\mathrm{band}}\leq B_0$ follows because candidate paths form a subset of all paths.

Next we bound the weight at every level. Recall that $x_j=\sum_{i=1}^j\lambda_i$. If $j_h = \max\{j:(h,j)\in\mathcal B\}$, the lower boundary of the band implies
\[
 \sum_{i \leq j_h} \lambda_i = x_{j_h}\leq c_h:=\min\left\{D,\frac{h+\beta D+W}{1+\beta}\right\}.
\]
Every candidate path ending at level $h$ uses distinct columns $1\leq i_1<\cdots<i_h\leq j_h$ and has weight $2^{-h}\prod_{r=1}^h\lambda_{i_r}$. Each product of $h$ distinct column weights appears $h!$ times when the $h$th power of their total weight is expanded. Hence
\begin{align*}
 \sum_jm^{\mathcal B}_{h,j}
 &\leq 2^{-h}\sum_{1\leq i_1<\cdots<i_h\leq j_h}\prod_{r=1}^h\lambda_{i_r}\\
 &\leq\frac{2^{-h}}{h!}\left(\sum_{i=1}^{j_h}\lambda_i\right)^h
 \leq\frac{2^{-h}c_h^h}{h!}.
\end{align*}

Using $B_0=\Theta((e/2)^D/\sqrt D)$, $\rho=2e^{-1/(1+\beta)}$, and Stirling's formula, the two inequalities below are equivalent up to constant factors:
\begin{align*}
    & 2^{-h}\frac{c_h^h}{h!}
      \leq O_{\beta,W}(1)B_0\rho^{D-h}
      =O_{\beta,W}(1)\frac{2^{-h}}{\sqrt D}
        e^{\frac{\beta D+h}{1+\beta}}
    \\ & \iff
    \sqrt{\frac Dh}\left(\frac{c_h}{h}\right)^h
      \leq O_{\beta,W}(1)e^{\frac{\beta}{1+\beta}(D-h)}.
\end{align*}
Put $\gamma:=\beta/(1+\beta)$ and $d:=D-h$. Since $c_h\leq h+\gamma d+W/(1+\beta)$,
\[
 \sqrt{\frac Dh}\left(\frac{c_h}{h}\right)^h e^{-\gamma d}
 \leq e^{W/(1+\beta)}\sqrt{\frac Dh}
       \left(1+\frac{\gamma d}{h}\right)^h e^{-\gamma d}.
\]
If $h\geq D/2$, the right-hand side is at most $\sqrt2\,e^{W/(1+\beta)}$, using $1+x\leq e^x$. If $h<D/2$, then $d/h>1$ and $\log(1+\gamma d/h)\leq(d/h)\log(1+\gamma)$, because $\log(1+\gamma t)/t$ is decreasing for $t>0$. Thus the right-hand side is at most
\[
 e^{W/(1+\beta)}\sqrt D\,
 e^{-[\gamma-\log(1+\gamma)]D/2}
 =O_{\beta,W}(1),
\]
since $\gamma-\log(1+\gamma)>0$.

This proves \eqref{eq:candidate-level-weight}. Since $\rho<1$, summing the geometric series bounds all candidate masses by $O_{\beta,W}(1)B_0$.
\end{proof}

\begin{restatable}{lemma}{pathconcentrationlemma} \label{lemma:path-concentration}
Draw $D$ independent uniform points in $[0,D]$, partitioned into column intervals $(x_{j-1},x_j]$, and let $F(s)$ count points in $[0,s]$. Then, with probability at least $3/4$,
\begin{equation}
 |F(s)-s|\leq\beta\min\{s,D-s\}+W
 \qquad\text{for every }s\in[0,D].
 \label{eq:uniform-counting-band-inside-lemma}
\end{equation}
Here the parameters $\beta$ and $W$ satisfy \eqref{eq:lower-bound-parameter-W}.
\end{restatable}
The proof appears in Appendix~\ref{app:path-concentration}.
\subsubsection*{Mass Pruning}

Mass pruning removes candidate locations whose reference expected late populations are too small for the concentration estimates in \eqref{eq:individual-node-deletion-upperbound}. We choose the cutoffs to ensure enough mass while guaranteeing that pruning preserves a constant fraction of the last-level mass, or, equivalently, bounds the mass lost along paths passing through the pruned locations.

Now, we explain how the mass thresholds for mass pruning are defined.
The mass thresholds we define have the following form:
\begin{equation}
 \varepsilon_j:=\delta\left(\theta^2\lambda_j\right)^{a\theta},
 \qquad
 \tau_j:=\frac{T}{u_j}\left(\frac{1}{\theta^2\lambda_j}\right)^{4a\theta},
 \qquad 1\leq j\leq N.
 \label{eq:mass-pruning-thresholds}
\end{equation}

Here the four parameters $\alpha, a, T, \delta$ are selected as follows:
\begin{align*}    
 \alpha:=(1-\beta)\log2-\tfrac12>0,
 \qquad \frac1{4\theta}<a<\frac\alpha8, \\
  0<\delta\leq\min\left\{1,\frac{\log2}{\theta^{-2}+a^{-1}}\right\},
 \qquad T\geq3/\delta.
\end{align*}
All four are constants independent of $N$. Such an $a$ exists for $\theta\geq64$ and $\beta\leq1/100$.
Thus \Cref{alg:rsd-integer-construction} keeps a tentative mass $m_{h,j}$ precisely when
\[
 qm_{h,j}u_j\geq T\left(\frac{1}{\theta^2\lambda_j}\right)^{4a\theta}
 =T\left(\frac{\delta}{\varepsilon_j}\right)^4.
\]

For the harmonic sequence $\lambda_i$, we have
\begin{equation}
k_j=\sum_{j < i \leq N} \lambda_i = \sum_{j < i \leq N} \frac{\theta}{\theta^3 + N - i }.
\end{equation}
By a simple integral estimate, we have

\begin{equation}
\begin{aligned}
    \theta (\log(\theta^3 + N - j) - \log(\theta^3)) = 
    \theta\log\frac{1}{\theta^2\lambda_j}
 \leq  \sum_{j < i \leq N} \lambda_i
 \\
    \text{and}
 \\
 \sum_{j < i \leq N} \lambda_i
 \leq\theta\log\frac{1}{\theta^2\lambda_j}+\theta^{-2} = \theta (\log(\theta^3 + N - j) - \log(\theta^3) + \frac{1}{\theta^3}).
\end{aligned}
 \label{eq:harmonic-clock-weight-comparison}
\end{equation}

These choices also satisfy the condition required by \Cref{lem:bounded-deadline-adjustment}, i.e., $\sum_i \lambda_i \varepsilon_i = O(1)$. More precisely,
\[
\begin{aligned}
 \sum_{j=1}^N\lambda_j\varepsilon_j
 =\delta \cdot \theta^{3a\theta + 1}
       \sum_{m=0}^{N-1}(\theta^3+m)^{-1-a\theta}\leq   &\ \delta \cdot \theta^{3a\theta + 1} (\frac{1}{a\theta}\cdot \frac{1}{\theta^{a 3\theta}} + \frac{1}{\theta^{3 + 3a\theta}})\\
  = & \  \delta(\theta^{-2}+a^{-1})\leq\log2.
\end{aligned}
\]
The upper bound obtained for $\sum_i \lambda_i \varepsilon_i$, together with \Cref{lem:bounded-deadline-adjustment}, implies $u_j/2\leq u_j^*\leq u_j$ for every $j$. Also, $\varepsilon_j\tau_j\geq\delta T\geq3\geq1+2\varepsilon_j$, so the surplus condition in \eqref{eq:adjusted-deadline-surplus} holds after rounding. 

Before mass pruning, we have the following value setup for the replication factor:
\begin{equation}
\begin{aligned}
 A&:=2^W\left(1+\frac{\theta^3}{4a\theta-1}\right),
 &Q&:=\max\{8,\,2AT\},\\
 q&:=\left\lceil\frac{2Q}{B_0u_N^*}\right\rceil.
\end{aligned}
\label{eq:mass-pruning-replication}
\end{equation}
This choice satisfies $q u_NB_{\mathrm{band}}\geq Q$, since $u_N^*\leq u_N$ and $B_{\mathrm{band}}\geq B_0/2$. The replication factor $q$ is then fixed throughout the construction. The constant $T$ can be chosen large enough for the later concentration estimates; increasing $T$ only increases the fixed constant $Q$.

\begin{lemma}[Control of removed mass]
\label{lem:control-removed-mass}
Use the specified values of $\lambda_i$ as in \eqref{eq:harmonic-schedule}, the candidate set in \Cref{def:candidate-set}, and the cutoffs and replication factor in \eqref{eq:mass-pruning-thresholds} and \eqref{eq:mass-pruning-replication}. Let $B_{\mathrm{mass}}:=\sum_jm_{D,j}$ be the last-level mass after \Cref{alg:rsd-integer-construction} finishes mass pruning. For all sufficiently large $N$,
\begin{equation}
 0\leq q u_N(B_{\mathrm{band}}-B_{\mathrm{mass}})\leq AT\leq Q/2.
 \label{eq:removed-last-level-mass}
\end{equation}
Consequently,
\begin{equation}
 B_{\mathrm{mass}}\geq\tfrac12B_{\mathrm{band}}\geq\tfrac14B_0,
 \qquad q u_NB_{\mathrm{mass}}\geq Q/2.
 \label{eq:retained-last-level-mass}
\end{equation}
In particular, after band pruning and mass pruning the last level retains at least a quarter of its initial mass.
\end{lemma}

\begin{proof}
For a candidate $v=(h,j)$, let $\widehat m_v$ denote its tentative mass when it is tested, before it is possibly set to zero (i.e., before applying mass pruning). This mass is the sum of the weights of all root-to-$v$ paths whose earlier vertices have survived mass pruning. If $v$ is discarded by mass pruning, then
\begin{equation}
 q u_j\widehat m_v<T\left(\frac{1}{\theta^2\lambda_j}\right)^{4a\theta}.
 \label{eq:discarded-prefix-threshold}
\end{equation}

Recall the independent Bernoulli variables $r_i$ with probabilities $p_i:=\lambda_i/(2+\lambda_i)$. For any success set $S\subseteq[j]$, its probability up to column $j$ is $u_jw(S)$. Therefore, $u_j\widehat m_v$ is the probability that the successes up to column $j$ pass through retained vertices (which survived mass and band pruning) and end at $v$.

Every candidate path ending at level $D$ that is lost to mass pruning has a unique first discarded vertex $v=(h,j)$. Charge its weight to that vertex. 
Mass pruning indicates that successful paths starting from $(0, 0)$ and ending at $v$ have small total probability. However, to see the effect on the residual weight of the last layer, we need to multiply the current probability by the probability of completion inside the candidate set. For this part, we use a naive upper bound without restricting the remaining locations to the candidate set $\mathcal{B}$. Let $Z_j:=\sum_{i>j}r_i$ be the independent suffix count.

\begin{align}
 u_N(B_{\mathrm{band}}-B_{\mathrm{mass}})
 & = \sum_{v=(h,j)\text{ mass pruned}}
          u_j\widehat m_v \Pr(\text{path from $v$ going through } \mathcal B \text{ ending at level } D) \\&\leq  \sum_{v=(h,j)\text{ mass pruned}}
          u_j\widehat m_v \Pr(Z_j=D-h).
 \label{eq:first-discarded-mass-charge}
\end{align}
This charges each lost path only once, even if several of its vertices are discarded.

Recall the notation $k_j=D-x_j=\sum_{i=j+1}^N\lambda_i$. For a candidate $(h,j)$, the upper boundary of the band gives $D-h\geq(1-\beta)k_j-W$. At a fixed column $j$, distinct levels $h$ give distinct values of $D-h$. Hence
\begin{equation}
\sum_{h:(h,j)\text{ discarded}}\Pr(Z_j=D-h)
 \leq\Pr\bigl(Z_j\geq(1-\beta)k_j-W\bigr). \label{eq:additional-completion-path-plus-mass-pruning}
\end{equation}
To upper bound the R.H.S. of \eqref{eq:additional-completion-path-plus-mass-pruning}, we will use the following form of the Chernoff inequality:
\begin{equation}
    \Pr(Z_j \geq \alpha) \leq \inf_{y>1} \frac{\E[y^{Z_j}]}{y^\alpha} \leq \frac{\E{2^{Z_j}}}{2^\alpha}
\end{equation}
Now, for the variable $Z_j$, we have:
\[
 \E[2^{Z_j}]=\prod_{i>j}(1+p_i)
 \leq\exp\left(\sum_{i>j}p_i\right)
 \leq e^{k_j/2}.
\]
Markov's inequality thus bounds the preceding tail by 
\begin{equation}
\Pr(Z_j \geq (1 - \beta) k_j - W) \leq \frac{e^{k_j/2}2^W}{2^{(1 - \beta)k_j}} = 2^W e^{k_j (1/2 - (1 - \beta)\log(2))} = e^{-\alpha k_j}2^W.
\end{equation}
Combining this with \eqref{eq:discarded-prefix-threshold}, \eqref{eq:first-discarded-mass-charge}, and \eqref{eq:harmonic-clock-weight-comparison} gives
\begin{equation}
\begin{aligned}
 q u_N(B_{\mathrm{band}}-B_{\mathrm{mass}})
 &\leq2^WT\sum_{j=1}^N
       \left(\frac{1}{\theta^2\lambda_j}\right)^{4a\theta}e^{-\alpha k_j}
 \leq2^WT\sum_{j=1}^N
       \left(\theta^2\lambda_j\right)^{(\alpha-4a)\theta},
\end{aligned}
 \label{eq:removed-mass-clock-sum}
\end{equation}
where we used $e^{-\alpha k_j} \leq (\theta^2 \lambda_j)^{\alpha \theta}$, which reduces to the following consequence of \eqref{eq:harmonic-clock-weight-comparison}:
\begin{align*}
e^{k_j} \geq (\theta^2 \lambda_j)^{-\theta} \iff k_j \geq \theta \log(1 + (N - j)/\theta^3 )
\end{align*}

Our choice of $a$ ensures $\alpha-4a>4a$ and $4a\theta>1$. Thus
\begin{align}
 \sum_{j=1}^N\left(\theta^2\lambda_j\right)^{(\alpha-4a)\theta}
 &\leq\sum_{m=0}^{\infty}(1+m/\theta^3)^{-4a\theta}\notag\\
 &\leq1+\int_0^{\infty}(1+s/\theta^3)^{-4a\theta}\,ds
 =1+\frac{\theta^3}{4a\theta-1}.
 \label{eq:harmonic-series-estimate}
\end{align}
Substitution into \eqref{eq:removed-mass-clock-sum} proves the upper bound $AT$ in \eqref{eq:removed-last-level-mass}. Finally, the choice of $q$ gives $q u_NB_{\mathrm{band}}\geq Q\geq2AT$, so at most half of $B_{\mathrm{band}}$ is lost. The bound $B_{\mathrm{band}}\geq B_0/2$ from \Cref{lemma:weight-of-candidate-set} yields \eqref{eq:retained-last-level-mass}.
\end{proof}

\subsection{Putting Everything Together}

Now, we have all the ingredients to prove that the instance produced by \Cref{alg:rsd-integer-construction} with the sequence $\lambda_i = \frac{\theta}{\theta^3 + N - i }$ for $1 \leq i \leq N$ has $n = \Theta(N^{\theta/2})$ agents (items) (\Cref{lemma:size-count-rsd-lowerbound}) and has distortion of $\Omega(\frac{n}{\sqrt{\log(n)}})$.

\begin{proof}[Proof of \Cref{thm:rsd-lowerbound}]
Pick $N \in \mathbf{N}$. Let's fix the sequence $\lambda_{i} = \frac{\theta}{\theta^3 + N - i}$ for $1 \leq i \leq N$. Set $D = \sum \lambda_i$.

Take the candidate set $\mathcal{B}$ as in \Cref{def:candidate-set} with parameters $\beta$ and $W$ satisfying \eqref{eq:lower-bound-parameter-W}, and set the cutoffs $\tau_j$ as in \eqref{eq:mass-pruning-thresholds}.

Let $\sigma_n$ be the output of \Cref{alg:rsd-integer-construction} with inputs $(N, D)$, candidate set $\mathcal{B}$ and cutoffs $\tau_j$ for $1 \leq j \leq N$ as in \eqref{eq:mass-pruning-thresholds}.

Recall the reference times $u_i$ and the adjusted times $u_i^*$ in \eqref{eq:adjusted-remaining-times}. Note that, for the specific choice of $\lambda_i$, there exists a sequence $\varepsilon_i$, specified in \eqref{eq:mass-pruning-thresholds}, such that:
\begin{itemize}
    \item $\sum_j\varepsilon_j\lambda_j\leq\log2$, so $u_N^*\geq u_N/2$ by \Cref{lem:bounded-deadline-adjustment}.
    \item After applying \textit{mass pruning} with cutoffs $\tau_j$, the residual mass at level $D$ is at least a quarter of the initial mass (\Cref{lem:control-removed-mass}). We refer to the residual mass at level $D$ as $B_{\mathrm{mass}}$.
\end{itemize}

Recall the replication factor in \eqref{eq:mass-pruning-replication}:
\begin{equation}
 q:=\left\lceil\frac{2Q}{B_0u_N^*}\right\rceil
 \label{eq:final-rsd-replication}
\end{equation}

This choice satisfies $q u_NB_{\mathrm{band}}\geq Q$, since $u_N^*\leq u_N$ and $B_{\mathrm{band}}\geq B_0/2$. The proof of \Cref{lem:control-removed-mass} uses only this lower bound on $q$, so it still gives $B_{\mathrm{mass}}\geq B_0/4$.

Combining \Cref{lem:baseline-path-probability} and \eqref{eq:reference-last-level-weight} with $u_N^*=\Theta(u_N)$ gives
\[
 B_0=\Theta\!\left(\frac{(e/2)^D}{\sqrt D}\right),
 \qquad
 u_N^*=\Theta(e^{-D/2}),
 \qquad
 q=\Theta\!\left(\sqrt D\,(2/\sqrt e)^D\right).
\]
Recall that showing the successful cascade inequality in \eqref{eq:successful-cascade} implies that $\rsd$ has distortion of at least $\frac{\E[(L - R)_+]2^{D- 1}}{q} \geq \frac{2^{D - 1}}{q}$, since the optimal value is $q$ and the penalty for any misplacement at level $D$ is $2^{D - 1}$.

If \eqref{eq:successful-cascade} holds, then with the current parameter setup we have:
\begin{equation}
\begin{aligned}
 \frac{2^{D-1}}q
 &=\Theta(2^D B_0u_N^*)
 =\Theta\!\left(\frac{2^D(e/2)^D e^{-D/2}}{\sqrt D}\right)\\
 &=\Theta\!\left(\frac{e^{D/2}}{\sqrt D}\right)
 =\Theta\!\left(\frac{n}{\sqrt{\log n}}\right).
\end{aligned}
\label{eq:rsd-final-distortion-scaling}
\end{equation}
The last equality uses $n=\Theta(e^{D/2})$ and $D=\Theta(\log n)$ from \Cref{lemma:size-count-rsd-lowerbound}. Equivalently, this is a lower bound of order $N^{\theta/2}/\sqrt{\log N}$ for fixed $\theta$.

We now show that the cascade is successful, as required in \eqref{eq:successful-cascade}. It suffices to prove
\[
 \E[L]\geq2
 \qquad\text{and}\qquad
 \E[R]\leq1,
\]
since then $\E[(L-R)_+]\geq\E[L]-\E[R]\geq1$.

For $\E[L]$, the retained last-level mass gives
\begin{equation}
 \E[L]=u_N^*\sum_jm^*_{D,j}
 \geq q u_N^*B_{\mathrm{mass}}
 \geq\frac{q u_N^*B_0}{4}
 \geq\frac Q2\geq2.
 \label{eq:rsd-final-late-population}
\end{equation}

For $\E[R]$, the coupling result \Cref{cor:rsd-deletion-coupling} and the individual deletion bound \Cref{lem:individual-node-deletion} reduce the claim to
\begin{equation}
 \sum_{(h,j)\in\mathcal B}\E[(V_{h,j}-U_{h,j})_+]\leq1.
 \label{eq:rsd-total-deletion-target}
\end{equation}
Locations discarded by pruning have $V_{h,j}=0$ and contribute nothing. We bound the deficit tails at each retained location and then sum over the band.

We use \Cref{lem:rsd-deficit-after-mass-pruning} below to bound the deficit at each retained location.
There are at most $2\beta k_j+2W+1$ candidate locations in column $j$. By \eqref{eq:harmonic-clock-weight-comparison} and $\log x\leq x^{a\theta}/(a\theta)$ for $x\geq1$,
\[
 2\beta k_j+2W+1
 \leq\left(2W+2+\frac{2\beta}{a}\right)
       \left(\frac{1}{\theta^2\lambda_j}\right)^{a\theta}.
\]
Choose $T\geq36/\delta^2$, and let
\[
 c:=\frac{\delta^2T}{12}\geq3,
 \qquad
 z_j:=(\theta^2\lambda_j)^{-2a\theta}\geq1.
\]
Since $z^3e^{-cz}$ is nonincreasing on $[1,\infty)$,
\[
 z_j e^{-cz_j}
 \leq e^{-c}z_j^{-2}
 =e^{-c}(\theta^2\lambda_j)^{4a\theta}.
\]
Applying this bound in \eqref{eq:rsd-deficit-after-mass-pruning} and summing over the candidate locations gives:
\begin{equation}
\begin{aligned}
 \sum_{(h,j)\in\mathcal B}\E[(V_{h,j}-U_{h,j})_+]
 &\leq\frac2\delta\left(2W+2+\frac{2\beta}{a}\right)
 e^{-\delta^2T/12}
 \sum_{j=1}^N\left(\theta^2\lambda_j\right)^{4a\theta}\\
 &\leq\frac2\delta\left(2W+2+\frac{2\beta}{a}\right)
 \left(1+\frac{\theta^3}{4a\theta-1}\right)e^{-\delta^2T/12}
 \leq1.
\end{aligned}
\label{eq:rsd-total-deletion-bound}
\end{equation}
The second line uses the harmonic-series estimate in \eqref{eq:harmonic-series-estimate}; it is finite because $4a\theta>1$. The last inequality holds for a sufficiently large fixed $T$. Choose $Q$ after $T$ as specified above. The coupling bound \eqref{eq:rsd-residual-from-deletions} now gives $\E[R]\leq1$. Together with \eqref{eq:rsd-final-late-population}, this proves \eqref{eq:successful-cascade} and completes the distortion lower bound in \eqref{eq:rsd-final-distortion-scaling}.

\end{proof}

\begin{lemma}[Deficit bound after mass pruning]
\label{lem:rsd-deficit-after-mass-pruning}
For every retained location $v=(h,j)$ with $h\geq1$, under the mass-pruning parameters in \eqref{eq:mass-pruning-thresholds},
\begin{equation}
 \E[(V_v-U_v)_+]
 \leq\frac2\delta
 \left(\frac{1}{\theta^2\lambda_j}\right)^{a\theta}
 \exp\!\left[-\frac{\delta^2T}{12}
       \left(\frac{1}{\theta^2\lambda_j}\right)^{2a\theta}\right].
 \label{eq:rsd-deficit-after-mass-pruning}
\end{equation}
\end{lemma}

\begin{proof}
Fix a retained $v=(h,j)$. The variables $U_v$ and $V_v$ count disjoint sets of agents, so they are independent sums of Bernoulli variables. Also $\E[V_v]=m^*_{h,j}u_j^*$ and $\E[U_v]\geq(1+\varepsilon_j)\E[V_v]$ by \eqref{eq:adjusted-deadline-surplus}. For the integer-valued deficit,
\[
 \E[(V_v-U_v)_+]
 =\sum_{\ell=1}^{\infty}\Pr(V_v-U_v\geq\ell).
\]
Markov's inequality with base $1+\varepsilon_j/2$, followed by summing the geometric series and using the Bernoulli moment bound $1+x\leq e^x$, gives
\begin{equation}
\begin{aligned}
 \E[(V_v-U_v)_+]
 &\leq\frac2{\varepsilon_j}
 \exp\!\left(\frac{\varepsilon_j}{2}\E[V_v]
       -\frac{\varepsilon_j}{2+\varepsilon_j}\E[U_v]\right)\\
 &\leq\frac2{\varepsilon_j}
 \exp\!\left(-\frac{\varepsilon_j^2}{4+2\varepsilon_j}\E[V_v]\right)\\
 &\leq\frac2{\varepsilon_j}
 \exp\!\left(-\frac{\varepsilon_j^2m^*_{h,j}u_j^*}{6}\right),
\end{aligned}
\label{eq:rsd-local-deficit-tail}
\end{equation}
where the last step uses $\varepsilon_j\leq1$.

The mass-pruning cutoff in \eqref{eq:mass-pruning-thresholds}, together with $u_j^*\geq u_j/2$, now yields \eqref{eq:rsd-deficit-after-mass-pruning}.
\end{proof}

\begin{lemma}[Size count of the scaled instance]
\label{lemma:size-count-rsd-lowerbound}
For the construction above, with fixed parameters and $q$ as in \eqref{eq:final-rsd-replication}, suppose $u_N^*=\Theta(u_N)$. Then
\begin{equation}
 n=\Theta(qB_0)=\Theta(e^{D/2})=\Theta(N^{\theta/2}),
 \qquad D=\Theta(\log n).
 \label{eq:rsd-instance-size}
\end{equation}
\end{lemma}

\begin{proof}
Every retained nonroot location satisfies $qm_{h,j}\geq\tau_j\geq1$, so rounding gives $qm_{h,j}\leq m^*_{h,j}\leq2qm_{h,j}$. Mass pruning only decreases the candidate masses. By \Cref{lemma:weight-of-candidate-set,lem:control-removed-mass},
\[
\begin{aligned}
 n&\geq qB_{\mathrm{mass}}\geq qB_0/4,\\
 n&=q+\sum_{(h,j)\in\mathcal B}m^*_{h,j}
 \leq2q\left(1+\sum_{(h,j)\in\mathcal B}m^{\mathcal B}_{h,j}\right)
 =O(qB_0).
\end{aligned}
\]
Hence $n=\Theta(qB_0)$. Since $B_0u_N^*=\Theta((\sqrt e/2)^D/\sqrt D)\to0$, the ceiling in \eqref{eq:final-rsd-replication} changes $q$ by at most a constant factor. Consequently,
\[
 qB_0=\Theta(1/u_N^*)=\Theta(e^{D/2}),
\]
which also implies $D=\Theta(\log n)$. Finally, comparison of the harmonic sum in \eqref{eq:harmonic-schedule} with its integral gives
\[
 D=\theta\log(1+N/\theta^3)+O_\theta(1)
  =\theta\log N+O_\theta(1).
\]
For fixed $\theta$, this yields $e^{D/2}=\Theta(N^{\theta/2})$ and proves \eqref{eq:rsd-instance-size}.
\end{proof}

\subsection{Coupling between Artificial and Actual $\rsd$} \label{subsection:Coupling-between-artificial-and-rsd}

We start with two deterministic serial dictatorship (SD) executions using the same agent order and the same preference rankings, including the same tie-breaking. In one execution, items may additionally be deleted at arbitrary times. Assume that this execution has enough dummy items, which are least preferred by every agent, and agents will be assigned to them if none of the initial items remain. We refer to the initial items as real items.
If $k$ real items have been deleted by time $t$, then we claim that the two sets of remaining real items differ by at most $k$ items at any time of the SD execution. Throughout this section, we assume agents are assigned distinct real numbers in $(0, 1)$.

\begin{lemma}[Stability of SD under item deletions]
\label{lem:sd-deletion-stability}
Let $S(t)$ and $\widetilde S(t)$ be the sets of real items remaining at time $0 \leq t \leq 1$ in an ordinary SD execution and an execution with deletions, respectively. The two executions start with the same real items and process the same agents in the same order with identical rankings. If $K(t)$ is the total number of real items deleted from the second execution by time $t$, then
\begin{equation}
\widetilde S(t)\subseteq S(t),
\qquad
\bigl|S(t)\mathbin{\triangle}\widetilde S(t)\bigr|
=\bigl|S(t)\setminus\widetilde S(t)\bigr|
\leq K(t).
\label{eq:sd-inventory-stability}
\end{equation}
\end{lemma}

\begin{proof}
Induct over the sequence of agent arrivals and deletion events. Initially, the sets of real items are equal and $K=0$. Deleting $r$ available real items from $\widetilde S$ preserves containment and increases both $|S\setminus\widetilde S|$ and $K$ by $r$.

Now consider an agent arrival. If the ordinary execution has no real item left, neither does the execution with deletions and all real items are assigned in both cases. Otherwise, let $x$ be the agent's favorite item in $S$. If $x\in\widetilde S$, it is also the favorite item in $\widetilde S$, so both executions remove $x$ and the difference is unchanged. If $x\notin\widetilde S$ and the second execution chooses a real item $y$, then $y\in\widetilde S\subseteq S$ and $y\neq x$. After the choices, containment still holds and the new difference set is
\[
\bigl((S\setminus\widetilde S)\setminus\{x\}\bigr)\cup\{y\},
\]
which has the same size as before. Finally, if the second execution takes a dummy, $\widetilde S$ is empty and removing $x$ from $S$ decreases the difference by one. Thus an agent arrival never increases the gap, while every deletion increases it by exactly the number deleted. This proves the claim at every time $0 \leq t \leq 1$.
\end{proof}

As an immediate corollary, we have the following implication for $\rsd$.
\begin{corollary}[Coupling $\rsd$ with the artificial process]
\label{cor:rsd-deletion-coupling}
Couple actual and artificial $\rsd$ using the same independent $\mathrm{Unif}[0,1]$ arrival times and the same preference rankings. Let $K(t)$ count all real items deleted by the artificial process by time $t$. Then,
\begin{equation}
0\leq \E[|(S(t)\setminus\widetilde S(t))|] \leq \E[K(t)].
\label{eq:rsd-subset-inventory}
\end{equation}
\end{corollary}

The above corollary justifies the first inequality of \eqref{eq:bounding-removal-imply-bounding-last-level-item}.
\begin{equation}
\E[R]\leq \E[K(t^*)]
\leq\sum_{(h,j)}\E[(V_{h,j}-U_{h,j})_+].
\label{eq:rsd-residual-from-deletions}
\end{equation}
Here $K(t^*)$ counts deletions across the entire instance until the deadline $t^* = 1 - u_N^*$.

\section{Probabilistic Serial Rule}\label{sec:ps}

The probabilistic serial (PS) rule, also referred to
as a simultaneous eating algorithm, is an ordinal mechanism
introduced by \cite{Bogomolnaia:2001aa}. 
Its operation is simple and intuitive. 
Under PS, each agent continuously consumes her most-preferred available 
item at unit rate; this occurs in parallel. 
Whenever that item is exhausted, she switches to her most-preferred remaining item. The process ends at time $1$, 
and each agent's fractional allocation records how much of each item she consumed.
A formal description is given in 
\Cref{alg:ps}.

\begin{center}
\myalg{alg:ps}{Probabilistic Serial Rule}{
\textbf{Input:} An instance $\sigma$.  \medskip \\
Set $t\gets0$ and $X(0) \gets 0_{n \times n}$. \\
Each item $b$ initially has capacity $1$, and is exhausted when $\sum_a X_{a,b}(t)=1$. \\
\textbf{While} $t<1$:
\begin{itemize}[noitemsep]
\item Each agent $a$ picks her favorite item $b_a$ among those not yet fully eaten.
\item All agents simultaneously eat their picked items at unit rate: for $s\ge t$,
    \[
      X_{a,b_a}(s)=X_{a,b_a}(t)+(s-t),
    \]
    with all other entries unchanged, until the first time $t'$ at which some item
    is fully eaten (or $t'=1$).
\item Set $t\gets t'$.
\end{itemize}
\textbf{Return} $X(1)$.
}
\end{center}
PS maintains a matrix $X$, where $X_{a, b}(t)$ denotes the fraction of item $b$ assigned to 
agent $a$ during the interval $[0, t]$. When $t = 1$,
the matrix is bistochastic and defines a perfect fractional matching.
By the Birkhoff--von Neumann theorem
\citep{Birkhoff:1946aa,Von-Neumann:1953aa}, the fractional
allocation can be implemented as a distribution over
perfect integral matchings.
For any instance $\sigma$ and metric $d$, we have
\begin{equation*}
\E_{M \sim \ps(\sigma)} [\cost(M, d)] = \sum_{a \in A} \sum_{b \in B} X_{a,b}(1)\ d(a, b) = \cost(X(1), d).
\end{equation*}

PS is a classic mechanism in the economics literature and several of its properties are understood. 
For example, it is known that PS produces a fractional allocation that 
is envy-free in the stochastic-dominance sense \cite{Bogomolnaia:2001aa}, 
and 
for a fixed set of object types, the assignment probabilities 
under PS and RSD converge as the number of copies of each type grows \citep*{Che:2010aa}. 
However, to our knowledge, no metric distortion guarantees for PS appear in the literature.
We address this gap, establishing both a lower
bound (in \Cref{sec:lb-ps}) and an upper bound 
(in \Cref{sec:ub-ps}).

\subsection{Distortion Lower Bound}\label{sec:lb-ps}

For many ordinal mechanisms, a natural starting point for establishing distortion lower bounds is 
to consider their behavior on line metrics
as in \cite*{Caragiannis:2016aa,Caragiannis:2024ab}. As we mention in the introduction,
the line instance used for the exponential lower bound for 
\emph{Serial Dictatorship} also gives a linear lower bound for PS;
see \Cref{fig:intro-example}.
In this section, we show that PS could perform even more poorly.
In fact, it can have exponential distortion.

\begin{restatable}{theorem}{thmpslb}
\label{thm:lb-ps}
For arbitrarily large $n$, there exists an instance $\sigma$ such that 
\[
\distortion(\ps, \sigma) = \Omega \left ( (10/9)^{n/2} \right ).
\] 
\end{restatable}

For each integer \(H\ge2\), we construct an ordinal instance \(\sigma_H\) and
a consistent pseudometric \(d_H\), induced by shortest-path distances in the following weighted graph.
We write each vertex as $(k,h)$, where $k$ is its column and $h$ is its height.
\begin{enumerate}[noitemsep]
    \item Column $-1$ contains a single vertex $(-1,0)$.
    \item Column $0$ contains a single vertex $(0,0)$, joined to $(-1,0)$ by an edge of length $1$.
    \item For each column $k$ with $1\le k<H$, there are two vertices $(k,0)$ and $(k,1)$. Each is joined to every vertex in column $k-1$ by an edge of length $2^{k-1}$.
    \item Column $H$ contains a single vertex $(H,0)$, joined to every vertex in column $H-1$ by an edge of length $2^{H-1}$.
\end{enumerate}
Vertex $(-1,0)$ contains a single item, and vertex $(0,0)$ contains a single agent.
Every other vertex contains one agent and one item.
We denote the agent and item at $(k,h)$, when present, by $a_{k,h}$ and $b_{k,h}$, respectively.
The instance has $n=2H$ agents and $n=2H$ items.
For a vertex in column $k$, we call the vertices in column $k+1$ its \textit{children}
and the vertices in column $k-1$ its \textit{parents}, whenever those columns exist.
In \Cref{fig:lb-ps-generic}, the height coordinate increases downward: height $0$ is the upper row and height $1$ is the lower row.
The drawing is schematic; distances are determined by the edge lengths.

\usetikzlibrary{arrows.meta,calc,backgrounds}
\definecolor{psblue}{RGB}{35,84,136}
\definecolor{pspale}{HTML}{D9ECF6}
\definecolor{psgray}{HTML}{61666C}
\newsavebox{\psladderreferencebox}
\newif\ifpsladderoverlay
\pgfkeys{
  /ps ladder/.is family,
  /ps ladder/overlay/.is if=psladderoverlay,
  /ps ladder/overlay/.default=true
}

\tikzset{
  ps consumed/.style={
    path picture={
      \pgfpointdiff
        {\pgfpointanchor{path picture bounding box}{center}}
        {\pgfpointanchor{path picture bounding box}{north}}
      \pgfgetlastxy{\ladderclockunused}{\ladderclockradius}
      \path[fill=black!12,draw=none]
        (path picture bounding box.center) --
        (path picture bounding box.north)
        arc[
          start angle=90,
          delta angle={-360*(#1)},
          radius=\ladderclockradius
        ] -- cycle;
    }
  }
}

\newcommand{\psladder}[3][]{%
\begingroup
\footnotesize
\pgfmathsetlengthmacro{\ladderlabelwidth}{width("$(H-1,0)$")}
\pgfmathsetlengthmacro{\ladderlabelheight}{height("$(H-1,0)$")}
\pgfmathsetlengthmacro{\ladderlabeldepth}{depth("$(H-1,0)$")}
\psladderoverlayfalse
\sbox{\psladderreferencebox}{\psladderpicture{#2}{#3}}%
\pgfkeys{/ps ladder/.cd,#1}%
\resizebox{0.95\linewidth}{!}{%
  \ifpsladderoverlay
    \makebox[\wd\psladderreferencebox][c]{\psladderpicture{#2}{#3}}%
  \else
    \usebox{\psladderreferencebox}%
  \fi
}%
\endgroup
}

\newcommand{\psladderpicture}[2]{%
\begin{tikzpicture}[
  x=1cm,y=-1cm,
  every node/.style={font=\small},
  vertex/.style={
    circle,draw=black,fill=none,
    line width=.6pt,
    font=\footnotesize,
    align=center,
    text width=\ladderlabelwidth,
    text height=\ladderlabelheight,
    text depth=\ladderlabeldepth,
    inner sep=2pt,
    outer sep=0pt
  },
  edge/.style={draw=black,line width=.6pt},
  edge label/.style={font=\small,inner sep=1pt},
  ps note/.style={font=\footnotesize,align=center,text=psgray},
  ps blue note/.style={ps note,text=psblue},
  ps eating/.style={edge,-{Latex[length=1.8mm,width=1.25mm]}},
  ps visiting/.style={ps eating,line width=1.25pt,
    -{Latex[length=2.4mm,width=1.65mm]}},
  ps ring/.style={draw=psblue,dashed,line width=.9pt}
]
  \ifpsladderoverlay
    \tikzset{vertex/.append style={fill=white}}
  \fi
  \node[vertex] (surplus) at (0,0) {$(-1,0)$};
  \node[vertex] (root) at (2.5,0) {$(0,0)$};

  \node[vertex,ps consumed=#1] (c1h0) at (5.4,0) {$(1,0)$};
  \node[vertex,ps consumed=#1] (c1h1) at (5.4,2.4) {$(1,1)$};
  \node[vertex,ps consumed=#1] (c2h0) at (8.3,0) {$(2,0)$};
  \node[vertex,ps consumed=#2] (c2h1) at (8.3,2.4) {$(2,1)$};
  \ifpsladderoverlay
    \node[vertex,ps consumed=#2] (c3h0) at (11.2,0) {$(3,0)$};
    \node[vertex,ps consumed=#2] (c3h1) at (11.2,2.4) {$(3,1)$};
    \coordinate (c4h0) at (14.1,0);
    \coordinate (c4h1) at (14.1,2.4);
  \else
    \node[vertex,ps consumed=#2] (penh0) at (11.7,0) {$(H-1,0)$};
    \node[vertex,ps consumed=#2] (penh1) at (11.7,2.4) {$(H-1,1)$};
    \node[vertex,ps consumed=#2] (terminal) at (14.6,0) {$(H,0)$};
  \fi

  \draw[edge] (surplus) -- node[edge label,midway,above] {$1$} (root);
  \draw[edge] (root) -- (c1h1);
  \draw[edge] (root) -- node[edge label,midway,above] {$1$} (c1h0);
  \draw[edge] (c1h1) -- (c2h1);
  \draw[edge] (c1h1) -- (c2h0);
  \draw[edge] (c1h0) -- (c2h1);
  \draw[edge] (c1h0) -- node[edge label,midway,above] {$2$} (c2h0);

  \ifpsladderoverlay
    \draw[edge] (c2h1) -- (c3h1);
    \draw[edge] (c2h1) -- (c3h0);
    \draw[edge] (c2h0) -- (c3h1);
    \draw[edge] (c2h0) -- node[edge label,midway,above] {$4$} (c3h0);
    \foreach \a/\b in {c3h0/c4h0,c3h0/c4h1,c3h1/c4h0,c3h1/c4h1} {
      \draw[edge] (\a) -- ($(\a.center)!.4!(\b)$);
      \path (\a.center) -- node[pos=.67,sloped] {$\cdots$} (\b);
    }
    \node[edge label,above] at (12.15,-.05) {$8$};
    \psladderannotations
  \else
    \draw[edge] (c2h1.east) -- (9.4,2.4);
    \draw[edge] (c2h0.east) -- (9.4,0);
    \node at (10,2.4) {$\cdots$};
    \node at (10,0) {$\cdots$};
    \draw[edge] (10.6,2.4) -- (penh1.west);
    \draw[edge] (10.6,0) -- (penh0.west);
    \draw[edge] (c2h1) -- (9.35,1.66);
    \draw[edge] (c2h0) -- (9.35,.74);
    \node[rotate=35] at (9.65,1.45) {$\cdots$};
    \node[rotate=-35] at (9.65,.95) {$\cdots$};
    \node[rotate=-35] at (10.35,1.45) {$\cdots$};
    \node[rotate=35] at (10.35,.95) {$\cdots$};
    \draw[edge] (10.65,1.66) -- (penh1);
    \draw[edge] (10.65,.74) -- (penh0);
    \draw[edge] (penh1) -- (terminal);
    \draw[edge] (penh0) -- node[edge label,midway,above] {$2^{H-1}$} (terminal);
  \fi
\end{tikzpicture}%
}

\newcommand{\psladderannotations}{%
  \begin{scope}[on background layer]
    \path[fill=psblue,fill opacity=.12,draw=psblue!45,line width=.6pt]
      (-.98,0)
      .. controls (-.98,-.65) and (-.65,-.9) .. (0,-.9)
      .. controls (1.7,-.9) and (3.7,-.9) .. (5.4,-.9)
      .. controls (6.05,-.9) and (6.35,-.55) .. (6.35,0)
      .. controls (6.35,.6) and (6.0,.95) .. (5.4,.95)
      .. controls (3.6,.95) and (1.8,.95) .. (0,.95)
      .. controls (-.65,.95) and (-.98,.6) .. (-.98,0)
      -- cycle;
  \end{scope}
  \node[ps blue note] at (2.0,1.45) {Unstable component};

  \pgfpointdiff{\pgfpointanchor{c1h0}{center}}{\pgfpointanchor{c1h0}{east}}
  \pgfgetlastxy{\laddervertexradius}{\ladderunused}
  \pgfmathsetlengthmacro{\ladderringradius}{\laddervertexradius+3pt}
  \draw[ps ring] (c1h1.center) circle[radius=\ladderringradius];
  \draw[ps ring] (c2h0.center) circle[radius=\ladderringradius];
  \node[ps blue note,anchor=north] at ([yshift=-10pt]c1h1.south)
    {Agent $a_{1,1}$\\leaves own item};
  \node[ps blue note,anchor=south] at ([yshift=10pt]c2h0.north)
    {Agent $a_{2,0}$\\leaves own item};

  \draw[ps eating] (root) -- (surplus);
  \draw[ps visiting] (c1h0) -- (c2h1);
  \draw[ps visiting] (c1h1) -- (c2h1);
  \draw[ps visiting] (c2h0) -- (c3h0);
  \node[ps note,anchor=north] at ([yshift=-10pt]c2h1.south) {Block $b_{2,1}$};
  \node[ps note,anchor=south] at ([yshift=10pt]c3h0.north) {Block $b_{3,0}$};
}

\begin{figure}[th]
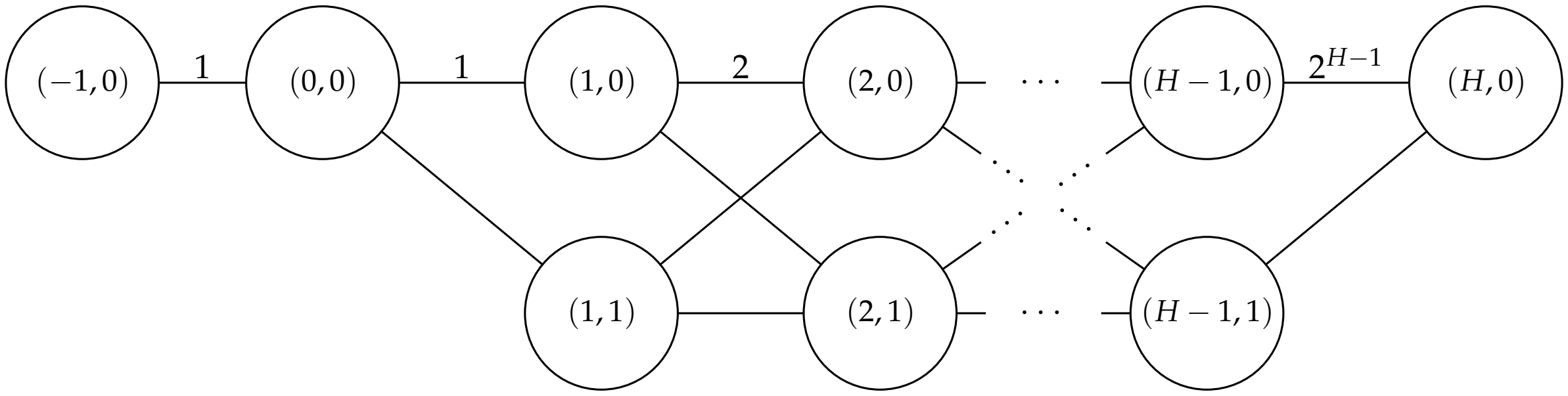

\centering
\psladder{0}{0}
\caption{Lower-bound instance for PS. The vertex $(0,0)$ has a single agent, vertex $(-1,0)$ has
only a surplus item, and every other vertex has one agent--item
pair.
A single length label applies to all edges between the indicated
adjacent columns.}
\label{fig:lb-ps-generic}
\end{figure}

For $0\le j<k\le H$, the distance between vertices in columns $j$ and $k$ is
$\sum_{\ell=j}^{k-1}2^\ell=2^k-2^j$.
Thus, for each agent $a_{k,h}$, the surplus item and her children's items, whenever present, are at distance $2^k$.
The other item in her column, when present, is also at distance $2^k$, attained by a path through a parent.
Her collocated item, when present, and all nonsurplus items in earlier columns are closer, while items beyond column $k+1$ are farther away.

Each agent ranks items by nondecreasing distance. Among items at distance $2^k$, agent $a_{k,h}$ ranks her children's items first, in increasing order of height, followed by $b_{-1,0}$ and then the other item in her column, omitting nonexistent items. All remaining distance ties are broken by increasing height.

\begin{claim}\label{obs:lb-ps-surplus}
Under PS on $\sigma_H$, every agent other than $a_{0,0}$ consumes her collocated item until it is exhausted. The surplus item $b_{-1,0}$ remains available at every time $t<1$.
\end{claim}
\begin{proof}
Each agent other than $a_{0,0}$ has a collocated item at distance zero, strictly closer than every other item. Hence she consumes it at unit rate until its exhaustion.
At time $t<1$, each nonsurplus item therefore has remaining capacity at most $1-t$.
Since all $n$ agents consume at unit rate, the total remaining capacity is $n(1-t)$.
There are $n-1$ nonsurplus items, so $b_{-1,0}$ has remaining capacity at least
$n(1-t)-(n-1)(1-t)=1-t>0$.
\end{proof}

Signposting, we briefly give 
some motivation behind the design of the 
lower bound instance. Although PS assigns agents in parallel, we
can demonstrate a cascade
similar to that used by \cite{Caragiannis:2016aa,Caragiannis:2024ab}. Agent $a_{0,0}$ should be assigned item $b_{-1, 0}$, but she only starts eating the item late into the mechanism's operation because of her slight preference for items at the children.
Similarly, other agents will prefer the items 
at their children over $b_{-1, 0}$. Inevitably, even the agent $a_{H,0}$ will have to eat part of item $b_{-1, 0}$; this portion is significant enough to induce large cost.

We now analyze the execution of PS on this lower bound instance.
For each item $b_{k,h}$, let $T_{k,h}$ denote its exhaustion time.

\begin{proposition}\label{prop:lb-ps-exhaustion}
Under PS on instance $\sigma_H$, the following recurrence holds:
\begin{equation}
T_{1,0}=\frac12,\qquad T_{1,1}=\frac34,
\end{equation}
\begin{align}
1-T_{k,0}
&=
\frac14\left(\frac59\right)^{k-2}
\qquad (2\le k\le H), \\
1-T_{k,1}
&=
\frac16\left(\frac59\right)^{k-2}
\qquad (2\le k\le H-1).
\end{align}
Moreover, for every $k$ such that $1 \leq k \leq H- 1$, we have
$T_{k,0}<T_{k,1}\le T_{k+1,0} $,
with equality in the second inequality only for $k=1$.
\end{proposition}

\begin{proof}

We proceed by induction on $k$.
For $2\le k\le H-1$, we will show that $T_{k,0}<1$ and
also that the following properties hold immediately after time $T_{k,0}$:

\begin{enumerate}[noitemsep]
\item Every item in columns $1,\ldots,k-1$ is exhausted. Moreover, the agents in the columns $0,\ldots,k-2$ are eating the
item at $(-1, 0)$.

\item
The agents at column $k-1$ are eating the item at $(k,1)$, and the newly
available agent at $(k,0)$ is now eating the item $(k+1,0)$. The agent at  $(k,1)$ and all agents in later columns are eating their own items.

\item Up to time $T_{k,0}$, the item at $(k,1)$ and every item in a later
column have been eaten only by their own collocated agents. Each therefore has
remaining capacity $1 - T_{k,0}$.
\end{enumerate}

We begin by analyzing the first phase of the mechanism.
The item $b_{1,0}$ is eaten by its collocated agent $a_{1,0}$ and agent $a_{0,0}$ for the time period $[0, \frac{1}{2})$. All other agents eat $\frac{1}{2}$ of their collocated item in this period. Thus, $T_{1,0} = \frac{1}{2}$.

Now consider the phase during time period $[\frac{1}{2}, \frac{3}{4})$. The item $b_{1,1}$ was partially eaten by its collocated agent during $[0, \frac{1}{2})$, and the agent $a_{0,0}$ also begins to eat this item at time $t = \frac{1}{2}$.
Meanwhile, the newly available agent $a_{1,0}$ begins to eat the item $b_{2,0}$ at time $t = \frac{1}{2}$, joining the agent $a_{2,0}$. All other agents continue to eat their collocated item. Thus,
$
T_{1,1}=T_{2,0}= \frac12 + \frac14 =  \frac34
$. An illustration of the resulting state
is given in \Cref{fig:lb-ps-consume}.

\begin{figure}[ht]
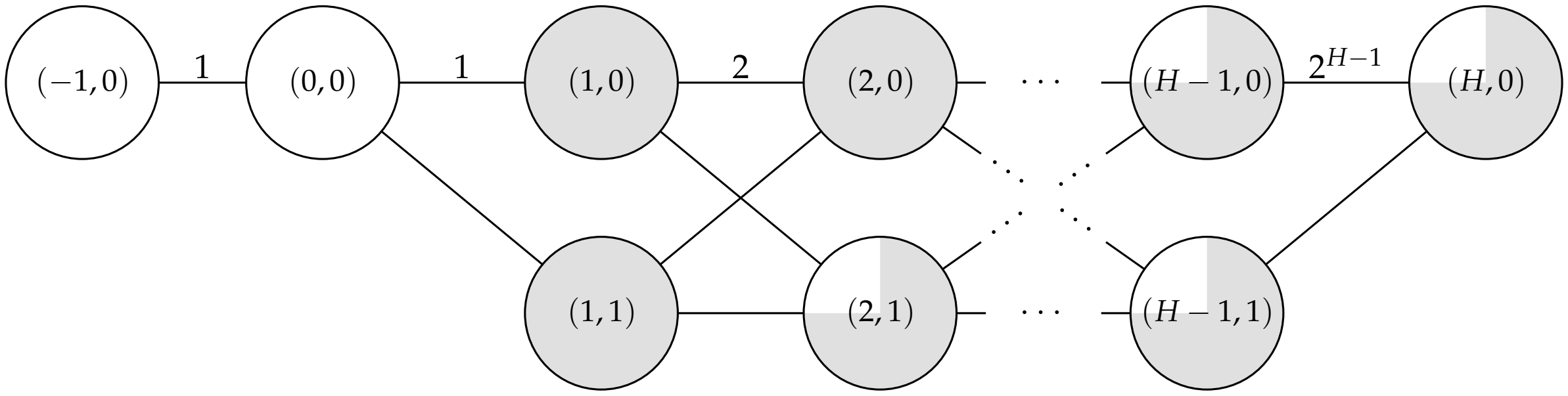

\centering
\psladder{1}{0.75}
\caption{Lower bound instance at time $t=3/4$. Gray sectors indicate the fraction of each item consumed; the unshaded
portion is its remaining capacity. The items $(1,0)$, $(1,1)$, and
$(2,0)$ are exhausted.
The item at $(-1,0)$ is untouched
and every other item is three-quarters
consumed.}
\label{fig:lb-ps-consume}
\end{figure}

If $H=2$, all nonsurplus items are exhausted at time $3/4$, and every agent consumes $b_{-1,0}$ for the rest of the process. The asserted exhaustion times and their order already follow in this case. For the remainder of the proof, assume $H\ge3$.

Consider the time period beginning at $t = T_{2,0}$.
At this moment, the agent $a_{0,0}$ will begin to eat the item $b_{-1,0}$ because the items at her children are exhausted. Furthermore, if $H > 2$,
then the agents in column $1$ will both begin to eat the item $b_{2,1}$ and the agent $a_{2,0}$ will begin to
eat the item $b_{3,0}$.
This establishes our three properties of the inductive hypothesis for $k = 2$.

Now suppose that the induction hypotheses hold for
some $k$ with $2 \leq k \leq H - 1$.
Recall that by definition, the item at vertex $(k,0)$ is already exhausted at time $ T_{k,0} < 1$.
For each item $b_{k,h}$, write $\delta_{k,h}=1-T_{k,h}$ for the time remaining after its exhaustion. By the third induction hypothesis, $b_{k,1}$ and every item in a later column have remaining capacity $\delta_{k,0}$ at time $T_{k,0}$.
We reason about the comparative rates of consumption of remaining items, assuming that a single agent eats at rate $1$.
The item at vertex $(k,1)$ is eaten at rate $3$; the two agents in column $k-1$ join the collocated agent $a_{k,1}$ in consumption. The item at vertex $(k+1,0)$ is consumed at rate $2$; a single agent at parent vertex $(k,0)$ joins the collocated agent $a_{k+1,0}$ in the consumption.
Every other available item except $b_{-1,0}$ is eaten at rate $1$. By \Cref{obs:lb-ps-surplus}, the surplus item cannot be exhausted before time $1$.
We discuss the implications below:

\begin{itemize}[noitemsep]
\item 
Since $b_{k,1}$ and every item in a later column have remaining capacity $\delta_{k,0}$ by the third assumption, we deduce
that item $(k,1)$ is the next to be exhausted, occurring at time
\[
T_{k,1} = T_{k,0} + \frac{\delta_{k,0}}{3} \quad \iff \quad  \delta_{k,1} = \frac{2}{3} \delta_{k,0}.
\]
\item 
At time $T_{k,1}$, the item at $(k+1,0)$ has remaining capacity
\[
\delta_{k,0}-2\left(\frac{\delta_{k,0}}{3}\right)=\frac{\delta_{k,0}}{3}>0.
\]
Now, the children of the agents at column $k-1$ are exhausted, so the agents in this column will proceed to eat the item $b_{-1, 0}$. However, the newly available
agent at vertex $(k,1)$ will join the agent at $(k,0)$
in the consumption of the item at $(k+1,0)$.
This
item is now eaten at rate $3$ and lasts another $\delta_{k,0}/9$ units of
time. Every other later item has remaining capacity $2\delta_{k,0}/3$ and
is still eaten only by its own collocated agent, so none is exhausted during
this interval.
\item 
Now the item at $(k+1,0)$ is eaten only by agent $a_{k+1,0}$ from time $0$, and by the agents of column $k$, from
their respective release times. Since all items in later columns are consumed at a slower rate, the item at $(k+1,0)$ is next to be exhausted.
Since the participating agents in column $k$ were previously eating the item at $(k,0)$ and $(k,1)$, we have
\[
T_{k+1,0}
+\bigl(T_{k+1,0}-T_{k,0}\bigr)
+\bigl(T_{k+1,0}-T_{k,1}\bigr)
=1
\quad
\iff
\quad 
\delta_{k+1,0}
=
\frac{\delta_{k,0}+\delta_{k,1}}{3}
=
\frac59\delta_{k,0}.
\]
\end{itemize}

Note that $T_{k+1,0} < 1$.
If $k+1<H$, the two agents of column $k$ now move to the item at $(k+1,1)$, and
the newly released agent $a_{k+1,0}$ moves to the item at $(k+2,0)$. All
earlier items are exhausted (with the exception of the item $b_{-1, 0}$). The agent $a_{0,0}$ and the agents of columns $1,\ldots,k-1$ are
eating the item $b_{-1, 0}$, while all other agents continue eating their collocated items. Moreover, the item at $(k+1,1)$ and every later item have been
eaten only by their own collocated agents and therefore have remaining
capacity $1-T_{k+1,0}$.
This establishes the induction step when $k+1<H$. If $k+1=H$, every nonsurplus item is now exhausted, so all agents consume $b_{-1,0}$ for the remaining time.

\paragraph{Establishing the Recurrence.}
Starting from $\delta_{2,0}=1/4$, our recurrence gives
\[
\delta_{k,0}
=
\frac14\left(\frac59\right)^{k-2}
\qquad (2\le k\le H).
\]
Together with $\delta_{k,1}=\frac23\delta_{k,0}$ for
$2\le k\le H-1$, this proves the stated formulas. For these columns,
\[
\delta_{k,0}>\delta_{k,1}>\delta_{k+1,0},
\]
whereas $\delta_{1,0}>\delta_{1,1}=\delta_{2,0}$. The asserted
exhaustion order follows.
\end{proof}

Having established the exhaustion times 
for the various items, we are now ready to establish the lower bound.

\begin{proof}[Proof of \Cref{thm:lb-ps}]
We consider the lower bound family $\{\sigma_H\}$ for a fixed integer $H \geq 2$.  Let $d=d_H$ be the consistent pseudometric defined above. It is straightforward to verify that the optimal matching in this metric $M^*$ assigns agents to their collocated items and  the remaining agent $a_{0,0}$ is assigned to item $b_{-1,0}$. Clearly, we have $\cost(M^*, d) = 1$. Every perfect matching assigns $b_{-1,0}$ to an agent at distance at least $1$, so no matching has smaller cost.

We now recall \Cref{prop:lb-ps-exhaustion}.
The agent $a_{H,0}$ eats her own item until $T_{H,0}$; her collocated item is her favorite and its exhaustion time is no earlier than that of any item in previous columns $1, \dots, H-1$. Since $T_{H,0} < 1$, the only remaining item for her to eat is the item $b_{-1, 0}$,  which is at distance $1+\sum_{j=0}^{H-1}2^j=2^H$. Lower bounding the cost of the fractional matching produced by PS, we have
\begin{equation}
\E_{M \sim \ps (\sigma_H)} \left [ \cost(M, d) \right ] \geq  2^H (1 - T_{H,0})  = 2^H\cdot\frac14\left(\frac59\right)^{H-2}
=
\left(\frac{10}{9}\right)^{H-2}.
\end{equation}

Since the number of agents is $n = 2H$ and the optimal matching has cost $1$, the desired distortion result immediately follows. 
\end{proof}

\begin{remark}
For completeness, we briefly discuss how our
instance can be adjusted to obtain a metric without
collocated agents or items. Let $d_0$ be the 
initial pseudometric defined on the $n$ agents
and $n$ items. For each agent $a$, let 
$\mathrm{rank}_a(b)\in\{0,\ldots,n-1\}$
be the number of items ranked above $b$, 
according to the specified
tie-breaking. Define
\[
R(a,b)=R(b,a)
=
1+\frac{\mathrm{rank}_a(b)}{n+1}.
\]
Set $R(x,y)=1$ between two distinct agents or two distinct items,
and $R(x,x)=0$. Every off-diagonal value lies in $[1,2)$, 
and in a triangle of distinct
points, any two sides sum to at least $2$, while the third is less
than $2$. Thus $R$ satisfies the triangle inequality.

Now for  $0<\varepsilon<1$, let $d_\varepsilon=d_0+\varepsilon R$.
This is clearly a metric on the $2n$ distinct agents and items. In particular,
every formerly collocated agent--item pair has positive distance.

The original distances are integers, so unequal agent--item
distances differ by at least $1$. The perturbation preserves their
order and resolves equal distances according to the specified
rankings. Thus every agent has pairwise distinct distances to the
items, and the entire PS trajectory, including all exhaustion
times, is unchanged.

Since $d_\varepsilon\ge d_0$, the cost in the metric $d_\eps$ cannot decrease. Moreover,
the matching of cost $1$ in the original instance has perturbed cost
at most $1+2n\varepsilon$. Writing $M^*$ for the optimal 
matching in the metric $d_\epsilon$, we have
\[
\frac{\E_{M \sim \ps(\sigma_H)} \left [ \cost(M, d_\epsilon) \right ]}{\cost(M^*, d_\eps)}
\ge
\frac{(10/9)^{H-2}}{1+2n\varepsilon}.
\]
Taking $\varepsilon$ arbitrarily small makes the
multiplicative loss arbitrarily small.
\end{remark}

\newcommand{\Nsmall}{N_{\mathrm{small}}}
\newcommand{\dcut}{\delta_{\mathrm{check}}}
\newcommand{\tcut}{t_{\mathrm{check}}}
\newcommand{\Aun}{A_{\mathrm{unstable}}}
\newcommand{\Bun}{B_{\mathrm{unstable}}}
\newcommand{\Afree}{A_{\mathrm{free}}}
\newcommand{\Bst}{B_{\mathrm{stable}}}
\newcommand{\budget}{\Phi}
\newcommand{\proxy}{\widetilde{\Phi}}

\subsection{Distortion Upper Bound}\label{sec:ub-ps}
In this section, we complement the exponential lower bound
with a matching upper bound.
\begin{theorem}\label{thm:ub-ps}
For every instance $\sigma$, $\distortion(\ps, \sigma) = O \left ( (10/9)^{n/2} \right )$.
\end{theorem}

In what follows, we fix an instance $\sigma$ and let
$d \in \rho(\sigma)$ be a consistent metric.
We aim to analyze how the matrix $X(t) = (X_{a,b}(t))$ evolves in relation to the
cost of an optimal perfect matching $M^*$ under $d$.
Abusing notation, we write $\cost(Z, d) = \sum_{a \in A} \sum_{b \in B} Z_{a, b} \cdot d(a, b)$ for any nonnegative matrix $Z = (Z_{a,b})$, such as $X(t)$.
We let $q_b(t)$ denote the fraction of item $b$ that remains
at time $t$.
We also let $b_a(t)$ denote the item that agent $a$
eats just after time $t$, for $t \in [0,1)$.
For ease of exposition, we also introduce some
terminology. We say that an \emph{event} is
a time at which an item is exhausted,
and a \emph{phase} is an interval $[s,t)$ between consecutive distinct event times, with the first phase starting at time $0$.

\subsubsection{Preliminary Lemmas}\label{sec:flow-formulation}
We will use two basic facts about PS throughout. First, once an agent starts eating an item, she keeps eating it until it is exhausted, since the set of available items only shrinks. Second, agents eat at total rate $n$ and the total capacity is $n$, so
\begin{equation}\label{eq:remaining}
 \sum_{b\in B} q_b(t) = n(1 - t)\qquad\text{for every } t \in [0,1];
\end{equation}
in particular, every item is exhausted by time $1$, and some item is exhausted exactly at time $1$.

We additionally state some lemmas that explain how the cost of
PS evolves over time.
Since the eating process can be viewed as agents receiving
units of mass from certain items during various time periods,
it is helpful to reason in terms of \textit{flows}.
Fix $t \in [0,1)$. We consider
a flow $f$ on the directed graph on vertices $A \cup B$ such that the following hold:
\begin{enumerate}[noitemsep]
\item Each item
$b$ sends $X_{a, b}(t)$ units of flow along edge $(b, a)$ for every $a \in A$.
\item Each agent $a$ sends $1$ unit of
flow along edge $(a, M^*(a))$.
\end{enumerate}
Observe that every agent has net outflow
$1 - t$, every item has net inflow $q_b(t)$, and
\begin{equation}\label{eq:flow-path}
 \sum_e f(e)\, d(e) = \sum_a d\big(a,M^*(a)\big) + \sum_{a,b} X_{a,b}(t)\, d(a,b)
 = \cost(M^*, d) + \cost(X(t), d).
\end{equation}
Here the sum ranges over all directed edges $e$ of the graph, and $d(e)$ denotes the distance between the endpoints of $e$.
By the flow decomposition theorem, we can write
\[
 f = \sum_{P\in\mathcal P} \alpha_P\, \mathbf{1}_P + \sum_{Q\in\mathcal Q} \beta_Q\, \mathbf{1}_Q,
 \qquad\alpha_P,\beta_Q \ge 0,
\]
where $\mathcal P$ is a set of directed paths, each from an agent to an item,
$\mathcal Q$ is a set of directed cycles, and $\mathbf{1}_P$ and $\mathbf{1}_Q$ are the arc-incidence vectors of $P$ and $Q$, respectively \citep{Gallai:1958aa,Schrijver:2003aa}.
Along the paths, the only sources are agents and the only sinks are
items, so we have
\[
 \sum_{P\text{ starts at }a} \alpha_P = 1 - t,\qquad
 \sum_{P\text{ ends at }b} \alpha_P = q_b(t).
\]
For an agent $a$ and an item $b$, let $\mathcal P_{ab} \subseteq \mathcal P$ be
the paths from $a$ to $b$, and define
\begin{equation*}
 F_{ab}(t) = \sum_{P\in\mathcal P_{ab}} \alpha_P .
\end{equation*}
Note that the resulting matrix $F(t) = (F_{ab}(t))$ has row sums $1 - t$ and column sums $q_b(t)$; in
particular, only columns corresponding to currently available items can contain positive entries.
Below, we show that its cost can
also be related to that of the optimal matching $M^*$
and the matrix $X(t)$.

\begin{lemma}\label{lem:transport}
For every time $t \in [0, 1)$, the matrix $F(t)$ satisfies
\begin{equation*}
\cost(F(t),d) \le \cost(M^*, d) + \cost(X(t), d).
\end{equation*}
\end{lemma}
\begin{proof}
For a path or cycle $W$, let $\ell(W) = \sum_{e\in W} d(e)$ be its length. Fix an agent $a$ and an item $b$.
We consider the paths $\mathcal P$ involved in the flow decomposition.
Every $P \in \mathcal P_{ab}$ goes from $a$ to $b$, so
\begin{align*}
 F_{ab}(t)\, d(a,b) = \sum_{P\in\mathcal P_{ab}} \alpha_P\, d(a,b)
 &\le \sum_{P\in\mathcal P_{ab}} \alpha_P\, \ell(P).
\end{align*}
Summing over all agents $a$ and items $b$,
\begin{align*}
 \cost(F(t),d)
 &\le \sum_{P\in\mathcal P} \alpha_P\, \ell(P) \\
 &\le \sum_{P\in\mathcal P} \alpha_P\, \ell(P) + \sum_{Q\in\mathcal Q} \beta_Q\, \ell(Q)\\
 &= \sum_e f(e)\, d(e) \\
 &= \cost(M^*, d) + \cost(X(t), d). \tag{by \eqref{eq:flow-path}}
\end{align*}
\end{proof}

In the following lemma, we also use the flow formulation to reason about the rate at which
the cost of $X(t)$ changes over time.
For ease of exposition, we let $\partial_t$ denote the right derivative with respect to $t$; away from events, this is the ordinary derivative.

\begin{lemma}[Rate inequality]\label{cor:rate}
At every time $t \in [0,1)$,
\[
  \partial_t \cost(X(t), d) = \sum_{a} d(a, b_a(t)) \le  \frac{\cost(M^*, d) + \cost(X(t), d)}{1 - t}.
\]
\end{lemma}

\begin{proof}
Just after time $t$, every agent $a$ eats $b_a(t)$ at unit rate, so the cost of $X(t)$ increases at rate $\sum_a d(a, b_a(t))$; this gives the equality. For the inequality, note that $b_a(t)$ is the favorite of agent $a$ among the items with $q_b(t) > 0$, and hence, since $d$ is consistent with her ranking, the closest of these items to $a$. The matrix $F(t)$ only uses items with $q_b(t) > 0$, and its row for $a$ sums to $1 - t$, so
\begin{align*}
 (1 - t)\, d(a,b_a(t)) = \sum_b F_{ab}(t)\, d(a,b_a(t))
 &\le \sum_b F_{ab}(t)\, d(a,b).
\end{align*}
Summing over $a$ gives $(1 - t)\, \partial_t \cost(X(t), d) \le \cost(F(t),d)$. Finally, apply \Cref{lem:transport}.
\end{proof}

Note that \Cref{cor:rate} is already
sufficient for establishing an exponential upper bound on the distortion, but not with our desired base of $\sqrt{10/9}$.
\begin{corollary}\label[corollary]{cor:exp-bound}
$\cost(X(1),d) \le (2^n - 1)\cost(M^*,d)$.
\end{corollary}
\begin{proof}
Let $t_1<\cdots<t_m=1$ be the distinct event times, and set $t_0=0$. Since each event exhausts at least one item, $m\le n$.
On each phase $[t_{i-1},t_i)$ every agent eats a fixed item, so
$\partial_s \cost(X(s),d)$ is constant there. We can therefore derive that
\begin{align*}
\cost(X(t_i),d) - \cost(X(t_{i-1}),d)
 &= \int_{t_{i-1}}^{t_i} \partial_s \cost(X(s),d)\, ds\\
 &\le \big(\cost(M^*,d) + \cost(X(t_{i-1}),d)\big)\int_{t_{i-1}}^{t_i} \frac{ds}{1 - t_{i-1}}
\tag{\Cref{cor:rate}}\\
 &= \big(\cost(M^*,d) + \cost(X(t_{i-1}),d)\big) \Big(1 - \frac{1 - t_i}{1 - t_{i-1}}\Big)\\
 &\le \cost(M^*,d) + \cost(X(t_{i-1}),d).
\end{align*}
That is, each phase at most doubles $\cost(M^*,d) + \cost(X(t),d)$. Chaining this
over the $m$ phases, and using $X(0) = 0$, we further have
\begin{align*}
\cost(M^*,d) + \cost(X(1),d)
 &\le 2\, \big(\cost(M^*,d) + \cost(X(t_{m-1}),d)\big)\\
 &\;\;\vdots\\
 &\le 2^m \left(\cost(M^*,d) + \cost(X(0),d)\right)\\
 &= 2^m\cost(M^*,d).
\end{align*}
Since $m\le n$, $\cost(X(1),d) \le (2^n - 1)\cost(M^*,d)$.
\end{proof}

The rate inequality also gives us a time-sensitive
bound on the cost of the matrix $X(t)$.
\begin{corollary}\label[corollary]{cor:prefix}
For every $t \in [0, 1)$,
\begin{align*}
\cost(X(t), d) \leq \left ( \frac{t}{1 - t}  \right )  \cost(M^*, d) \qquad \text{and} \qquad
\sum_{a} d(a, b_a(t)) \leq \frac{\cost(M^*, d)}{(1 - t)^2}.
\end{align*}
\end{corollary}
\begin{proof}
We define the function $g(t) = (1 - t) \left (\cost(X(t), d) + \cost(M^*, d)\right )$. Note that this function is continuous on $[0, 1]$, since $\cost(X(t), d)$ is. On every phase each agent eats a fixed item, so $g$ is differentiable in the interior of each phase. Taking derivatives with respect to $t$ and recalling \Cref{cor:rate}, we get
\begin{equation*}
\partial_t g(t) = (1 - t)\, \partial_t \cost (X(t), d) - \left ( \cost(X(t), d) + \cost(M^*, d) \right) \leq 0.
\end{equation*}
Since there are only finitely many events, $g$ is nonincreasing on $[0, 1)$, and thus $g(t) \leq g(0) = \cost(M^*, d)$, as $X(0) = 0$. Dividing by $1 - t$ gives
\begin{equation}\label{eq:prefix-bound}
\cost(M^*, d) + \cost(X(t), d) \le \frac{\cost(M^*, d)}{1 - t},
\end{equation}
and the first inequality follows by rearranging.

For the second inequality, \Cref{cor:rate} and \eqref{eq:prefix-bound} give
\[
 \sum_a d(a, b_a(t)) \le \frac{\cost(M^*, d) + \cost(X(t), d)}{1 - t} \le \frac{\cost(M^*, d)}{(1 - t)^2}.
\]
\end{proof}

\Cref{cor:prefix} reveals a striking fact. If we were to run PS
for a constant amount of time (say, until $t = 0.99$), then
the cost of the maintained partial matching $X(t)$ would be at most a constant multiple of the optimal cost. Moreover, if one were to run the mechanism until time $t = 1 - \frac{1}{n}$,
the cost would be at most a linear multiple of the optimal cost.
This suggests that much of the cost is amplified in the
late stages of the mechanism's operation.
Much of our analysis in subsequent sections is
devoted to these stages.

If $\cost(M^*,d) = 0$, then \Cref{cor:prefix} gives $\cost(X(t),d) = 0$ for every $t < 1$, and hence $\cost(X(1),d) = 0$ by continuity. For the remainder of the analysis, we therefore assume that $\cost(M^*,d) > 0$.

\subsubsection{Late Stage Components of PS}
Fix a checkpoint $\tcut = 1 - \dcut$, where the remaining time is
$\dcut = \Nsmall^{-\Nsmall}$ for an integer $\Nsmall \ge 1$ that we choose at the end of the analysis.
We are interested in analyzing how the cost of $X(t)$ amplifies
beyond this checkpoint. By the checkpoint, some agents will have interacted
with one another in the sense of eating the same items.
It is helpful to group agents based on these
interactions.

We revisit the flow $f$ introduced in \Cref{sec:flow-formulation}
and consider the undirected graph on $A \cup B$,
containing the edges $\{a, M^*(a)\}$ for $a \in A$ and the edges $\{a, b\}$ with $X_{a, b}(\tcut) > 0$.
In this graph, we further include the edges $\{a, b_a(\tcut)\}$ if
they are not already present. Call the resulting graph $G$.
We work with the (maximal) connected components of $G$ at time $\tcut$. For ease of exposition, we will just refer to these sets as \emph{components}.
We first state and justify four relevant properties of components, and then describe
certain types of components that we will track in our analysis.
\begin{claim}\label{clm:components}
Every connected component $\Gamma$ of $G$ satisfies the following properties.
\begin{enumerate}[label=(P\arabic*),leftmargin=*, noitemsep]
 \item\label{P:closure} Up to the checkpoint, the agents of $\Gamma$ have eaten only items of $\Gamma$, and the items of $\Gamma$ have been eaten only by agents of $\Gamma$; just after the checkpoint, every agent of $\Gamma$ eats an item of $\Gamma$.
 \item\label{P:restriction} Running PS on $\Gamma$ alone reproduces the actual run on $\Gamma$ up to $\tcut$, including the choices made just after $\tcut$.
 \item\label{P:balance} $|\Gamma \cap A| = |\Gamma \cap B|$, and
 $\displaystyle\sum_{b\in\Gamma\cap B} q_b(\tcut) = |\Gamma \cap A|\, \dcut$.
 \item\label{P:diameter} For all $x,y \in \Gamma$,
 $\displaystyle d(x,y) \le \left(\frac{2}{\dcut^{2}} + 1\right)\cost(M^*,d)$.
\end{enumerate}
\end{claim}
\begin{proof}
We prove each item separately.

\ref{P:closure} immediately follows from the definition of $G$.

Up to the checkpoint, the remaining capacities of the items of $\Gamma$ only change through the eating by agents of $\Gamma$, by \ref{P:closure}. Each agent eats her favorite available item, which lies in $\Gamma$ and is therefore also her favorite among the available items of $\Gamma$; one can inductively recover the run of PS on $\Gamma$, event by event. This proves \ref{P:restriction}.

For \ref{P:balance}, every agent $a$ lies in the same component as  her optimal assignment $M^*(a)$, and thus $\Gamma$ has as many items as agents. By \ref{P:closure}, the items of $\Gamma$ have been eaten only by the agents of $\Gamma$, each of whom has eaten $\tcut$ units. Since each item has initial capacity $1$, we get $\sum_{b\in\Gamma\cap B} q_b(\tcut) = |\Gamma \cap B| - \tcut\, |\Gamma \cap A| = |\Gamma \cap A|\, \dcut$. 

It remains to prove \ref{P:diameter}.
First note that $d(a,b_a(t))$ is nondecreasing in $t$: agent $a$ always eats her closest available item, and the set of available items only shrinks. So every edge $\{a,b\}$ of $G$ other than $\{a,M^*(a)\}$ satisfies $d(a,b) \le d(a,b_a(\tcut))$. Now take a simple path in $G$ from $x$ to $y$. It uses at most two edges at each agent, and its edges of the form $\{a,M^*(a)\}$ are distinct edges of $M^*$. By the triangle inequality and \Cref{cor:prefix},
\[
 d(x,y) \le 2\sum_a d\big(a,b_a(\tcut)\big) + \cost(M^*,d) \le \left(\frac{2}{\dcut^{2}} + 1\right)\cost(M^*,d). \qedhere
\] 
\end{proof}

\paragraph{Stable components.}
For an item $b$, let $N_b$ be the number of agents eating it just after the checkpoint.
We say that a component $\Gamma$ is \emph{stable} if every item $b \in \Gamma$ that is available
just after $\tcut$ satisfies $q_b(\tcut) = N_b\dcut$.
Otherwise, we say that the component is \emph{unstable}.

In a stable component, the $N_b$ agents eating an available item $b$ would deplete it at exactly time $1$. So if the component were run on its own, which by \ref{P:restriction} agrees with the actual run up to the checkpoint, none of its agents would switch items again before time $1$. In the actual run, this behavior can be disrupted only by an agent from outside the component eating one of its items.
Such an agent could either come from an unstable component,
or from a stable component that was itself disturbed earlier.

As the preceding discussion shows, the number of unstable components and their sizes have a significant effect
on the late stages of the mechanism.
Below, we show that unstable components must be large and thus, there
cannot be too many of them. The key is the following fact about the times at which items run out.

\begin{lemma}\label{lem:event-gap}
Consider the operation of PS on any instance with $N$ agents and $N$ items. Every event at a time $\tau < 1$ satisfies $1 - \tau \ge N^{-N}$.
\end{lemma}
\begin{proof}
Initially, the time and all capacities are integers. Suppose that at some event, the time and all remaining capacities are fractions with a common denominator $Q$. The next event occurs when some item with $c \le N$ eaters runs out, after $q/c$ further units of time, where $q$ is its remaining capacity; afterwards, the time and all remaining capacities have a common denominator dividing $cQ$. Each item runs out once, and by \eqref{eq:remaining} some item runs out at time $1$, so there are at most $N - 1$ events before time $1$. Hence every such event $\tau$ is a fraction with denominator at most $N^{N-1}$, and so is the positive number $1 - \tau$. Therefore $1 - \tau \ge N^{-(N-1)} \ge N^{-N}$.
\end{proof}

\begin{corollary}\label[corollary]{cor:small-stable}
Every component with at most $\Nsmall$ agents is stable; equivalently, every unstable component has more than $\Nsmall$ agents.
\end{corollary}
\begin{proof}
Consider a component $\Gamma$ with $N = |\Gamma \cap A| \le \Nsmall$.
By \ref{P:closure}, every agent of $\Gamma$ eats exactly one item of $\Gamma$ just after the checkpoint, so $\sum_{b\in\Gamma\cap B} N_b = N$, and by \ref{P:balance},
\[
 \sum_{b\in\Gamma\cap B} q_b(\tcut) = N\dcut = \sum_{b\in\Gamma\cap B} N_b\, \dcut .
\]
If $\Gamma$ were unstable, some item $b \in \Gamma$ would therefore have $q_b(\tcut) < N_b\, \dcut$, and in particular $N_b \ge 1$. When
PS is run on $\Gamma$ alone, which agrees with the actual run up to and just after the
checkpoint by \ref{P:restriction}, these $N_b$ agents keep eating $b$ until the next event, so some item runs out at a time $\tau$ with $\tcut < \tau \le \tcut + q_b(\tcut)/N_b < 1$. Then
\[
 1 - \tau < \dcut = \Nsmall^{-\Nsmall} \le N^{-N},
\]
contradicting \Cref{lem:event-gap}.
\end{proof}

The distinction between stable and unstable components splits the agents and items into two groups. We let $\Aun$ and $\Bun$ denote the sets of agents and items in unstable components. 

\paragraph{Residents and visitors.}
In a stable component, every item $b$ that is available at the checkpoint is being eaten by $N_b$ agents and has exactly $N_b\dcut$ left. We call such an item a \emph{block}, and these $N_b$ agents its \emph{residents}. By \ref{P:closure}, every agent of a stable component is a resident of exactly one block.
After the checkpoint, a resident keeps eating her block until it is exhausted, at which point we say that she is \emph{freed}. For $t \in [\tcut,1]$, we let $\Afree(t)$ denote the set of \emph{free} agents at time $t$: the agents of $\Aun$, together with the residents freed by time $t$. Thus $\Afree(t)$ grows from $\Afree(\tcut) = \Aun$ to $\Afree(1) = A$, since every block is exhausted by time $1$. 

An agent who eats a block without being one of its residents is a \emph{visitor}; she is necessarily free, since a resident only leaves her block once it is exhausted. 
For $t \in [\tcut,1]$, let $V_b(t)$ denote the amount of block $b$ eaten by visitors during $[\tcut,t]$. Since its residents continue eating at total rate $N_b$ until it is exhausted, we have
\begin{equation}\label{eq:block-capacity}
 V_b(t) = N_b(1 - t) - q_b(t)
\end{equation}
up to and including the exhaustion time of $b$.

Let $\Bst(t)$ denote the set of blocks that are still available at time $t$. Since the items of stable components that are not blocks were exhausted by the checkpoint, every available item lies in $\Bun$ or in $\Bst(t)$, and the agents that are not free at time $t$ are exactly the residents of the blocks in $\Bst(t)$. Rearranging \eqref{eq:block-capacity}, summing over $\Bst(t)$, and comparing with \eqref{eq:remaining}, we get, for every $t \in [\tcut,1]$,
\begin{equation}\label{eq:unstable-capacity}
\sum_{b\in\Bun} q_b(t) = |\Afree(t)|\, (1 - t) + \sum_{b\in\Bst(t)} V_b(t)\ \ge\ |\Aun|\, (1 - t).
\end{equation}
In particular, if $\Aun \ne \varnothing$, some item of an unstable component remains available until time $1$.
From now on, we assume that $\Aun \ne \varnothing$; the case $\Aun = \varnothing$ is handled directly in the proof of \Cref{thm:ub-ps}.

\begin{figure}[ht]
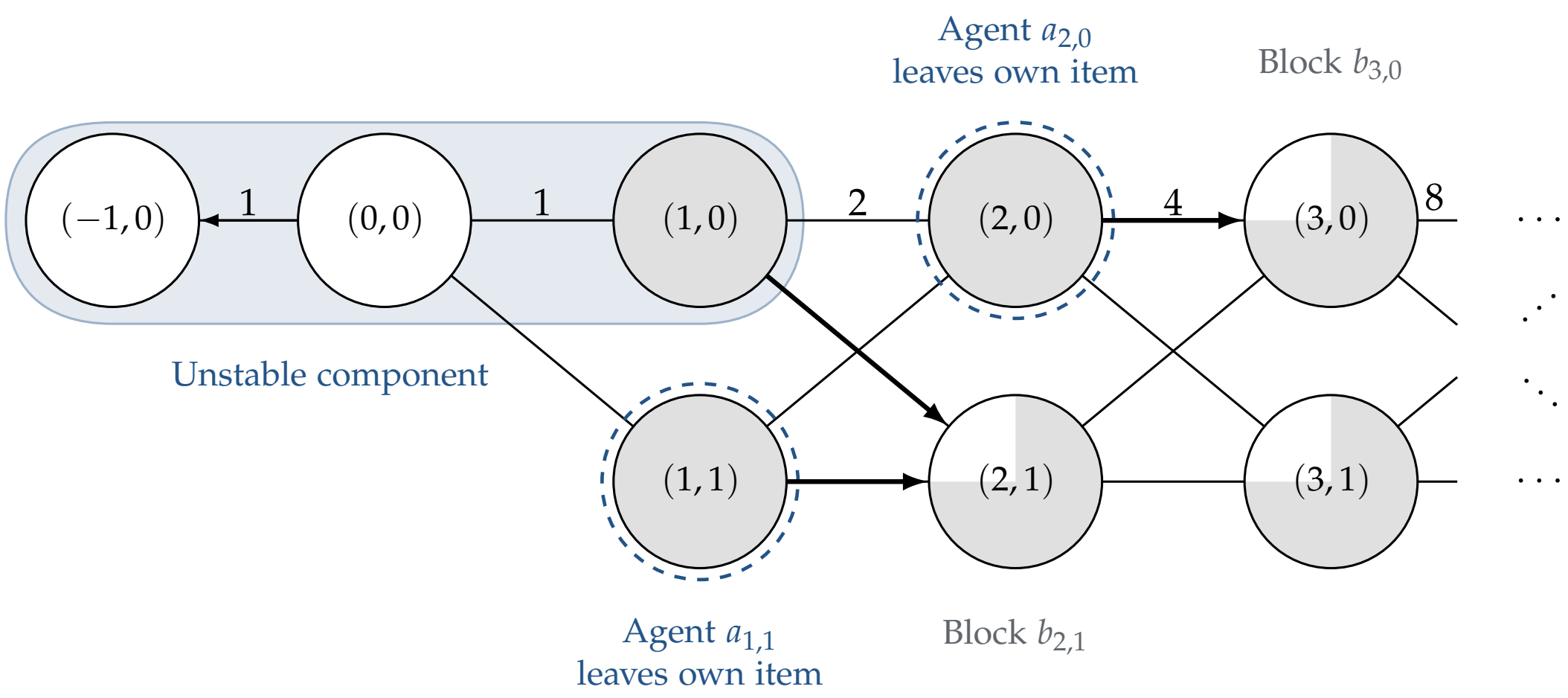

\centering
\psladder[overlay]{1}{0.75}
\caption{Unstable components, visitors, and newly freed agents
     in the layered graph ($H\ge5$).
     For illustration, take $N_{\mathrm{small}}=1$, so $\tcut=0$,
     and consider the state just after $t=3/4$.
     The agents of the fixed unstable component
     (shaded light blue) are $a_{0,0}$ and $a_{1,0}$.
     Dashed rings mark the exhaustion of $b_{1,1}$ and $b_{2,0}$,
     which frees their residents $a_{1,1}$ and $a_{2,0}$ at this time.
     Arrows show current eating choices; thick arrows indicate visits to stable blocks.
     Gray sectors show consumption.}
\label{fig:ps-components-overlay}
\end{figure}

\Cref{fig:ps-components-overlay} illustrates unstable components, visitors, and newly freed residents in the layered graph instance.

\paragraph{Anchors and epochs.}
For every unstable component $\Gamma$, we fix an \emph{anchor} $y_\Gamma$: an item of $\Gamma$ that is exhausted last among the items of $\Gamma$. Thus, whenever some item of $\Gamma$ is available, so is $y_\Gamma$.
For $t \in [\tcut,1)$, we let $Y(t)$ denote the set of anchors that are available at time $t$; by the previous paragraph, $Y(t) \ne \varnothing$.
Note that the set $Y(t)$ only loses elements as $t$ increases, and it starts with one anchor per unstable component. We call a maximal time interval on which $Y(t)$ is constant an \emph{epoch}, and write the epochs as $[u_{j-1},u_j)$ for $j = 1,\dots,s$, where $\tcut = u_0 < u_1 < \dots < u_s = 1$. Each epoch ends when at least one anchor is exhausted, so there are at most as many epochs as unstable components. By \Cref{cor:small-stable}, each unstable component contains at least $\Nsmall + 1$ agents. Hence
\begin{equation}\label{eq:epochs}
 s \le \frac{|\Aun|}{\Nsmall + 1}.
\end{equation}

\subsubsection{The Cost of an Epoch}\label{sec:epoch-cost}
Recall that free agents are the only source of disturbance after the checkpoint $\tcut$. Since each of them eats her closest available item, we can bound what she pays by comparing her with the anchors, which remain available. For $t \in [\tcut,1)$, we write $d(x,Y(t)) = \min_{y\in Y(t)} d(x,y)$ for the distance from $x$ to the nearest available anchor; this is well-defined because $Y(t)$ is nonempty.

\begin{claim}\label{clm:anchors}
The following properties hold for every $t \in [\tcut,1)$.
\begin{enumerate}[label=(P\arabic*),start=5,leftmargin=*, noitemsep]
 \item\label{P:close} Every item $b \in \Bun$ that is available at time $t$ satisfies $d(b,Y(t)) \le \left(\frac{2}{\dcut^{2}} + 1\right)\cost(M^*,d)$.
 \item\label{P:nearest} Every agent $a$ satisfies $d(a,b_a(t)) \le d(a,Y(t))$ and $d(b_a(t),Y(t)) \le 2\, d(a,Y(t))$.
\end{enumerate}
\end{claim}
\begin{proof}
For \ref{P:close}, if $b \in \Bun$ is available at time $t$, then so is the anchor $y_\Gamma$ of its component $\Gamma$, so $d(b,Y(t)) \le d(b,y_\Gamma)$, and the bound follows from \ref{P:diameter}.

For \ref{P:nearest}, the anchors in $Y(t)$ are available at time $t$, and $b_a(t)$ is the closest available item to $a$.  Hence, $d(a,b_a(t)) \le d(a,Y(t))$. The triangle inequality then gives $d(b_a(t),Y(t))  \le 2\,d(a,Y(t))$.
\end{proof}

We first separate the cost rate into contributions from residents that are not yet free and agents that are free. A resident $a$ that is not yet free still eats $b_a(\tcut)$, whereas every free agent eats at distance at most $d(a,Y(t))$ by \ref{P:nearest}. Thus \Cref{cor:rate} and \Cref{cor:prefix} give, for every $t \in [\tcut,1)$,
\begin{equation}\label{eq:cost-split}
\begin{split}
 \partial_t \cost(X(t),d)&= \sum_{a\notin\Afree(t)} d(a,b_a(\tcut)) + \sum_{a\in\Afree(t)} d(a,b_a(t))\\
 &\le \frac{\cost(M^*,d)}{\dcut^{2}} + \sum_{a\in\Afree(t)} d(a,Y(t)).
\end{split}
\end{equation}
Now fix an epoch $[u_{j-1},u_j)$. Its anchor set is constant, so $d(x,Y(t)) = d(x,Y(u_{j-1}))$ throughout the epoch. Let $\Delta_j = |\Afree(u_j)| - |\Afree(u_{j-1})|$, and let $u_a$ be the release time of each agent $a \in \Afree(u_j)\setminus\Afree(u_{j-1})$. Our convention is to count releases in $(u_{j-1},u_j]$: releases at the right endpoint are included, while agents already free at the left endpoint are excluded.

We distinguish three contributions to the cost of the epoch:
\begin{itemize}[leftmargin=*]
 \item \textit{Cost before release.} By \eqref{eq:cost-split}, residents that are not yet free contribute at most $\cost(M^*,d)/\dcut^2$ to the instantaneous cost rate. Their contribution over the epoch is therefore at most $(u_j-u_{j-1})\cost(M^*,d)/\dcut^2$.
 \item \textit{Cost from agents already free.} Each $a \in \Afree(u_{j-1})$ pays at most $d(a,Y(u_{j-1}))$ per unit time. Since the epoch lasts at most $1-u_{j-1}$, her cost during it is at most $(1-u_{j-1})d(a,Y(u_{j-1}))$.
 \item \textit{Cost after new releases.} An agent released at $u_a \in (u_{j-1},u_j]$ pays at most $(1-u_a)d(a,Y(u_{j-1}))$ after her release and before the epoch ends. Her cost before release is covered by the first contribution (see first bullet).
\end{itemize}
The first contribution has a uniform cost-rate bound, and the second is bounded using the state at the start of the epoch. The third requires more care because a newly freed agent can visit other blocks and free further residents. To control this cascade, we compare a resident's distance to the anchors with that of a visitor helping to free her.

\paragraph{Effective radius.}
Suppose that a free agent $p$ eats a block $b$ just after time $t$ in the epoch, and let $a$ be a resident of $b$. By \ref{P:nearest}, $d(b,Y(t)) \le 2d(p,Y(t))$. The resident $a$ need not be located at $b$, but \Cref{cor:prefix} gives
\[
 d(a,b) = d(a,b_a(\tcut)) \le \frac{\cost(M^*,d)}{\dcut^2} \le D,
 \qquad D = \left(\frac{2}{\dcut^2}+1\right)\cost(M^*,d),
\]
where $D$ is the component-diameter bound from \ref{P:diameter}. By triangle inequality, we then have $ d(a,Y(t))  \le 2d(p,Y(t))+D$.
Adding $D$ to both distances turns this bound into a factor of two, which motivates the following definition.

\begin{definition}[Effective radius]\label{def:effective-radius}
For $t \in [\tcut,1)$ and $x \in A\cup B$, the \emph{effective radius} of $x$ at time $t$ is
\[
 R_t(x) \coloneqq d(x,Y(t)) + D.
\]
\end{definition}
The effective radius is constant throughout each epoch. From the earlier discussion, we see that whenever a free agent $p$ visits the block of a resident $a$,
\begin{equation}\label{eq:doubling}
 R_t(a) \le 2d(p,Y(t))+2D = 2R_t(p).
\end{equation}

We now relate an agent's release time to the visitors that helped free her. Let $a$ be released at $u_a \in (u_{j-1},u_j]$, and let $b$ be her block. At its exhaustion, \eqref{eq:block-capacity} gives $V_b(u_a)=N_b(1-u_a)$, so
\begin{equation}\label{eq:release-split}
 1-u_a = \frac{V_b(u_{j-1})}{N_b} + \frac{V_b(u_a)-V_b(u_{j-1})}{N_b}.
\end{equation}
The first term is her \emph{head start}, her share of what visitors ate from her block before the epoch; the second is her share of what visitors eat from it during the epoch.

Since $d(a,Y(u_{j-1})) \le R_{u_{j-1}}(a)$, the third bullet bounds each newly freed agent's cost \emph{after her release and before the epoch ends} by $R_{u_{j-1}}(a)(1-u_a)$. Her cost before release is accounted for separately in the first bullet.  Multiplying \eqref{eq:release-split} by $R_{u_{j-1}}(a)$ and summing over these agents, the head starts contribute at most

\[
 \sum_{b\in\Bst(u_{j-1})} \frac{V_b(u_{j-1})}{N_b}\sum_{a\text{ resident of }b} R_{u_{j-1}}(a).
\]
In the second term of \eqref{eq:release-split}, each unit that a visitor $p$ eats from a block $b$ exhausted by $u_j$ adds $1/N_b$ to each resident's share. Since $b$ has $N_b$ residents, \eqref{eq:doubling} bounds the weighted contribution of that unit by
\begin{equation}\label{eq:visitor-charge}
 \frac{1}{N_b}\sum_{a\text{ resident of }b}R_{u_{j-1}}(a)
 \le \frac{1}{N_b}\sum_{a\text{ resident of }b}2R_{u_{j-1}}(p)
 = 2R_{u_{j-1}}(p).
\end{equation}

Each visitor $p$ free at the start of the epoch eats at unit rate, so she eats at most $1-u_{j-1}$ units before it ends. Their total weighted contribution is therefore at most
\[
 2(1-u_{j-1})\sum_{p\in\Afree(u_{j-1})}R_{u_{j-1}}(p).
\]
The remaining visitors are themselves freed during the epoch; this is the cascade.

Apart from the cascade, these bounds depend only on the state at the start of the epoch, and we collect them in the \emph{budget}: for $t \in [\tcut,1)$,
\begin{equation}\label{eq:budget}
 \budget(t) = 2(1 - t)\sum_{a\in\Afree(t)} R_t(a) + \sum_{b\in\Bst(t)} \frac{V_b(t)}{N_b}\sum_{a\text{ resident of }b} R_t(a).
\end{equation}
The following lemma, the heart of the analysis, controls the cascade; we prove it in \Cref{sec:displacement}.

\begin{lemma}[Displacement inequality]\label{lem:displacement}
Let $\lambda = \sqrt{10/9}$ and $\kappa = 16(10 + 9\lambda)^2 < 6080$. For every epoch $[u_{j-1},u_j)$, with $\Delta_j$ and $u_a$ as above,
\[
 \sum_{a\in\Afree(u_j)\setminus\Afree(u_{j-1})} R_{u_{j-1}}(a)\, (1 - u_a) \le \frac{\kappa}{2}\left(\lambda^{\Delta_j} - 1\right)\budget(u_{j-1}).
\]
\end{lemma}

\begin{corollary}\label[corollary]{cor:epoch-cost}
For every epoch $[u_{j-1},u_j)$,
\[
 \int_{u_{j-1}}^{u_j} \sum_{a\in\Afree(t)} R_t(a)\, dt \le \frac12\left(\kappa\lambda^{\Delta_j} - 1\right)\budget(u_{j-1}).
\]
\end{corollary}
\begin{proof}

During the epoch, $R_t = R_{u_{j-1}}$. Agents already free at $u_{j-1}$ contribute to the integral throughout the epoch; agents released at $u_a$ contribute only from $u_a$ until $u_j$. Thus
\begin{equation}\label{eq:epoch-radius-accounting}
\begin{aligned}
 \int_{u_{j-1}}^{u_j}\sum_{a\in\Afree(t)}R_t(a)\,dt
 &\le (1-u_{j-1})\sum_{a\in\Afree(u_{j-1})}R_{u_{j-1}}(a)
 + \sum_{a\in\Afree(u_j)\setminus\Afree(u_{j-1})}(1-u_a)R_{u_{j-1}}(a),
\end{aligned}
\end{equation}
where the inequality uses $u_j\le1$. The first term in the last expression is at most $\frac12\budget(u_{j-1})$ by \eqref{eq:budget}, since the head-start term of the budget is nonnegative. By \Cref{lem:displacement}, the second term is at most $\frac{\kappa}{2}(\lambda^{\Delta_j}-1)\budget(u_{j-1})$. Their sum is therefore at most
\[
 \frac12(1+\kappa\lambda^{\Delta_j}-\kappa)\budget(u_{j-1})
 \le \frac12(\kappa\lambda^{\Delta_j}-1)\budget(u_{j-1}),
\]
since $\kappa\ge2$.
\end{proof}

\subsubsection{Putting it All Together}\label{sec:proxy}
\Cref{cor:epoch-cost} bounds what the free agents pay during an epoch in terms of the budget at its start. To chain the epochs, we need a quantity that we can track across them. The budget itself is measured from the current anchors, so when an anchor is exhausted, the radii, and hence the budget, can jump. 
For $t \in [\tcut,1]$, we instead define and work with the \emph{proxy}
\begin{equation}\label{eq:proxy}
 \proxy(t) = \cost(M^*,d) + \cost(X(t),d) + (1 - t)\frac{\cost(M^*,d)}{\dcut^{2}} + 2D\sum_{b\in\Bun} q_b(t),
\end{equation}
which does not depend on the anchors. 
The following properties let us convert each epoch-cost bound into a multiplicative bound on the proxy, while controlling its initial and final values.

\begin{claim}\label{clm:proxy}
The proxy satisfies the following properties.
\begin{enumerate}[label=(P\arabic*),start=7,leftmargin=*, noitemsep]
 \item\label{P:proxy-rate} $\proxy$ is continuous on $[\tcut,1]$, and $\partial_t \proxy(t) \le \sum_{a\in\Afree(t)} R_t(a)$ for every $t \in [\tcut,1)$.
 \item\label{P:proxy-budget} $\budget(t) \le 2\, \proxy(t)$ for every $t \in [\tcut,1)$.
 \item\label{P:proxy-ends} $\proxy(\tcut) \le (2 + 6|\Aun|)\, \frac{\cost(M^*,d)}{\dcut}$, and $\proxy(1) = \cost(M^*,d) + \cost(X(1),d)$.
\end{enumerate}
\end{claim}
\begin{proof}
We prove each item separately.

For \ref{P:proxy-rate}, $\proxy$ is continuous, since $\cost(X(t),d)$ and the remaining capacities are. Since remaining capacities only decrease,
\begin{align*}
 \partial_t \proxy(t)
 \le \partial_t \cost(X(t),d)-\frac{\cost(M^*,d)}{\dcut^2}
 \le \sum_{a\in\Afree(t)}d(a,Y(t))
 \le \sum_{a\in\Afree(t)}R_t(a),
\end{align*}
where the final two inequalities use \eqref{eq:cost-split} and the definition of the effective radius.

We next prove \ref{P:proxy-budget} by a charging argument. Since $d(a,Y(t)) \le d(a,b) + d(b,Y(t))$ for every agent $a$ and item $b$, each unit of mass that $F(t)$ moves from $a$ to $b$ costs at least $d(a,Y(t)) - d(b,Y(t))$. By \Cref{lem:transport} and the row and column sums of $F(t)$, we thus have

\begin{equation}\label{eq:charging}
\begin{split}
 \cost(M^*,d) + \cost(X(t),d)
 &\ge (1-t)\sum_{a\in A}d(a,Y(t)) - \sum_{b\in B}q_b(t)\,d(b,Y(t)).
\end{split}
\end{equation}

Recall that every agent is free or a resident of a block in $\Bst(t)$, and that every available item lies in $\Bun$ or in $\Bst(t)$. For a block $b \in \Bst(t)$, we have $(1 - t)N_b - q_b(t) = V_b(t)$ by \eqref{eq:block-capacity}, and $d(a,Y(t)) \ge d(b,Y(t)) - d(a,b_a(\tcut))$ for each resident $a$ of $b$. Since $q_b(t) \le N_b(1 - t)$, the terms of $b$ in~\eqref{eq:charging} contribute

\begin{align*}
 &(1-t)\sum_{a\text{ resident of }b}d(a,Y(t)) - q_b(t)\,d(b,Y(t))\\
 &\qquad = \frac{V_b(t)}{N_b}\sum_{a\text{ resident of }b}d(a,Y(t))
  + \frac{q_b(t)}{N_b}\sum_{a\text{ resident of }b}\bigl(d(a,Y(t))-d(b,Y(t))\bigr)\\
 &\qquad \ge \sum_{a\text{ resident of }b}\Bigl(\frac{V_b(t)}{N_b}\,d(a,Y(t))-(1-t)\,d(a,b_a(\tcut))\Bigr).
\end{align*}

The available items of $\Bun$ contribute at least $-D\sum_{b\in\Bun} q_b(t)$ by \ref{P:close}. Combining this with the block calculation and \eqref{eq:charging}, and then applying \Cref{cor:prefix}, gives

\begin{align*}
 &(1-t)\sum_{a\in\Afree(t)}d(a,Y(t))
   +\sum_{b\in\Bst(t)}\frac{V_b(t)}{N_b}
     \sum_{a\text{ resident of }b}d(a,Y(t))\\
 &\quad\le \cost(M^*,d)+\cost(X(t),d)
   +(1-t)\sum_{a\notin\Afree(t)}d(a,b_a(\tcut))
   +D\sum_{b\in\Bun}q_b(t)
   \tag*{\eqref{eq:charging}, \ref{P:close}}\\
 &\quad\le \cost(M^*,d)+\cost(X(t),d)
   +(1-t)\frac{\cost(M^*,d)}{\dcut^2}
   +D\sum_{b\in\Bun}q_b(t).
   \tag{\Cref{cor:prefix}}
\end{align*}

Replacing each distance on the left-hand side by its effective radius adds

\begin{align*}
 D\left((1-t)|\Afree(t)|+\sum_{b\in\Bst(t)}V_b(t)\right)
 &=D\sum_{b\in\Bun}q_b(t).
 \tag*{\eqref{eq:unstable-capacity}}
\end{align*}

Combining these two displays with the definition of $\proxy(t)$, we obtain
\[
 (1 - t)\sum_{a\in\Afree(t)} R_t(a) + \sum_{b\in\Bst(t)} \frac{V_b(t)}{N_b}\sum_{a\text{ resident of }b} R_t(a) \le \proxy(t).
\]

Doubling the displayed left-hand side dominates the budget in \eqref{eq:budget}, since the head-start term is nonnegative. Hence $\budget(t) \le 2\,\proxy(t)$.

Finally, for \ref{P:proxy-ends}, note that $1 - \tcut = \dcut$, and that $\sum_{b\in\Bun} q_b(\tcut) = |\Aun|\, \dcut$ by \ref{P:balance}. Evaluating~\eqref{eq:proxy} at the checkpoint, we get

\begin{align*}
 \proxy(\tcut)
 &= \cost(M^*,d) + \cost(X(\tcut),d) + \frac{\cost(M^*,d)}{\dcut} + 2D\dcut\, |\Aun|\\
 &\le 2\frac{\cost(M^*,d)}{\dcut} + 2D\dcut\, |\Aun| \tag{\Cref{eq:prefix-bound}}\\
 &= \bigl(2 + (4 + 2\dcut^{2})\, |\Aun|\bigr)\, \frac{\cost(M^*,d)}{\dcut} \tag{definition of $D$}\\
 &\le (2 + 6|\Aun|)\, \frac{\cost(M^*,d)}{\dcut}. \tag{$\dcut \le 1$}
\end{align*}

At time $1$, the third term of~\eqref{eq:proxy} vanishes because $1 - t = 0$, and the last because every item is exhausted; hence $\proxy(1) = \cost(M^*,d) + \cost(X(1),d)$.
\end{proof}

We are now ready to prove our main result.

\begin{proof}[Proof of \Cref{thm:ub-ps}]
We choose $\Nsmall = 180$, and note that $\kappa < 6080 < (10/9)^{90} = \lambda^{\Nsmall}$. 
If $\Aun = \varnothing$, then no resident is freed before time $1$: until the first resident is freed, no agent is free, so no block has a visitor, and by \eqref{eq:block-capacity} every block has remaining capacity $N_b(1 - t) > 0$ for $t < 1$. So after the checkpoint every agent keeps eating $b_a(\tcut)$, and by \Cref{cor:rate} and \Cref{cor:prefix},
\[
 \cost(X(1),d) = \cost(X(\tcut),d) + \dcut\sum_a d(a,b_a(\tcut)) \le \frac{2}{\dcut}\cost(M^*,d).
\]

Otherwise, for every epoch $[u_{j-1},u_j)$, we have
\begin{align*}
 \proxy(u_j) - \proxy(u_{j-1})
 &\le \int_{u_{j-1}}^{u_j} \sum_{a\in\Afree(t)} R_t(a)\, dt \tag{\ref{P:proxy-rate}}\\
 &\le \frac12\left(\kappa\lambda^{\Delta_j} - 1\right)\budget(u_{j-1}) \tag{\Cref{cor:epoch-cost}}\\
 &\le \left(\kappa\lambda^{\Delta_j} - 1\right)\proxy(u_{j-1}). \tag{\ref{P:proxy-budget}}
\end{align*}
That is, each epoch multiplies the proxy by at most $\kappa\lambda^{\Delta_j}$. Chaining this over the $s$ epochs, as in the proof of \Cref{cor:exp-bound}, gives $\proxy(1) \le \kappa^s\lambda^{\Delta_1+\dots+\Delta_s}\, \proxy(\tcut)$, where the exponents telescope to $|\Afree(1)| - |\Afree(\tcut)| = n - |\Aun|$. 

The size bound on unstable components limits the number of epochs, allowing us to control the factor $\kappa^s$. Indeed, using $\kappa\le\lambda^{\Nsmall}$ and \eqref{eq:epochs},
\[
 \kappa^s\lambda^{n-|\Aun|}
 \le \lambda^{n-|\Aun|+\frac{\Nsmall|\Aun|}{\Nsmall+1}}
 = \lambda^n\lambda^{-\frac{|\Aun|}{\Nsmall+1}}.
\]
Writing $z = \lambda^{-\frac{1}{\Nsmall+1}} \in (0,1)$, the remaining exponential decay $z^{|\Aun|}$ will let us bound its product with the linear factor $2+6|\Aun|$ by an absolute constant. Thus, by \ref{P:proxy-ends},
\begin{align*}
 \cost(X(1),d)
 &\le \kappa^s\lambda^{n-|\Aun|}\, (2 + 6|\Aun|)\, \Nsmall^{\Nsmall}\, \cost(M^*,d)\\
 &\le (2 + 6|\Aun|)\, z^{|\Aun|}\, \lambda^n\, \Nsmall^{\Nsmall}\, \cost(M^*,d)\\
 &\le \frac{6}{1 - z}\, \Nsmall^{\Nsmall}\, \lambda^n\, \cost(M^*,d),
\end{align*}
 where the last step uses $(2 + 6m)z^m \le 6(m + 1)z^m \le 6\sum_{i=0}^m z^i \le \frac{6}{1-z}$ with $m=|\Aun|$. This bound also covers the case $\Aun = \emptyset$. Since $\lambda^n = (10/9)^{\frac{n}{2}}$ and the constant $\frac{6\Nsmall^{\Nsmall}}{1-z}$ is independent of $n$, $\sigma$, and $d$, this proves \Cref{thm:ub-ps}.
\end{proof}

\subsection{Truncated Probabilistic Serial Rule}\label{ssec:tps}
We have seen thus far that PS could have exponential distortion on 
some instances. At the same time, our analysis that led to \Cref{thm:ub-ps}
suggested that much of the cost is amplified in the very late stages
of the mechanism. In light of this we ask, what happens if 
we terminate the mechanism at a specified cutoff $T$? 
We call this modification \emph{TruncatedPS}. 

The obvious caveat of a premature termination is that we will no
longer have a perfect fractional matching. 
To make matters worse, if this cutoff is set too early, 
then the maintained matrix
might even fail to partially assign an item.
For example, when all agents have identical preferences, the 
final item will only get partially assigned after time $1 - 1/n$. 
Below, we show that setting $T$ to be the slightly larger $1 - 1/(2n)$
allows us to complete the matrix to a perfect fractional matching. 

\begin{lemma}\label{lem:ps-doubling}
Let $X(t)$ be the matrix maintained by PS on an instance $\sigma$.
For any $t \in \left [1 - \frac{1}{2n}, 1 \right ]$, there 
exists a bistochastic matrix $Y = (Y_{a, b})$ such that 
$Y \leq 2 X(t)$ entrywise.
\end{lemma}

\begin{proof}
We reason in terms of flows.
Fix $t\in[1-1/(2n),1]$.
We consider a directed network with the following configuration:
\begin{enumerate}[noitemsep]
\item For each agent $a$, there is an edge from the source $a^\dagger$ to $a$ of capacity $1$.
\item For each item $b$, there is an edge from $b$ to the sink $b^\dagger$ of capacity $1$.
\item For each pair $(a,b)\in A\times B$, there is an edge from $a$ to $b$ of capacity $2X_{a,b}(t)$.
\end{enumerate}
It is not hard to see that a flow of $n$ units
to the sink corresponds to our desired matrix $Y$.
We will show that such a flow exists by the max-flow min cut theorem (cf. \citep{Schrijver:2003aa}). In particular, it suffices to show that every
$a^\dagger$-$b^\dagger$ cut has capacity at least $n$.

Fix subsets $U\subseteq A$ and $V\subseteq B$, and consider the cut with source side $\{a^\dagger\}\cup U\cup V$. Its outgoing edges are the $n-|U|$ edges from $a^\dagger$ to agents outside $U$, the $|V|$ edges from items in $V$ to $b^\dagger$, each of capacity $1$, and the edges from $a\in U$ to $b\notin V$, of capacity $2X_{a,b}(t)$.
Hence, the cut has capacity 
\begin{equation}
(n - |U|) +|V| + 2 \sum_{a \in U, b \notin V} X_{a, b}(t) 
\end{equation}
If $|U|\leq |V|$, then this capacity is at least $n$.
Otherwise, since $|U|\leq n$ and $t\geq1-1/(2n)$, we have

\begin{equation*}
|V| \leq |U| -1 \leq \left (1 - \frac{1}{n} \right )|U| \leq (2t - 1)|U|.
\end{equation*}
Since every row of $X(t)$ sums to $t$ and every column sums to at most $1$, this implies that
\begin{equation*}
    2\sum_{a \in U,b \notin V} X_{a,b}(t)
    =
    2\left(t|U|-\sum_{a \in U,b \in V} X_{a,b}(t)\right)
    \geq 2t|U|-2|V|
    \geq |U|-|V|.
\end{equation*}
This again shows that the capacity is at least $n$, as desired.

\end{proof}
\begin{remark}
The lower bound of the interval used in \Cref{lem:ps-doubling} is tight. To see this, consider an instance in which every agent has identical preferences. At any time  $t < 1 - \frac{1}{2n}$, less than $1/2$ of the final item has been consumed, so we cannot obtain our desired matrix $Y$ without increasing entries by a factor larger than $2$.
\end{remark}

\Cref{lem:ps-doubling} gives a procedure that we call \emph{doubling}, as every entry 
in the support of the matrix is at most doubled. If we run TruncatedPS until time $T = 1 - 1/(2n)$
and postprocess with the doubling procedure, we arrive at 
\Cref{alg:ps-truncated}.

\begin{center}
\myalg{alg:ps-truncated}{TruncatedPS (with doubling)}{
\textbf{Input:} An instance $\sigma$. \medskip \\
Let $X(T)$ be the matrix maintained by PS at time  $T = 1 - 1/(2n)$. \\
Let $Y$ be any bistochastic matrix with $Y \leq 2X(T)$ (guaranteed by \Cref{lem:ps-doubling}).\\
Return $Y$. 
}
\end{center}

We write $\tps$ for TruncatedPS with doubling and, as with PS, interpret its output $Y$ as a distribution over perfect integral matchings with marginals $Y$.

Below, we show that this simple modification of PS significantly improves the distortion.
\begin{reptheorem}{thm:ub-ps-truncated}
For every instance $\sigma$, \Cref{alg:ps-truncated} has distortion $O(n)$.
\end{reptheorem}
\begin{proof}[Proof of \Cref{thm:ub-ps-truncated}]
Fix an instance $\sigma$ and consistent metric $d$ with optimal assignment $M^*$.
Letting $X$ and $Y$ be the associated matrices, we have

\begin{align*}
\E_{M \sim \tps(\sigma)} \left [ \cost (M, d) \right ]
&= \sum_{a \in A} \sum_{b \in B} Y_{a, b} \cdot d(a, b) \\
&\leq 2 \sum_{a \in A} \sum_{b \in B} X_{a, b}(T) \cdot d(a, b) \tag{\Cref{lem:ps-doubling}} \\
&\leq 2 \left(\frac{T}{1-T}\right) \cost(M^*,d) \tag{\Cref{cor:prefix}} \\
&= 2 \left(\frac{1-1/(2n)}{1-(1-1/(2n))}\right) \cost(M^*,d) \\
&= (4n-2)\cost(M^*,d),
\end{align*}
which implies our stated distortion guarantee.
\end{proof}

Intriguingly, we show that the modification has a similar distortion lower bound as RSD.

\begin{theorem} \label{thm:lower-bound-truncated-ps}
For arbitrarily large $n$, there exists an instance $\sigma$ such that 
\[
\distortion(\tps, \sigma) = \Omega \left (\frac{n}{\sqrt{\log(n)}} \right),
\] 
where $\tps$ refers to the mechanism described in \Cref{alg:ps-truncated}.
\end{theorem}

\begin{proof}

Use the same instance and parameters as in the proof of \Cref{thm:rsd-lowerbound}. We prove by strong induction on the column index $j$ that, under PS, all items at every retained vertex $(h,j)$ are exhausted by the deadline $1-u_j^*$. The case $j=0$ is immediate, since column $0$ contains no items.
For $j\geq1$, assume the claim for every column $i<j$, and fix a retained vertex $(h,j)$. Suppose, for a contradiction, that its items are not exhausted by $1-u_j^*$.
The co-located agents eat at this vertex throughout $[0,1 - u_j^*]$. By the induction hypothesis, all items in earlier columns are exhausted by $1-u_{j-1}^*$. Thus \eqref{eq:preference} forces every parent agent of $(h,j)$ to eat items at $(h,j)$ throughout $[1-u_{j-1}^*,1-u_j^*]$.

We show that if the items at $(h, j)$ are not exhausted by $1 - u_j^*$, then more than their capacity has been eaten, leading to a contradiction.
By \eqref{eq:adjusted-deadline-surplus}, the parent agents, together with the co-located agents, eat at least
\[
 (u_{j-1}^*-u_j^*)\sum_{i<j}m^*_{h-1,i} + (1 - u_j^*)m^*_{h, j}
 \geq(1+\varepsilon_j)m^*_{h,j}u_j^* + (1 - u_j^*)m^*_{h, j}
 >m^*_{h,j}.
\]
This contradiction completes the induction.

Take $v=(D,j)$, one of the locations at level $D$ with a non-zero number of agents.
Choose an item $b$ at $v$ and break ties within this location so that co-located agents rank $b$ last and all other agents rank it first. The mass-pruning cutoffs in \eqref{eq:mass-pruning-thresholds} and $u_j^* \geq u_j/2$ give $m^*_{D,j}u_j^* \geq \tau_j u_j^* \geq \tau_j u_j/2 \geq 3/2>1$. Until $b$ is exhausted, outside agents eat no other item at $v$. Thus co-located agents must eat the other $m^*_{D,j}-1$ items themselves before reaching $b$, which takes at least $1-1/m^*_{D,j}>1-u_j^*$. Therefore, it is fully eaten by outside agents. Moreover, $b$ is exhausted before the cutoff:
\begin{align*}
    1-u_j^* &\leq 1-u_N^* < 1-\frac{1}{2n} \iff 2n \cdot u_N^* > 1 \\
    \text{since}\qquad n u_N^*
    &\geq qB_{\mathrm{mass}}u_N^*
    \geq\frac{qB_0u_N^*}{4}
    \geq\frac Q2\geq2.
\end{align*}
Thus, any matching produced from the support of $\tps$ will cost at least $2^{D - 1}$ for just matching the item $b$.
 Since the optimal cost is $q$, \eqref{eq:rsd-final-distortion-scaling} gives
\[
 \distortion(\tps,\sigma)
 \geq\frac{2^{D-1}}q
 =\Omega\!\left(\frac{n}{\sqrt{\log n}}\right).
\]
\end{proof}

\section{Truthful Mechanisms}\label{sec:truth}

Next, we turn to \emph{truthfulness}, one of the central considerations in designing matching mechanisms. 

\subsection{General lower bound}

We start by giving the first lower bound for truthful mechanisms beyond the $\Omega(\log n)$ lower bound of \citep*{Anari:2023aa} that applies to all mechanisms.

\begin{reptheorem}{thm:tie-sqrtlog}
Every truthful-in-expectation matching mechanism has distortion at least $2^{\Omega(\sqrt{\log n})}$.
\end{reptheorem}

Since this is the first lower bound that crucially uses truthfulness, it helps to start by building some informal intuition for the conflict between truthfulness and low distortion that we can exploit for a better lower bound.

Once again, consider the metric which shows that Serial Dictatorships can have exponentially large distortion \citep*{Caragiannis:2016aa,Caragiannis:2024ab}. To review, agents and items sit on a line, with one item at $0$, one agent at $1$, and one of each at every point $2^t$ for $1 \leq t \leq n - 1$. Agents always break ties in favor of items to the right. This causes a domino effect when a Serial Dictatorship allows agents to choose from left to right. Each agent at $2^t$ ``steals'' the item at $2^{t + 1}$, leading to distortion $2^n - 1$. 

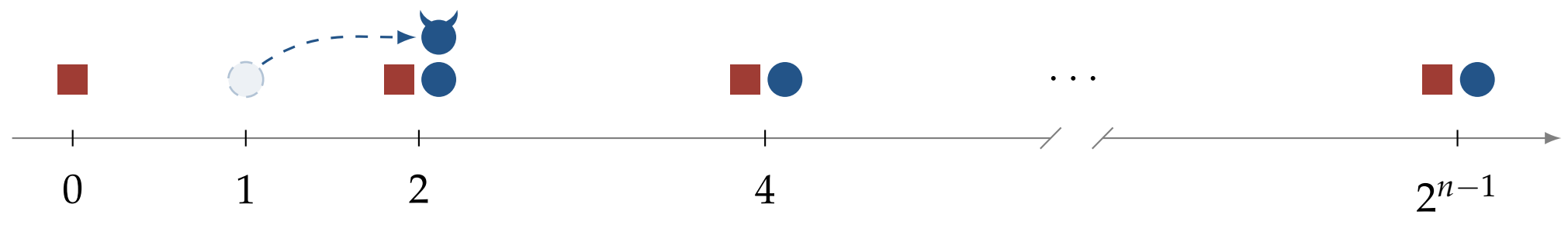
\begin{figure}[ht]
\centering
\pgfmathsetlengthmacro{\lineunit}{0.8*\textwidth/8.95}
\begin{tikzpicture}[x=\lineunit,y=1cm,>=latex,font=\small]
  \definecolor{agentblue}{RGB}{35,84,136}
  \definecolor{itemred}{RGB}{159,60,52}

  \draw[gray] (-.35,0) -- (5.65,0);
  \draw[gray,->] (5.95,0) -- (8.6,0);
  \draw[gray] (5.59,-.09) -- (5.71,.09)
              (5.89,-.09) -- (6.01,.09);

  \foreach \x/\lab in {0/0,1/1,2/2,4/4,8/{2^{n-1}}} {
    \draw (\x,-.07) -- (\x,.07);
    \node[below=5pt] at (\x,0) {$\lab$};
  }

  \node[rectangle,fill=itemred,inner sep=0pt,minimum size=7.245pt]
    at (0,.5) {};

  \node[circle,fill=agentblue,fill opacity=.08,
        draw=agentblue,draw opacity=.35,densely dashed,line width=.5pt,
        inner sep=0pt,minimum size=8.4525pt]
    (a0-old) at (1,.5) {};

  \node[rectangle,fill=itemred,inner sep=0pt,minimum size=7.245pt]
    at (1.885,.5) {};
  \node[circle,fill=agentblue,inner sep=0pt,minimum size=8.4525pt]
    (a0-new) at (2.115,.86) {};

  \begin{scope}[shift={(a0-new.center)},x=1.2075pt,y=1.2075pt]
    \fill[agentblue]
      (-2.5,2.1) .. controls (-3.6,2.9) and (-4.0,4.3) .. (-3.7,5.5)
      .. controls (-3.2,4.3) and (-2.4,3.6) .. (-1.1,3.0) -- cycle;
    \fill[agentblue]
      (2.5,2.1) .. controls (3.6,2.9) and (4.0,4.3) .. (3.7,5.5)
      .. controls (3.2,4.3) and (2.4,3.6) .. (1.1,3.0) -- cycle;
  \end{scope}

  \node[circle,fill=agentblue,inner sep=0pt,minimum size=8.4525pt]
    at (2.115,.5) {};

  \draw[->,agentblue,dashed,line width=.6pt,
        shorten <=1pt,shorten >=1pt]
    (a0-old.north east) to[out=35,in=180]
    (a0-new.west);

  \foreach \x in {4,8} {
    \node[rectangle,fill=itemred,inner sep=0pt,minimum size=7.245pt]
      at (\x-.115,.5) {};
    \node[circle,fill=agentblue,inner sep=0pt,minimum size=8.4525pt]
      at (\x+.115,.5) {};
  }

  \node at (5.8,.5) {$\cdots$};
\end{tikzpicture}
\caption{The agent at $1$ pretends to move to $2$.}
\label{fig:truthful-line-misreport-small}
\end{figure}

As a stand-in for a low distortion mechanism, consider a mechanism which appears to magically know that this is the underlying metric, and chooses the optimum matching.\footnote{This may seem silly, but low-distortion mechanisms \emph{do} appear magical in this kind of way.} Such a mechanism will always match the agent at $1$ to the item at $0$. How could this agent lie and increase her chance of getting a higher ranked item? If any such increase is possible, it would violate truthfulness by \Cref{prop:equiv} (even though the expected cost does not decrease in this particular metric). One way is to impersonate the agent at $2$, as in \Cref{fig:truthful-line-misreport-small}. In this modified instance, the mechanism can no longer distinguish between the two agents at $2$, and so the lying agent may now have a $50\%$ chance of getting the item at $2$, which she prefers. (Since we can assume that the mechanism is anonymous by \Cref{prop:anonymization}, this kind of reasoning can be operationalized rigorously.)

More generally, an agent at $2^t$ is incentivized to pretend to be at $2^{t + 1}$ as long as her probability of getting matched to the item at $0$ is larger than her probability of failing to get matched to an item at $2^{t + 1}$. We can drive down the latter probability by having groups of agents at each point that grow geometrically.\footnote{\cite{Caragiannis:2016aa,Caragiannis:2024ab} also use geometrically growing groups in their $\Omega(n^{0.29})$ lower bound for RSD.} (Intuitively, larger groups to the right make it easier for an agent to ``slip in'' undetected and steal an item in the next group.) Low distortion constrains the mechanism to have a high probability of matching agents to collocated items. Altogether, we can argue that for each of the groups except the rightmost (which has no next group), the probability that any agent matches to the item at $0$ is at most $O(D/r)$, where $D$ is the distortion and $r$ is the growth rate of the groups. With $k$ of these groups, this leaves probability mass $1 - O(kD/r)$ for the rightmost group, which forces distortion $D$ at least $2^k(1 - O(kD/r))$. Solving for $D$ and balancing the parameters, we get (roughly) $r \approx k2^k$, a total of $n\approx 2^{k^2}$ agents, and  $D\geq \Omega(2^k)$, giving \Cref{thm:tie-sqrtlog}. 

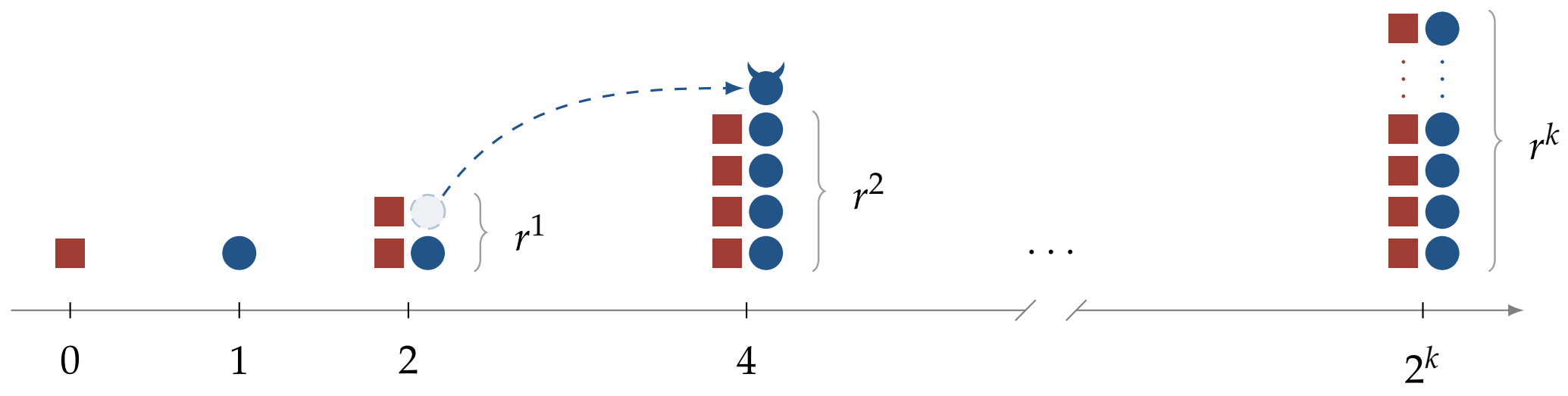
\begin{figure}[ht]
\centering
\pgfmathsetlengthmacro{\lineunit}{0.8*\textwidth/8.95}
\begin{tikzpicture}[x=\lineunit,y=1cm,>=latex,font=\small]
  \path[use as bounding box] (-.45,-.55) rectangle (9.05,2.79);
  \definecolor{agentblue}{RGB}{35,84,136}
  \definecolor{itemred}{RGB}{159,60,52}
  \tikzset{
    agent/.style={circle,fill=agentblue,inner sep=0pt,minimum size=8.4525pt},
    item/.style={rectangle,fill=itemred,inner sep=0pt,minimum size=7.245pt},
    countbrace/.style={decorate,decoration={brace,amplitude=3pt},draw=gray!75,line width=.45pt}
  }

  \draw[gray] (-.35,0) -- (5.65,0);
  \draw[gray,->] (5.95,0) -- (8.6,0);
  \draw[gray] (5.59,-.09) -- (5.71,.09)
              (5.89,-.09) -- (6.01,.09);
  \foreach \x/\lab in {0/0,1/1,2/2,4/4,8/{2^k}} {
    \draw (\x,-.07) -- (\x,.07);
    \node[below=5pt] at (\x,0) {$\lab$};
  }

  \node[item] at (0,.5) {};
  \node[agent] at (1,.5) {};

  \node[agent,fill opacity=.08,draw=agentblue,draw opacity=.35,
        densely dashed,line width=.5pt] (a-old) at (2.115,.86) {};
  \node[agent] (a-new) at (4.115,1.94) {};

  \begin{scope}[shift={(a-new.center)},x=1.2075pt,y=1.2075pt]
    \fill[agentblue]
      (-2.5,2.1) .. controls (-3.6,2.9) and (-4.0,4.3) .. (-3.7,5.5)
      .. controls (-3.2,4.3) and (-2.4,3.6) .. (-1.1,3.0) -- cycle;
    \fill[agentblue]
      (2.5,2.1) .. controls (3.6,2.9) and (4.0,4.3) .. (3.7,5.5)
      .. controls (3.2,4.3) and (2.4,3.6) .. (1.1,3.0) -- cycle;
  \end{scope}

  \foreach \y in {.5,.86} {
    \node[item] at (1.885,\y) {};
  }
  \node[agent] at (2.115,.5) {};

  \foreach \y in {.5,.86,1.22,1.58} {
    \node[item] at (3.885,\y) {};
    \node[agent] at (4.115,\y) {};
  }

  \foreach \y in {.5,.86,1.22,1.58,2.46} {
    \node[item] at (7.885,\y) {};
    \node[agent] at (8.115,\y) {};
  }
  \foreach \y in {1.87,2.02,2.17} {
    \fill[itemred] (7.885,\y) circle[radius=.45pt];
    \fill[agentblue] (8.115,\y) circle[radius=.45pt];
  }

  \draw[countbrace] (2.39,1.02) -- (2.39,.34)
    node[midway,right=6pt,text=black] {$r^1$};
  \draw[countbrace] (4.39,1.74) -- (4.39,.34)
    node[midway,right=6pt,text=black] {$r^2$};
  \draw[countbrace] (8.39,2.62) -- (8.39,.34)
    node[midway,right=6pt,text=black] {$r^k$};

  \node at (5.82,.5) {$\cdots$};

  \draw[->,agentblue,dashed,line width=.6pt,shorten <=1pt,shorten >=1pt]
    (a-old.north east) to[out=55,in=180] (a-new.west);
\end{tikzpicture}
\caption{An agent slips into the next, larger group.}
\label{fig:truthful-line-misreport-big}
\end{figure}

Now we present the proof formally.

\begin{proof}[Proof of \Cref{thm:tie-sqrtlog}]
Suppose we have a truthful-in-expectation matching mechanism $\mathcal M$ with distortion at most $D$. The proof proceeds in three steps. First, we specify the instance (described above) and the induced preferences. Then, we consider what happens when an agent misreports and show that she gets an item in the next group with high probability. Finally, we leverage truthfulness to argue that all agents are unlikely to get matched to the item furthest to the left, which gives a contradiction if $D$ is too small.

\paragraph{The instance.}
For integer parameters $k, r$ to be fixed later, imagine we have an instance $\sigma$ derived from the following metric. The agents and items are grouped into sets $A_0, A_1, \dots, A_k$ and $B_0, B_1, \dots, B_k$ respectively, with $|A_t| = |B_t| = r^t$ for each $0 \leq t \leq k$. The agents and items occupy points on a line. In general, the agents and items in $A_t$ and $B_t$ are collocated at position $2^t$, with one exception: the single item $b_0 \in B_0$ is at $0$.

In terms of ordinal rankings, each agent $a \in A_t$ has the preference
$$B_t \succ B_{t - 1} \succ \cdots \succ B_1 \succ B_{t + 1} \succ b_0 \succ B_{t + 2} \succ \dots \succ B_k,$$
where the prefix $B_t \succ B_{t - 1} \succ \cdots \succ B_1$ is empty for $t=0$ and the suffix $B_{t+1},\dots, B_k$ is empty for $t = k$.
For $t<k$, note that in the metric,
$a$ has the same distance (specifically, $2^t$) to $B_{t + 1}$ and $b_0$, and we are always breaking the tie in favor of the former. The way agents break ties within a group is unimportant, but to be explicit, one can assume that they all have a common default ranking for items in the same group. 
The key feature of the preference is that, for $t<k$, an agent $a \in A_t$ has the prefix $S_t := \cup_{j = 1}^{t + 1} B_j$, which she prefers to $b_0$.

\paragraph{An agent infiltrates the next group.} Fix some $t < k$ and $a \in A_t$, and consider the instance $\sigma'$ obtained by changing $a$'s preference to match the agents in $A_{t + 1}$. We will argue that if the mechanism has small distortion, then $a$ must now be likely to get an item in $B_{t + 1}$.

By anonymity, every agent in $\{a\} \cup A_{t + 1}$ has the same probability $p_t$ of getting an item outside $B_{t + 1}$. Consider the metric consistent with $\sigma'$ which is identical to the metric defining $\sigma$, but moves $a$ to be collocated with $A_{t + 1} \cup B_{t + 1}$ at $2^{t + 1}$.
The optimum matching has cost $2^{t} + 1$, and a mechanism with probability $p_t$ of matching agents in $\{a\} \cup A_{t + 1}$ outside $B_{t + 1}$ has cost at least $p_t(r^{t + 1} + 1)2^t$. Since the mechanism has distortion at most $D$, it follows that
$$p_t(r^{t + 1} + 1)2^t \leq D(2^{t} + 1).$$
Sloppily bounding $2^{t} + 1\leq 2^{t + 1}$, we get
$$p_t \leq \frac{2D}{r^{t + 1} + 1}.$$
Thus, the smaller $D$ is and the larger $r$ and $t$ are, the more likely $a$ is to get an item in $B_{t + 1}$.

\paragraph{Was the lie worth it?} We know that if $a$ pretends to be in $A_{t + 1}$ (changing the instance from $\sigma$ to $\sigma'$), then she is not likely to get an item worse than one in $B_{t + 1}$. Intuitively, the mechanism is truthful only if the same is true in the original instance $\sigma$, meaning she is not likely to get $b_0$ in $\sigma$. 

Formally, for $t < k$ and $a \in A_t$, \Cref{prop:equiv} gives us
\begin{align*}
\Pr_{M \sim \mathcal M(\sigma)}[M(a) = b_0]
&\leq \Pr_{M \sim \mathcal M(\sigma)}[M(a) \notin S_t] \\
&\leq \Pr_{M \sim \mathcal M(\sigma')}[M(a) \notin S_t] \\
&\leq \Pr_{M \sim \mathcal M(\sigma')}[M(a) \notin B_{t+1}]\\
&\leq p_t \leq \frac{2D}{r^{t+1} + 1}.
\end{align*}
For $A_k$,  we simply have
$$\sum_{a\in A_k}\Pr_{M \sim \mathcal M(\sigma)}[M(a) = b_0] \leq \frac{D}{2^k}$$
since $d(a, b_0) = 2^k$, and the distortion is at most $D$. Summing over all agents $a$, we get
$$1 = \sum_{a \in A} \Pr_{M \sim \mathcal M(\sigma)}[M(a) = b_0] \leq \frac{D}{2^k} + \sum_{t = 0}^{k - 1} \frac{2D r^t}{r^{t + 1} + 1} \leq D\left(\frac{1}{2^k} + \frac{2k}{r}\right).$$
Balancing parameters, we can set $r = k2^{k+1}$ so that $D \geq 2^{k-1}$. The total number of agents is
$$n = \sum_{t=0}^k r^t \le (k+1)r^k = (k+1) k^k 2^{k(k+1)} = 2^{O(k^2)}$$
and so $D \geq 2^{\Omega(\sqrt{\log n})}$. \qedhere

\end{proof}

We briefly note that the same lower bound and essentially the same proof work for mechanisms that are \emph{envy-free} rather than truthful. The intuition is that the construction shows that if a mechanism has low distortion $D = 2^{o(\sqrt{\log n})}$, then some agent can benefit from impersonating the group to the right. But the same argument shows that this agent would \emph{envy} an agent in the next group, and so the same logical sequence follows. 

\subsection{Deterministic Pareto optimal mechanisms}

Next, we consider whether we can obtain stronger lower bounds when introducing further constraints on the mechanism
such as (ex-post) Pareto optimality. 

\begin{reptheorem}{thm:det-po}
Every deterministic truthful mechanism that is Pareto optimal has distortion at least $2^n - 1$.
\end{reptheorem}

\begin{proof}

Once again, the lower bound will use the instance described in \Cref{fig:truthful-line-misreport-small}. We have items at positions $0, 2, 4, \dots, 2^{n-1}$ and agents at positions $1, 2, 4, \dots, 2^{n-1}$. Let $\sigma$ be the particular instance that canonically labels the items $b_0, \dots, b_{n-1}$ and agents $a_0, \dots, a_{n - 1}$ from left to right, and has agents break ties towards the right. In particular, in $\sigma$, $a_t$ has the preference
$$b_t\succ b_{t- 1} \succ \cdots \succ b_1 \succ b_{t + 1} \succ b_0 \succ b_{t + 2} \succ \cdots \succ b_{n - 1}$$
except $a_0$ prefers $b_1$ to $b_0$ and $a_{n-1}$'s preference is just right to left.

We will show that for any deterministic, truthful, Pareto optimal mechanism, there is some relabeling of the agents such that $b_0$ is matched to the rightmost agent at $2^{n-1}$. Any matching that makes this choice has distortion at least $2^{n} - 1$, since an optimal matching for any instance on the line recursively matches the leftmost item to the leftmost agent (see \Cref{fig:pareto-domino}).

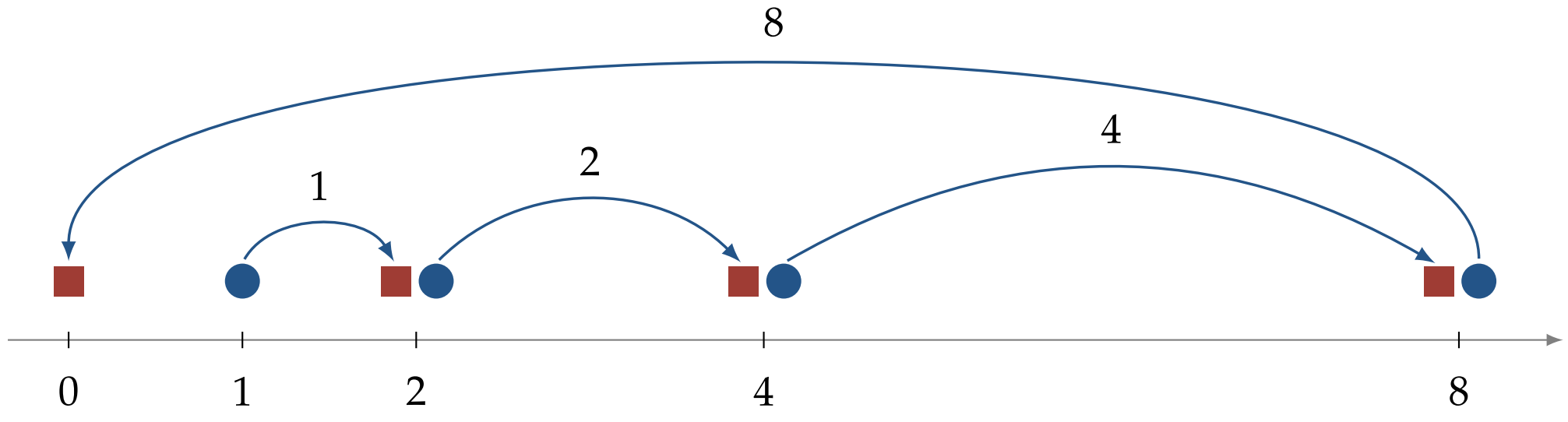
\begin{figure}[ht]
\centering
\pgfmathsetlengthmacro{\lineunit}{0.8*\textwidth/8.95}
\begin{tikzpicture}[x=\lineunit,y=1cm,>=latex,font=\small]
  \path[use as bounding box] (-.45,-.5) rectangle (8.75,2.65);
  \definecolor{agentblue}{RGB}{35,84,136}
  \definecolor{itemred}{RGB}{159,60,52}
  \tikzset{
    agent/.style={circle,fill=agentblue,inner sep=0pt,minimum size=8.4525pt},
    item/.style={rectangle,fill=itemred,inner sep=0pt,minimum size=7.245pt},
    matching/.style={->,agentblue,line width=.65pt,shorten <=1pt,shorten >=1pt}
  }
  \draw[gray,->] (-.35,0) -- (8.6,0);
  \foreach \x in {0,1,2,4,8} {
    \draw (\x,-.07) -- (\x,.07);
    \node[below=5pt] at (\x,0) {$\x$};
  }
  \node[item] (b0) at (0,.5) {};
  \node[agent] (a0) at (1,.5) {};
  \foreach \x/\i in {2/1,4/2,8/3} {
    \node[item] (b\i) at (\x-.115,.5) {};
    \node[agent] (a\i) at (\x+.115,.5) {};
  }
  \draw[matching] (a0.north) to[out=60,in=120]
    node[midway,above=2pt,text=black] {$1$} (b1.north);
  \draw[matching] (a1.north) to[out=45,in=135]
    node[midway,above=2pt,text=black] {$2$} (b2.north);
  \draw[matching] (a2.north) to[out=30,in=150]
    node[midway,above=2pt,text=black] {$4$} (b3.north);
  \draw[matching] (a3.north) .. controls +(0,2.25) and +(0,2.25) ..
    node[midway,above=3pt,text=black] {$8$} (b0.north);
\end{tikzpicture}
\caption{Our instance $\sigma$ with $n = 4$, showing that matching the leftmost item to the rightmost agent forces cost at least $2^{n} - 1$.}
\label{fig:pareto-domino}
\end{figure}

Though we are considering a deterministic mechanism, it will be helpful to anonymize it by averaging over all relabelings, as in \Cref{prop:anonymization}. The anonymized mechanism is only truthful-in-expectation, but if it has \emph{any} positive probability of matching $b_0$ to $a_{n - 1}$ in $\sigma$, then there exists a relabeling on which the original mechanism has distortion $2^{n} - 1$. The core property of truthfulness-in-expectation that we will use is the following immediate corollary of \Cref{prop:equiv}. 
If $\mathcal{M}$ is truthful-in-expectation, and $\tau$ and $\tau'$
are two instances that differ only in $a$'s ranking, such that $a$
has the same set $S$ of top $|S|$ items in both rankings
(possibly in a different internal order), then
\begin{equation}\label{eq:prefix-truth}
\Pr_{M\sim \mathcal{M}(\tau)}[M(a) \in S] = \Pr_{M\sim \mathcal{M}(\tau')}[M(a) \in S].
\end{equation}

Like in the proof of \Cref{thm:tie-sqrtlog}, we will apply this property to adjacent instances in which one agent pretends to move to the right. We can show that the probability mass of matching with $b_0$ moves to the right step by step, until some reaches $a_{n - 1}$. While \Cref{thm:tie-sqrtlog} only considered instances where only one agent's preference differs from the original, here it will be helpful to consider a \emph{family} of connected instances that we can analyze inductively. We call these \emph{admissible} instances.

In every admissible instance, the items are fixed exactly as they are in $\sigma$, but agents may occupy different points in $\{2^0, \dots, 2^{n-1}\}$. We require that for each $j \geq 0$, there are at most $j + 1$ agents in the positions $\{2^0, \dots, 2^j\}$ (intuitively, agents only move to the right from $\sigma$). For $0\leq s < t \leq n$, we say an admissible instance has \emph{type} $(s, t)$ if we have singletons $a_0, \dots, a_{s-1}$ at $2^0, \dots, 2^{s-1}$ respectively and a cluster $a_{s}, \dots, a_{t - 1}$ all collocated at $2^{t - 1}$. (The preferences are derived from the metric as usual, ordering by distance with agents breaking ties towards the right.)

The central claim is as follows.

\begin{claim}
For any anonymous, ex-post Pareto optimal, truthful-in-expectation mechanism $\mathcal{M}$, and any admissible instance $\tilde{\sigma}$ of type $(s, t)$, 
$$\Pr_{M\sim \mathcal{M}(\tilde{\sigma})}[M(b_0) \notin \{a_0, \dots, a_{s-1}\}] = \frac{t - s}{ts!}.$$
\end{claim}

The claim is sufficient to establish the theorem. Since $\sigma$ is an admissible instance of type $(n-1, n)$ it implies
$$\Pr_{M\sim \mathcal{M}(\sigma)}[M(b_0) = a_{n-1}] = \Pr_{M\sim \mathcal{M}(\sigma)}[M(b_0) \notin \{a_0, \dots, a_{n-2}\}] = \frac{1}{n!} > 0$$
as needed. Thus, it only remains to prove the claim. Before doing so, we mention one key property of Pareto optimal mechanisms which we use repeatedly: In any admissible instance, an agent at $2^j$ can only match to one of $b_0, \dots, b_{j + 1}$. If she matches to $b_k$ with $k \geq j + 2$, then since at most $k - 1$ other agents are to the left of $2^k$, some agent at or to the right of $2^k$ gets matched to an item in $\{b_0, \dots, b_{k-1}\}$. She and the original agent (matched to $b_k$) would both prefer to swap, which violates Pareto optimality. We call this the \emph{no-skipping} property, since an agent cannot skip over an item to the right.

\paragraph{Proof of the claim.} We will proceed by induction on $s$, simultaneously for all $t>s$. The cases where $s = 0$ are immediate; the left side becomes $\Pr_{M\sim \mathcal{M}(\tilde{\sigma})}[M(b_0) \notin \varnothing] = 1$, and the right side is $\frac{t}{t} = 1$. 

Now suppose that the claim holds for all $(s, t)$ with some fixed $s$, and consider any admissible instance $\tilde{\sigma}$ of type $(s + 1, t)$. Our goal is to show that 
\begin{equation}\label{eq:induct-step}
\Pr_{M\sim \mathcal{M}(\tilde{\sigma})}[M(b_0) \notin \{a_0, \dots, a_{s}\}] = \frac{t - s-1}{t(s+1)!}.
\end{equation}
We can also view $\tilde{\sigma}$ as an admissible instance of type $(s, s + 1)$, by viewing $a_0, \dots, a_{s - 1}$ as the singletons and $a_{s}$ as the cluster. Thus the inductive hypothesis already establishes
$$\Pr_{M\sim \mathcal{M}(\tilde{\sigma})}[M(b_0) \notin \{a_0, \dots, a_{s-1}\}] = \frac{1}{(s+1)!}.$$
Therefore, to show \eqref{eq:induct-step}, it suffices to find the probability that $\mathcal{M}$ matches $b_0$ to $a_s$ in $\tilde{\sigma}$. That is,
\begin{equation}\label{eq:induct-step-2}
\Pr_{M\sim \mathcal{M}(\tilde{\sigma})}[M(b_0) = a_s] = \frac{1}{(s+1)!} - \frac{t - s-1}{t(s+1)!} = \frac{1}{ts!}.
\end{equation}
To prove this, we will use truthfulness to compare with another admissible instance $\sigma'$ of type $(s, t)$. It is identical to $\tilde{\sigma}$, but we move $a_s$ to join the cluster at $2^{t - 1}$ (see \Cref{fig:cluster-move}). It turns out to be tricky to compare these instances directly, but we can consider moving $a_s$ step by step to each position $2^s, 2^{s + 1}, \dots, 2^{t - 1}$, and compare the changes between each step. The intuition is that in each pair of adjacent instances, the preferences of $a_s$ share a long common prefix (which excludes $b_0$). Thus, we can use truthfulness (specifically, \eqref{eq:prefix-truth}) and Pareto optimality to determine the probability that $a_s$ matches to $b_0$ in all these intermediate instances. 

\begin{figure}[ht]
\centering
\begin{tikzpicture}[x=.94cm,y=.82cm,>=latex,font=\small]
  \definecolor{agentblue}{RGB}{35,84,136}
  \definecolor{itemred}{RGB}{159,60,52}
  \tikzset{
    agent/.style={circle,fill=agentblue,inner sep=0pt,minimum size=8.4525pt},
    item/.style={rectangle,fill=itemred,inner sep=0pt,minimum size=7.245pt}
  }
  \draw[gray] (-.35,0) -- (7.12,0);
  \draw[gray,->] (7.48,0) -- (15.9,0);
  \draw[gray] (7.06,-.09) -- (7.18,.09)
              (7.42,-.09) -- (7.54,.09);
  \foreach \x/\lab in {0/0,.8/1,1.9/2,3.6/{2^s},5.9/{2^{s+1}},8.7/{2^{t-2}},12/{2^{t-1}}} {
    \draw (\x,-.07) -- (\x,.07);
    \node[below=5pt] at (\x,0) {$\lab$};
  }
  \node[item] at (0,.5) {};
  \node[agent] at (.8,.5) {};
  \node[item] at (1.74,.5) {};
  \node[agent] at (2.06,.5) {};
  \node at (2.75,.5) {$\cdots$};
  \node[item] at (3.44,.5) {};
  \node[agent,fill opacity=.08,draw=agentblue,draw opacity=.35,
    densely dashed,line width=.5pt] (old) at (3.76,.5) {};
  \node[above=5pt] at (old.north) {$a_s$};
  \node[item] at (5.9,.5) {};
  \node at (7.3,.5) {$\cdots$};
  \node[item] at (8.7,.5) {};
  \node[item] at (11.84,.5) {};
  \foreach \y in {.5,.92,1.34} {\node[agent] at (12.16,\y) {};}
  \node[agent] (new) at (12.16,1.76) {};
  \begin{scope}[shift={(new.center)},x=1.2075pt,y=1.2075pt]
    \fill[agentblue]
      (-2.5,2.1) .. controls (-3.6,2.9) and (-4.0,4.3) .. (-3.7,5.5)
      .. controls (-3.2,4.3) and (-2.4,3.6) .. (-1.1,3.0) -- cycle;
    \fill[agentblue]
      (2.5,2.1) .. controls (3.6,2.9) and (4.0,4.3) .. (3.7,5.5)
      .. controls (3.2,4.3) and (2.4,3.6) .. (1.1,3.0) -- cycle;
  \end{scope}
  \draw[decorate,decoration={brace,amplitude=3pt},gray]
    (12.5,1.96) -- (12.5,.3)
    node[midway,right=6pt,text=black] {$t-s$ agents};
  \draw[->,agentblue,dashed,line width=.6pt,shorten <=1pt,shorten >=1pt]
    (old.north east) .. controls +(.47,.92) and +(-.47,.92) .. (5.84,.72);
  \draw[->,agentblue,dashed,line width=.6pt,shorten <=1pt,shorten >=1pt]
    (5.98,.72) .. controls +(.7,1.07) and +(-.7,1.07) .. (8.64,.72);
  \draw[->,agentblue,dashed,line width=.6pt,shorten <=1pt,shorten >=1pt]
    (8.78,.72) .. controls +(.55,.95) and +(-1.2,0) .. (new.west);
  \node[text=black] at (7.95,1.99) {one position at a time};
  \node at (14.8,.5) {$\cdots$};
\end{tikzpicture}
\caption{The last singleton joins the cluster, changing type $(s+1,t)$ to type $(s,t)$.}
\label{fig:cluster-move}
\end{figure}
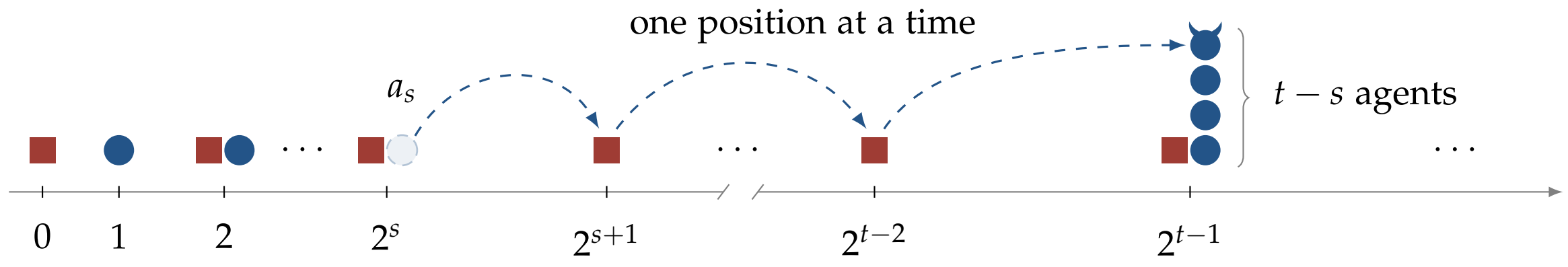

More formally, let $\sigma_j$ denote the instance for which $a_s$ is at $2^j$, so that $\sigma_s = \tilde{\sigma}$ and $\sigma_{t - 1} = \sigma'$. At each intermediate position $2^j$ with $s<j<t-1$, Pareto optimality prohibits $a_s$ from matching to the right. By no-skipping, the only item to the right she could receive is $b_{j+1}$. If she did, then some agent to the right of $a_s$ would match with one of $b_1,\dots,b_j$, since there are only $s<j$ agents to the left of $a_s$. This agent and $a_s$ would both prefer to swap, violating Pareto optimality. Thus, for each $s<j<t-1$, in the intermediate instance $\sigma_j$, $a_s$ only has positive probability of matching to $\{b_0,\dots,b_j\}$. At the initial position $2^s$ (instance $\tilde{\sigma}$), no-skipping guarantees that $a_s$ matches to one of $\{b_0,\dots,b_{s+1}\}$.

Now, consider moving $a_s$ from $2^j$ to $2^{j+1}$, for $s\leq j<t-1$. In the corresponding instances $\sigma_j$ and $\sigma_{j+1}$, the rankings of $a_s$ share the common prefix $\{b_1,\dots,b_{j+1}\}$. Thus, \eqref{eq:prefix-truth} implies that
\begin{equation}\label{eq:hops}
\Pr_{M\sim \mathcal{M}(\sigma_j)}[M(a_s)\notin\{b_1,\dots,b_{j+1}\}] = \Pr_{M\sim \mathcal{M}(\sigma_{j+1})}[M(a_s)\notin\{b_1,\dots,b_{j+1}\}].
\end{equation}
No-skipping implies that $a_s$ only matches to $\{b_0,\dots,b_{j+1}\}$, and we showed that this is further restricted to $\{b_0,\dots,b_{j}\}$ when $s < j < t - 1$. Thus, \eqref{eq:hops} says that the probability that $\mathcal{M}$ matches $a_s$ to $b_0$ is the same in $\sigma_s, \sigma_{s + 1}, \dots, \sigma_{t-2}$. Chaining these together (including the final application of \eqref{eq:hops} at $j = t - 2$), we get
\begin{equation}\label{eq:chain}
\Pr_{M\sim \mathcal{M}(\tilde{\sigma})}[M(a_s)=b_0] = \Pr_{M\sim \mathcal{M}(\sigma')}[M(a_s)\notin\{b_1,\dots,b_{t-1}\}].
\end{equation}
Now, consider how the items $b_1, \dots, b_{t-1}$ can be matched in $\sigma'$. For $t<n$, no-skipping implies that the $t$ agents $a_0, \dots, a_{t - 1}$ can only get matched to the $t + 1$ items $b_0, \dots, b_t$. If one of $b_1, \dots, b_{t-1}$ is matched to an agent $a_j$ with $j \geq t$, then it means that one of the agents in the cluster $\{a_s, \dots, a_{t - 1}\}$ must get matched to $b_t$. However, this agent and $a_j$ would prefer to swap, violating Pareto optimality. Thus, $b_1, \dots, b_{t-1}$ can only match with $a_0, \dots, a_{t - 1}$, and exactly one special agent matches to a different item, which must be $b_0$ or $b_t$. The same is trivially true when $t = n$, except $b_0$ is the only possible different item. 

In either case, the special agent is in the cluster $\{a_s, \dots, a_{t - 1}\}$ if and only if $b_0$ is not matched to the singletons $\{a_0, \dots, a_{s-1}\}$, since the only item outside $\{b_1,\dots,b_{t-1}\}$ that a singleton can receive is $b_0$, by no-skipping. Thus, 
\begin{align*}
\Pr_{M\sim \mathcal{M}(\sigma')}[M(b_0)\notin \{a_0, \dots, a_{s-1}\}] &= \sum_{j = s}^{t - 1}\Pr_{M\sim \mathcal{M}(\sigma')}[M(a_j)\notin\{b_1,\dots,b_{t-1}\}] \\
&= (t - s)\Pr_{M\sim \mathcal{M}(\tilde{\sigma})}[M(a_s)=b_0]. \tag{anonymity and \eqref{eq:chain}}
\end{align*}
But the left side is $\frac{t - s}{ts!}$ by the inductive hypothesis. Dividing by $t - s$, we get \eqref{eq:induct-step-2}, which concludes the proof. \qedhere

\end{proof}

\section{Discussion}

The primary focus of this paper has been to develop a deeper understanding of Random Serial Dictatorship (RSD) and the Probabilistic Serial rule (PS) through the metric distortion framework. While the distortion bounds often involved intricate technical arguments, we find it encouraging how often the results mirror our intuition about these mechanisms and what makes them fair. 

To illustrate these deep-seated intuitions, consider the housing lottery controversy at Princeton from 2019. After a suspicious number of students \href{https://www.dailyprincetonian.com/article/2019/04/students-react-to-room-draw-error}{noticed} that they had received unlucky draw numbers two years in a row, Princeton \href{https://princetonhousing.github.io/housing_email.pdf}{offered a \$1,000 housing credit} to affected rising seniors in the bottom half of their class's draw. An analysis by \href{https://medium.com/@yangsong20/princeton-room-draw-is-not-at-all-random-2368bd377ab8}{two} \href{https://princetonhousing.github.io/}{students} found that draw groups with unchanged membership got \emph{exactly} the same relative order both years, as if the draw had never been reshuffled (perhaps because the software reused the same random seed). 

While we understand intuitively why random priorities are more fair than fixed ones, and how disastrous it can be to get a low priority in RSD, the metric distortion framework makes these intuitions concrete and puts the strengths and weaknesses of these mechanisms under a microscope. Doing so allows us to develop a fine-grained understanding of their behavior, and even motivates how to make adjustments that mitigate the issues we find, and improve the mechanisms. We also find it encouraging that tweaks like \emph{Cluster Completion} feel very intuitive in hindsight. When we think of the most ``at risk'' agents (the lowest priority in RSD), it feels natural that a mechanism ought to take special care and protect them with reasonable fallback options. The fact that the metric distortion framework has helped us uncover these ideas makes us excited about what else we might learn as we work to close the gap between $\Omega(\log n)$ and $O(n^{\log_4 3})$. Besides this open problem, we briefly mention a few potential avenues for future work.

\paragraph{Empirical explorations.} Since matching is a problem that goes far beyond theoretical interest, it would be exciting to explore whether a convincing empirical case can be made for the mechanisms we proposed. One setting that seems a natural fit is assigning reviewers to papers in large publication venues. Here, a ``cluster'' may manifest as a community of researchers and the papers in their area of expertise. It could be interesting to use data from some of these venues and explore how large these clusters tend to be, how well they track with our intuition about what they represent, and whether the matchings our algorithms choose tend to give reviewers papers they prefer. 

\paragraph{Truthful mechanisms.} Now that we have shown how truthfulness can be explicitly leveraged in distortion lower bounds, we are optimistic about closing the gaps for these mechanisms. We can interpret \Cref{thm:det-po} as showing that Serial Dictatorships are the best ``reasonable'' deterministic truthful mechanisms, and we conjecture that a similar result can be proved for RSD.

\begin{conjecture}\label{conj:truth}
Every truthful-in-expectation, ex-post Pareto optimal mechanism has distortion $\tilde{\Omega}(n)$. 
\end{conjecture}

We suspect that this conjecture may be challenging to prove, since it would imply a version of \Cref{thm:rsd-lowerbound}, which was already fairly involved. It is also natural to ask what happens if we remove the requirement of Pareto optimality, but certain artificial truthful mechanisms can be difficult to contend with, and the conjecture above feels like the most natural next step.

\paragraph{Retaining axiomatic properties.} 
RSD and PS owe much of their appeal to axiomatic properties, such as truthfulness or envy-freeness, that provide meaningful safeguards in real-world deployments. 
Our modifications show how to improve their distortion guarantees, and it would be exciting to explore whether the same improvements can be obtained while retaining these axiomatic properties, possibly in an approximate sense. For example, 
\citep{Connor:2026aa} show that under RSD, an agent’s expected utility for another agent’s allocation is at most \(\sqrt{2}\) times her expected utility for her own. What tradeoffs are possible between distortion and approximate guarantees of this kind?

\subsection*{AI Methodology}

For the $O(n^{\log_4 3})$ distortion bound, the authors built on the observation in~\cite{Anari:2023aa} that truncated RSD has low cost. They then asked GPT-6 Astra to explore ways to complete the resulting partial matching, potentially by coming up with some structure that approximates an optimal matching. The AI developed this idea into the Cluster Completion algorithm and extended the procedure to fractional partial matchings following subsequent requests from the authors.

Regarding the sublinear distortion bound of $n^{\alpha}$, where $\alpha \approx 0.95$, the authors initially developed an approach based on induction over clusters that yielded a distortion bound of $O(n/\sqrt{\log n})$ where AI was only used in the calculations, unexpectedly matching the $\operatorname{RSD}$ lower bound at the time. Following this initial result, GPT-6 Astra generated the algorithm with distortion $n^{\alpha}$ without significant additional input from the authors. The current write-up substantially simplifies the original presentation and incorporates moderate modifications to the algorithm.

For the $\operatorname{RSD}$ lower bound, the authors asked the AI to strengthen the existing bound. Through several iterations,  GPT-6 Astra produced a construction yielding a distortion lower bound of $\Omega(n/\sqrt{\log n})$. The construction presented here differs slightly from the initial version and includes several simplifications; the write-up has also been substantially revised.

Our results for PS came from varying degrees of AI usage. The first result, the analysis of
TruncatedPS with the stopping time of $1 - 1/2n$ (in \Cref{ssec:tps}), was discovered without use of AI. 
With use of GPT-6 Astra, we then obtained the exponential lower bound of $\Omega( (10/9)^{n/2} )$. Using ideas
from the analysis of Truncated PS, the authors derived an upper bound of $O(2^n)$. To obtain the matching upper
bound, the authors interactively used GPT-6 Astra to obtain increasingly refined guarantees for the late stages 
of the mechanism. The current write-up improves the presentation of key ideas.

GPT-6.1 Sol and Claude Opus 5.5 were used for proofreading the manuscript and assisting with the creation of figures.

\subsection*{Acknowledgments}

JH was supported by the Simons Foundation Collaboration
on the Theory of Algorithmic Fairness and the Simons Foundation
Investigators Award~17351. YH was supported in part by
NSF grant CNS-2107078. AH was supported by NSF grant MFAI 2501597.

MC and PR thank Meenal Gupta for many discussions on the distortion of RSD. PR also thanks Max Dupré la Tour, Ndiamé Ndiaye, Abner Turkieltaub, and Adrian Vetta for early discussions on the metric distortion of PS at the Bellairs Research Institute in 2024, which included the $\Omega(n)$ lower bound via \Cref{fig:intro-example}, as well as Ariel Procaccia and Shirley Zhang for helpful conversations about matching in general. JH thanks Ndiamé Ndiaye and Agn\`es Totschnig for discussions on the axiomatic properties of various matching mechanisms, including RSD and PS.

\bibliography{mybib}
\bibliographystyle{alpha-all}

\appendix
\section{Technical Details for the RSD Lower Bound}\label{app:rsd-details}

\subsection{Integer Total Weight}\label{app:rsd-integral-weight}

We show how to make the total harmonic weight integral while preserving the estimates used in \Cref{sec:rsd}. Fix the parameters $\theta,\beta,W,a,\delta,T$ as in that section.

Write
\[
 S_M:=\theta\sum_{m=0}^{M-1}\frac{1}{\theta^3+m}.
\]
The sequence $S_M$ increases to infinity and satisfies $S_{M+1}-S_M=\theta/(\theta^3+M)$. For each sufficiently large integer $D$, choose
\[
 N:=\max\{M\in\mathbb Z_{\geq1}:S_M\leq D\},
 \qquad
 \eta_N:=D-S_N.
\]
Then $N$ tends to infinity with $D$, and
\[
 0\leq\eta_N<\frac{\theta}{\theta^3+N}.
\]
Define the original and corrected weights by
\[
 \bar\lambda_i:=\frac{\theta}{\theta^3+N-i},
 \qquad
 \lambda_1:=\bar\lambda_1+\eta_N,
 \qquad
 \lambda_i:=\bar\lambda_i\quad(2\leq i\leq N).
\]
Thus
\[
 \sum_{i=1}^N\lambda_i=D\in\mathbb Z,
 \qquad
 \bar\lambda_1\leq\lambda_1<2\bar\lambda_1,
 \qquad
 \eta_N=O_\theta(N^{-1}).
\]

All masses, reference and adjusted times, cutoffs, and the replication factor below are defined from the corrected weights using the definitions in \Cref{sec:rsd}. We will establish the following bounds for sufficiently large $N$ in an unbounded sequence, then verify them in the order listed.
\begin{enumerate}[leftmargin=*]
\item \emph{Integer total and small weights.} We require
\[
 \sum_{i=1}^N\lambda_i=D\in\mathbb Z,
 \qquad
 \sum_i\lambda_i^2\leq\frac{\theta^2}{\theta^3-1}\leq\frac18,
 \qquad
 0<\theta^2\lambda_j\leq1.
\]
These are the integrality condition in \Cref{alg:rsd-integer-construction} and the squared-weight bounds used in \Cref{lem:baseline-path-probability,lemma:weight-of-candidate-set}; the last bound controls the surplus parameters in \Cref{eq:mass-pruning-thresholds}.

\item \emph{Clock and candidate-count bounds.} For every $1\leq j\leq N$, we need
\[
 \begin{aligned}
 k_j&\geq\theta\log\frac{1}{\theta^2\lambda_j},\\
 2\beta k_j+2W+1
 &\leq\left(2W+2+\frac{2\beta}{a}\right)
       (\theta^2\lambda_j)^{-a\theta}.
 \end{aligned}
\]
The first is the lower bound in \Cref{eq:harmonic-clock-weight-comparison}, used in the proof of \Cref{lem:control-removed-mass} to obtain \Cref{eq:removed-mass-clock-sum}. The second is the candidate-count estimate used in the proof of \Cref{thm:rsd-lowerbound} immediately before \Cref{eq:rsd-total-deletion-bound}.

\item \emph{Summability.} For each fixed exponent $p>1$ used below, we need
\begin{equation}
 \sum_{j=1}^N(\theta^2\lambda_j)^p
 \leq1+\frac{\theta^3}{p-1}.
 \label{eq:rsd-perturbed-power-sum}
\end{equation}
This is the integral estimate underlying \Cref{eq:harmonic-series-estimate}. We use it with $p=1+a\theta$ for the deadline adjustment and $p=4a\theta$ for the mass-loss and deletion bounds.

\item \emph{Deadline adjustment and rounding surplus.} We require
\[
 \sum_j\lambda_j\varepsilon_j\leq\log2,
 \qquad
 u_j/2\leq u_j^*\leq u_j,
\]
together with $\varepsilon_j\leq1$ and $\varepsilon_j\tau_j\geq1+2\varepsilon_j$. These are the bounds established after \Cref{eq:harmonic-clock-weight-comparison} and used in \Cref{lem:bounded-deadline-adjustment}, the derivation of \Cref{eq:adjusted-deadline-surplus}, and the proof of \Cref{lem:rsd-deficit-after-mass-pruning}.

\item \emph{Mass retention and deletions.} We need the same global estimates as in \Cref{eq:removed-last-level-mass,eq:rsd-total-deletion-bound}:
\[
 \begin{aligned}
 q u_N(B_{\mathrm{band}}-B_{\mathrm{mass}})&\leq AT,\\
 \sum_{(h,j)\in\mathcal B}\E[(V_{h,j}-U_{h,j})_+]&\leq1.
 \end{aligned}
\]
These retain $B_{\mathrm{mass}}\geq B_0/4$ and give the residual-item bound through \Cref{cor:rsd-deletion-coupling}.

\item \emph{Size and distortion.} Finally, we need the scaling in \Cref{eq:rsd-instance-size,eq:rsd-final-distortion-scaling}:
\[
 \begin{aligned}
 D&=\theta\log N+O_\theta(1),\\
 n&=\Theta(qB_0)=\Theta(e^{D/2})=\Theta(N^{\theta/2}),\\
 \frac{2^{D-1}}q&=\Theta\!\left(\frac{n}{\sqrt{\log n}}\right).
 \end{aligned}
\]
\end{enumerate}

\paragraph{1. Integer total and small weights.}
The integrality condition follows from the setup. The correction to the squared-weight sum is $2\bar\lambda_1\eta_N+\eta_N^2=O_\theta(N^{-2})$. Since
\[
 \theta^2\sum_{m=0}^{\infty}(\theta^3+m)^{-2}
 <\theta^2\int_{-1}^{\infty}(\theta^3+s)^{-2}\,ds
 =\frac{\theta^2}{\theta^3-1},
\]
the strict slack in this bound absorbs the correction for sufficiently large $N$, giving the same estimate
\[
 \sum_i\lambda_i^2\leq\frac{\theta^2}{\theta^3-1}
 \leq\frac2\theta\leq\frac18.
\]
Also, $\lambda_1<2\theta/(\theta^3+N-1)\leq\theta^{-2}$ for sufficiently large $N$, so $0<\theta^2\lambda_j\leq1$ for every $j$. The baseline path-probability and band-pruning lemmas therefore apply to the corrected weights.

\paragraph{2. Clock and candidate-count bounds.}
Every suffix sum is unchanged:
\[
 k_j=\sum_{i>j}\lambda_i=\sum_{i>j}\bar\lambda_i,
 \qquad 1\leq j\leq N.
\]
For $j\geq2$, \Cref{eq:harmonic-clock-weight-comparison} therefore holds exactly as stated. For $j=1$, the inequality $1\leq\lambda_1/\bar\lambda_1<2$ instead gives
\[
 \theta\log\frac{1}{\theta^2\lambda_1}
 \leq k_1
 \leq\theta\log\frac{1}{\theta^2\lambda_1}
       +\theta\log2+\theta^{-2}.
\]
The lower clock bound is thus preserved for every $j$, and in particular
\[
 e^{-\alpha k_j}\leq(\theta^2\lambda_j)^{\alpha\theta}.
\]
For the candidate-count bound, set $y_j:=(\theta^2\lambda_j)^{-1}\geq1$. For $j\geq2$, the upper bound in \Cref{eq:harmonic-clock-weight-comparison} gives
\[
 \begin{aligned}
 2\beta k_j+2W+1
 &\leq2\beta\theta\log y_j+2\beta\theta^{-2}+2W+1\\
 &\leq\frac{2\beta}{a}y_j^{a\theta}+2W+2\\
 &\leq\left(2W+2+\frac{2\beta}{a}\right)y_j^{a\theta}.
 \end{aligned}
\]
Here we used $\log y\leq y^{a\theta}/(a\theta)$ for $y\geq1$ and $2\beta\theta^{-2}\leq1$.

For $j=1$, we use $k_1\leq\theta\log y_1+\theta\log2+\theta^{-2}$. Since $\lambda_1<2\bar\lambda_1$, we have
\[
 y_1\geq\frac{\theta^3+N-1}{2\theta^3}.
\]
Choose $N$ sufficiently large that
\[
 \left(\frac{\theta^3+N-1}{2\theta^3}\right)^{a\theta}
 \geq1+2\beta\theta\log2,
\]
which is possible because $a\theta>0$. Then $y_1^{a\theta}\geq1+2\beta\theta\log2$, and
\[
 \begin{aligned}
 2\beta k_1+2W+1
 &\leq2\beta\theta\log y_1+2\beta\theta\log2
       +2\beta\theta^{-2}+2W+1\\
 &\leq\frac{2\beta}{a}y_1^{a\theta}+2W+2+2\beta\theta\log2\\
 &\leq\left(2W+2+\frac{2\beta}{a}\right)y_1^{a\theta}.
 \end{aligned}
\]
The last step uses $2W+2\geq1$. Thus, for every $1\leq j\leq N$, the number of candidate locations in column $j$ is at most
\[
 2\beta k_j+2W+1
 \leq\left(2W+2+\frac{2\beta}{a}\right)
       (\theta^2\lambda_j)^{-a\theta}.
\]
Combining this count with the local deficit estimate in \Cref{eq:rsd-deficit-after-mass-pruning} gives the counting step in \Cref{eq:rsd-total-deletion-bound}, in the proof of \Cref{thm:rsd-lowerbound}. Separately, the lower clock bound $e^{-\alpha k_j}\leq(\theta^2\lambda_j)^{\alpha\theta}$ gives the second inequality in \Cref{eq:removed-mass-clock-sum}, in the proof of \Cref{lem:control-removed-mass}.

\paragraph{3. Summability.}
For any fixed $p>1$,
\[
 \sum_{j=1}^N(\theta^2\lambda_j)^p
 =\sum_{m=0}^{N-1}(1+m/\theta^3)^{-p}
   +O_{\theta,p}(N^{-p}).
\]
The right-hand side converges to a value strictly smaller than
\[
 1+\int_0^\infty(1+s/\theta^3)^{-p}\,ds
 =1+\frac{\theta^3}{p-1}.
\]
Consequently, \Cref{eq:rsd-perturbed-power-sum} holds for sufficiently large $N$, with the same constant as the harmonic-series estimate in \Cref{eq:harmonic-series-estimate}.

\paragraph{4. Deadline adjustment and rounding surplus.}
Using \Cref{eq:rsd-perturbed-power-sum} with $p=1+a\theta$, the definitions in \Cref{eq:mass-pruning-thresholds} give
\[
 \begin{aligned}
 \sum_j\lambda_j\varepsilon_j
 &=\frac{\delta}{\theta^2}
   \sum_j(\theta^2\lambda_j)^{1+a\theta}\\
 &\leq\delta(\theta^{-2}+a^{-1})\leq\log2.
 \end{aligned}
\]
Hence $u_j/2\leq u_j^*\leq u_j$ still holds. Since $\theta^2\lambda_j\leq1$, we also retain $\varepsilon_j\leq\delta\leq1$ and $\varepsilon_j\tau_j\geq\delta T\geq3\geq1+2\varepsilon_j$. Thus the rounding surplus and local deficit estimates remain valid with the same constants.

\paragraph{5. Mass retention and deletions.}
Since $\alpha-4a>4a$ and $\theta^2\lambda_j\leq1$, the first-discarded-vertex argument in the proof of \Cref{lem:control-removed-mass}, using the lower clock bound from item~2 and \Cref{eq:rsd-perturbed-power-sum} with $p=4a\theta$, gives
\[
 \begin{aligned}
 q u_N(B_{\mathrm{band}}-B_{\mathrm{mass}})
 &\leq2^WT\sum_{j=1}^N
       (\theta^2\lambda_j)^{(\alpha-4a)\theta}\\
 &\leq2^WT\left(1+\frac{\theta^3}{4a\theta-1}\right)=AT.
 \end{aligned}
\]
This reproduces \Cref{eq:removed-last-level-mass}, so the choice of $q$ in \Cref{eq:mass-pruning-replication} still ensures $B_{\mathrm{mass}}\geq B_0/4$. Similarly, \Cref{lem:rsd-deficit-after-mass-pruning}, the candidate-count bound from item~2, and \Cref{eq:rsd-perturbed-power-sum} reproduce \Cref{eq:rsd-total-deletion-bound}:
\[
 \begin{aligned}
 \sum_{(h,j)\in\mathcal B}\E[(V_{h,j}-U_{h,j})_+]
 &\leq\frac2\delta\left(2W+2+\frac{2\beta}{a}\right)
       \left(1+\frac{\theta^3}{4a\theta-1}\right)
       e^{-\delta^2T/12}\\
 &\leq1.
 \end{aligned}
\]
Thus the coupling bound and the late-population estimate in \Cref{eq:rsd-final-late-population} establish the same successful cascade for the corrected instance.

\paragraph{6. Size and distortion.}
The perturbation changes the total harmonic weight by $\eta_N=O_\theta(N^{-1})$, so
\[
 D=S_N+\eta_N
  =\theta\log(1+N/\theta^3)+O_\theta(1)
  =\theta\log N+O_\theta(1).
\]
The estimates in \Cref{lem:baseline-path-probability,lem:bounded-deadline-adjustment,lemma:size-count-rsd-lowerbound} therefore yield
\[
 \begin{aligned}
 n&=\Theta(qB_0)=\Theta(e^{D/2})=\Theta(N^{\theta/2}),\\
 \frac{2^{D-1}}q
 &=\Theta\!\left(\frac{e^{D/2}}{\sqrt D}\right)
  =\Theta\!\left(\frac{n}{\sqrt{\log n}}\right).
 \end{aligned}
\]
This gives the claimed distortion lower bound for arbitrarily large instances with an integral number of levels, using the same fixed parameters.

\subsection{Path Concentration}\label{app:path-concentration}

We use the following second-moment consequence of Wellner's inequality, stated without proof.

\begin{theorem}[Wellner's inequality, specialized form {\citep[Theorem~1]{Wellner:1977aa}}]
\label{thm:wellner-weighted-empirical-process}
Let $D\geq1$ be an integer, and let $F(s)$ count how many of $D$ independent uniform points in $[0,D]$ lie in $[0,s]$. Suppose $w:[0,D]\to(0,\infty)$ is continuous, satisfies $w(s)=w(D-s)$, and is nondecreasing on $[0,D/2]$. Then
\[
 \Pr\bigl(\exists s\in[0,D/2]:|F(s)-s|>w(s)\bigr)
 \leq16\int_0^{D/2}\frac{ds}{w(s)^2}.
\]
\end{theorem}

\pathconcentrationlemma*

\begin{proof}[Proof of \Cref{lemma:path-concentration}]
Apply \Cref{thm:wellner-weighted-empirical-process} with
\[
 w(s)=\beta\min\{s,D-s\}+W,
\]
which is positive, continuous, symmetric, and nondecreasing on $[0,D/2]$. The probability of leaving the band on the lower half is at most
\[
 16\int_0^{D/2}\frac{ds}{(\beta s+W)^2}
 \leq\frac{16}{\beta W}.
\]
Reflecting the uniform points about $D/2$ gives the same bound on the upper half. A union bound therefore yields
\[
 \Pr\bigl(\exists s\in[0,D]:
 |F(s)-s|>\beta\min\{s,D-s\}+W\bigr)
 \leq\frac{32}{\beta W}\leq\frac14,
\]
where the last inequality follows from \Cref{eq:lower-bound-parameter-W}. This proves \Cref{eq:uniform-counting-band-inside-lemma}.

\end{proof}

\section{Parameter Analysis for Coordinated Merging}\label{app:single-layer-parameter-inequality}

\criticalparameterlemma*

\begin{proof}[Proof of \Cref{lemma:inequality-of-single-layer-critical-exponent}]
For $0<a<1$, define
\[
x_a=\left(\frac29\right)^{1/a},
\qquad
h(a)=(1-x_a)^{-a}+a(1-x_a)-2.
\]
The function $x_a$ increases with $a$, and $(1-x)^{-a}+a(1-x)-2$ increases with both $a$ and $x$. Thus $h$ is continuous and strictly increasing. Since $h(0.95)<0<h(0.951)$, its unique zero is $\alpha_\star$. The definitions of $x_\star$ and $\beta_\star$ then give
\begin{equation}\label{eq:single-layer-parameter-bounds}
0.95<\alpha_\star<0.951,\qquad
0.205<x_\star<0.206,\qquad
0.032<\beta_\star<0.033.
\end{equation}

Write $a=\alpha_\star$, $b=\beta_\star$, and $c=9b/2$. In particular,
\[
a>\frac{19}{20},\qquad b<\frac1{30},\qquad c<\frac3{20}.
\]
Fix $k\geq2$ and $0<x<1$, and set
\[
t=\frac1k,\qquad u=1-x,\qquad
f=u^{-a},\qquad R=x^{-a},\qquad v=(1-t)^{-a}.
\]
Clearing the positive denominator in \Cref{eq:single-layer-parameter-inequality}, it suffices to prove $P_t(x)\geq0$, where
\[
P_t(x)=(1-ct)f
+u\left(\frac{v-1}{t}-cv\right)-2+b(u+2t)R.
\]

We compare $P_t$ with the limiting expression
\[
\Phi(x)\coloneqq(1-x)^{-a}+a(1-x)-2
+b(1-x)\left(x^{-a}-\frac92\right).
\]
The critical parameter definitions give $\Phi(x_\star)=\Phi'(x_\star)=0$, while
\[
\Phi''(x)
=a(a+1)(1-x)^{-a-2}
+ba x^{-a-2}\bigl(2x+(a+1)(1-x)\bigr)>0.
\]
Hence $\Phi$ is minimized at $x_\star$, and $\Phi(x)\geq0$ throughout $(0,1)$. We now handle the correction for finite $k$ in two cases.

If $x\leq1/2$, put $\lambda=a(a+1)/2$. The second derivative bound for $(1-t)^{-a}$ gives $v\geq1+at+\lambda t^2$. Since $1/t-c>0$,
\begin{align*}
P_t(x)-\Phi(x)
&\geq t\left[-cf+u(\lambda-ca-c\lambda t)+2bR\right]\\
&>t\left[-\frac3{10}+\frac{27}{80}\right]
=\frac3{80}t>0.
\end{align*}
For the second line, we used $f<2$, $u\geq1/2$, and
\[
\lambda-ca-c\lambda t
>\frac9{10}-\frac3{20}-\frac3{40}
=\frac{27}{40},
\]
which follows from $9/10<\lambda<1$, $a<1$, and $t\leq1/2$. Thus $P_t(x)\geq\Phi(x)\geq0$.

If $x\geq1/2$, the simpler bound $v\geq1+at$ gives
\[
P_t(x)\geq\Psi_t(x)\coloneqq
(1-ct)f+u\bigl(a-c(1+at)\bigr)-2+b(u+2t)R.
\]
For fixed $x$, $\Psi_t(x)$ is affine in $t$, with $\Psi_0(x)=\Phi(x)\geq0$. It therefore suffices to check $\Psi_{1/2}(x)\geq0$. As a function of $x$, $\Psi_{1/2}$ is strictly convex. At $x=1/2$, the parameter bounds above and $19/10<2^a<2$ give
\begin{align*}
\Psi_{1/2}(1/2)
&=\left(1-\frac{3b}{4}\right)2^a+\frac a2-2
-\frac c2\left(1+\frac a2\right)>\frac15,\\
\Psi_{1/2}'(1/2)
&=a2^{a+1}\left(1-\frac{15b}{4}\right)-a+c\left(1+\frac a2\right)-b2^a
>\frac12.
\end{align*}
Convexity now gives $\Psi_{1/2}(x)>0$ for $x\geq1/2$. Affinity in $t$ then gives $\Psi_t(x)\geq0$ for $0\leq t\leq1/2$, completing the proof of \Cref{eq:single-layer-parameter-inequality}. Its left-hand side has endpoint limits $-\infty$ as $x\downarrow0$ and $9b/(2k)<1$ as $x\uparrow1$.

Finally, suppose the same inequalities hold for some exponent $a\in(0,1)$ and fixed $b>0$, even if required only on the grids $x=s/k$. Letting $k\to\infty$ along a sequence with $s/k\to x_a$, the limiting inequality gives $h(a)\geq0$, because $x_a^{-a}=9/2$ makes the coefficient of $b$ vanish. Since $h$ is strictly increasing, $a\geq\alpha_\star$.
\end{proof}

\newcommand{\Anew}{A_j}
\newcommand{\lev}{\mathrm{lev}}
\newcommand{\Lev}[1]{L_{#1}}
\newcommand{\help}{\eta}
\newcommand{\debt}{\beta}
\newcommand{\settle}[1]{\hat u_{#1}}
\newcommand{\ncomer}{\nu}

\section{Displacement Inequality for PS Upper Bound}\label{sec:displacement}
In this section, we prove \Cref{lem:displacement}, which is the most technical step of our analysis.
Throughout, we fix an epoch $[u_{j-1},u_j)$ and write $R=R_{u_{j-1}}$ for the effective radius, which does not change during the epoch.
We also let $\Anew=\Afree(u_j)\setminus\Afree(u_{j-1})$ denote the set of agents freed during the epoch, so that $|\Anew|=\Delta_j$, and recall that every $a\in\Anew$ is freed at time $u_a$.
We need to show that
\begin{equation}\label{eq:displacement-goal}
 \sum_{a\in\Anew}R(a)\,(1-u_a)\le\frac{\kappa}{2}\left(\lambda^{\Delta_j}-1\right)\budget(u_{j-1}).
\end{equation}

Once freed, a resident behaves like any other free agent, and helps free further residents, whose effective radius may be twice her own by \eqref{eq:doubling}.
Counted naively, this doubles our bound with every generation of freed agents, much as in \Cref{cor:exp-bound}; to avoid this, we group agents according to their radius, and keep track of on whose behalf every unit is eaten.
In \Cref{sec:disp-sources}, we make this precise; this lets us split the time left of every freed agent into her own time and her debt.
We bound the debts in \Cref{sec:disp-solvency} and the own time in \Cref{sec:disp-growth}, and we combine the two bounds in \Cref{sec:disp-proof}.

\subsection{Levels and Sources}\label{sec:disp-sources}
We first group agents by the scale of their effective radius.

\begin{definition}[Levels]\label{def:levels}
For every point $x$, let $\lev(x)=\lceil\log_2R(x)\rceil$ be its \emph{level}; so $2^{\lev(x)-1}<R(x)\le2^{\lev(x)}$.
For an integer $\ell$, we let $\Lev\ell=\{a\in\Anew:\lev(a)=\ell\}$.
\end{definition}

Whenever a free agent eats a block $b$, every resident of $b$ receives \emph{help} from her at rate $1/N_b$; so, counting from the checkpoint, a resident of $b$ has received $V_b(t)/N_b$ units of help by time $t$.

If $a\in\Anew$ is a resident of $b$, then $b$ is exhausted at time $u_a$, and \eqref{eq:block-capacity} gives $V_b(u_a)=N_b(1-u_a)$.
So every $a\in\Anew$ receives exactly $1-u_a$ units of help, all of it by time $u_a$.
Not all of this help drives the cascade, however.
We attribute every unit eaten to a source, as follows.

\begin{definition}[Sources]\label{def:sources}
Every unit that a free agent eats, from the checkpoint until the end of the epoch, has a \emph{source}, which is either the budget or a level.
\begin{enumerate}[label=(\roman*),noitemsep]
 \item Everything that agents of $\Afree(u_{j-1})$ eat has its source at the budget.
 \item Let $a\in\Anew$. For a level $\ell$, let $\help_a^\ell(t)$ be the help with source $\ell$ that $a$ receives by time $t$, and let $\help_a^\budget(t)$ be the help that she receives with source the budget; write $\help_a^\ell=\help_a^\ell(u_a)$ and $\help_a^\budget=\help_a^\budget(u_a)$.
 The \emph{debt} of $a$ is $\debt_a=\sum_{\ell\ge\lev(a)}\help_a^\ell$, and her \emph{settling time} is $\settle a=u_a+\debt_a$.
 During $[u_a,\settle a)$, she passes on her debt: what she eats has source $\ell$ at rate $\help_a^\ell/\debt_a$, for every level $\ell\ge\lev(a)$.
 From $\settle a$ on, she eats on her own behalf: what she eats has source $\lev(a)$.
\end{enumerate}
\end{definition}

What an agent $a\in\Anew$ eats after her release depends only on the help that she received before it, so the definition determines all sources by induction over the releases, and all rates are piecewise constant.
Note that sources start afresh in every epoch, since the agents free at its start eat with source the budget, and that $u_a\le\settle a\le1$, since $\debt_a\le1-u_a$.

\begin{claim}\label{clm:sources}
The following properties hold.
\begin{enumerate}[label=(P\arabic*),start=10,leftmargin=*,noitemsep]
 \item\label{P:levels} If a free agent $p$ eats from the block of a resident $a$ during the epoch, then $\lev(a)\le\lev(p)+1$.
 \item\label{P:own-level} Everything that an agent $p\in\Anew$ eats has its source at a level of at least $\lev(p)$.
 \item\label{P:kept} Every $a\in\Anew$ satisfies $\help_a^\ell=0$ for all levels $\ell<\lev(a)-1$, and
 \[
  1-\settle a=\help_a^\budget+\help_a^{\lev(a)-1}.
 \]
\end{enumerate}
\end{claim}
\begin{proof}
We prove each property separately.
For \ref{P:levels}, \eqref{eq:doubling} gives $R(a)\le2R(p)$, so $\log_2R(a)\le\log_2R(p)+1$, and we take ceilings.
Property \ref{P:own-level} is immediate from \Cref{def:sources}, as the sources of a debt of $p$ are levels of at least $\lev(p)$.
For \ref{P:kept}, help with a level $\ell$ as its source comes from some $p\in\Anew$, since agents of $\Afree(u_{j-1})$ give help with source the budget, and $\ell\ge\lev(p)\ge\lev(a)-1$ by \ref{P:own-level} and \ref{P:levels}.
So the $1-u_a$ units of help that $a$ receives consist of $\help_a^\budget$, $\help_a^{\lev(a)-1}$ and $\debt_a$, and $1-\settle a=(1-u_a)-\debt_a$.
\end{proof}

For a level $\ell$, we let $E_\ell=\sum_{a\in\Lev\ell}\help_a^\budget$ be the help that agents of level $\ell$ receive from the budget, and we let $\widehat E=\sum_\ell2^{\ell-1}E_\ell$.
Below, we show that this input is paid for by the budget.

\begin{claim}\label{clm:input}
The input satisfies $\widehat E\le\budget(u_{j-1})$.
\end{claim}
\begin{proof}
The help with source the budget consists of what visitors ate before $u_{j-1}$, and of what agents of $\Afree(u_{j-1})$ eat during the epoch.
Every $a\in\Anew$ is a resident of a block $b\in\Bst(u_{j-1})$, and receives $V_b(u_{j-1})/N_b$ units of the former; so the former contributes at most
\[
 \sum_{b\in\Bst(u_{j-1})}\frac{V_b(u_{j-1})}{N_b}\sum_{a\text{ resident of }b}2^{\lev(a)-1}\le\sum_{b\in\Bst(u_{j-1})}\frac{V_b(u_{j-1})}{N_b}\sum_{a\text{ resident of }b}R(a).
\]
An agent $p\in\Afree(u_{j-1})$ eats at most $1-u_{j-1}$ units during the epoch, and by \ref{P:levels}, each of them helps only residents of levels $\ell\le\lev(p)+1$, for whom $2^{\ell-1}\le2^{\lev(p)}<2R(p)$; so the latter contributes at most $2(1-u_{j-1})\sum_{p\in\Afree(u_{j-1})}R(p)$.
By \eqref{eq:budget}, the two contributions add up to at most $\budget(u_{j-1})$.
\end{proof}

\subsection{Debts and Solvency}\label{sec:disp-solvency}
By the definition of settling times, the time left of every $a\in\Anew$ splits as
\[
 1-u_a=\left(1-\settle a\right)+\debt_a
\]
into her own time after settling, and her debt.
We bound the two parts separately, starting with the debts.
For a time $t$, let $\Anew(t)=\{a\in\Anew:\settle a<t\}$ denote the set of agents that settle before $t$; note that $\Anew(t)=\Anew$ for $t>1$.
In this subsection, we show that debts are paid for by agents who settled earlier.

\begin{lemma}[Debts are small]\label{lem:debts}
Every $a\in\Anew$ satisfies
\[
 2^{\lev(a)}\debt_a\le\sum_{g\in\Anew(u_a)}2^{\lev(g)}\left(1-\settle g\right).
\]
\end{lemma}

Both this lemma and the inequalities that we need in \Cref{sec:disp-growth} follow from the next lemma.
By \ref{P:kept}, its left-hand side is the help with source $\ell$ that agents keep, together with the help with source $\ell$ that is still to be passed on at time $t$; the lemma bounds both by the time that agents of level $\ell$ have spent eating on their own behalf.

\begin{lemma}[Solvency]\label{lem:solvency}
For every level $\ell$ and every $t\in[u_{j-1},u_j]$,
\[
 \sum_{a\in\Lev{\ell+1}}\help_a^\ell(t)+\sum_{a\in\Anew:\,\lev(a)\le\ell,\ u_a\ge t}\help_a^\ell(t)\le\sum_{g\in\Lev\ell}\left(t-\settle g\right)_+.
\]
\end{lemma}
\begin{proof}
Let $C_\ell(t)$ be the amount eaten during $[u_{j-1},t]$ with source $\ell$.
Only agents of $\Anew$ eat with a level as their source, and an agent $g\in\Lev\ell$ eats on her own behalf only from $\settle g$ on, for at most $(t-\settle g)_+$ units of time by time $t$.
Any other eating with source $\ell$ is done by an agent $a\in\Anew$ with $\lev(a)\le\ell$ who passes on her debt during $[u_a,\settle a)$, at rate $\help_a^\ell/\debt_a$; this amounts to at most $\help_a^\ell$, and to nothing by time $t$ unless $u_a<t$.
Hence
\[
 C_\ell(t)\le\sum_{g\in\Lev\ell}\left(t-\settle g\right)_++\sum_{a\in\Anew:\,\lev(a)\le\ell,\ u_a<t}\help_a^\ell.
\]
On the other hand, everything that free agents eat from a block is help for its residents, and the residents of the blocks exhausted during the epoch are the agents of $\Anew$.
By \ref{P:kept}, help with source $\ell$ goes to agents of level $\ell+1$ or of level at most $\ell$, and an agent freed before $t$ receives all of her help by then, so
\[
 C_\ell(t)\ge\sum_{a\in\Lev{\ell+1}}\help_a^\ell(t)+\sum_{a\in\Anew:\,\lev(a)\le\ell,\ u_a<t}\help_a^\ell+\sum_{a\in\Anew:\,\lev(a)\le\ell,\ u_a\ge t}\help_a^\ell(t).
\]
Comparing the two bounds proves the lemma.
\end{proof}

\begin{proof}[Proof of \Cref{lem:debts}]
For every level $\ell\ge\lev(a)$, \Cref{lem:solvency} at time $t=u_a$ gives $\help_a^\ell\le\sum_{g\in\Lev\ell}(u_a-\settle g)_+$, as $a$ receives all of her help by time $u_a$.
Multiplying by $2^{\lev(a)}\le2^\ell$ and summing over $\ell\ge\lev(a)$ gives
\[
 2^{\lev(a)}\debt_a\le\sum_{\ell\ge\lev(a)}2^\ell\sum_{g\in\Lev\ell}\left(u_a-\settle g\right)_+\le\sum_{g\in\Anew(u_a)}2^{\lev(g)}\left(1-\settle g\right).\qedhere
\]
\end{proof}

Next, we show that the help that agents of level $\ell$ keep from level $\ell-1$ is bounded by the time that agents of level $\ell-1$ spend on their own behalf, as measured by settling times.

\begin{lemma}[Settling inequalities]\label{lem:settling}
Let $t$ be a time and $\ell$ a level, and list the agents of $\Lev\ell\cap\Anew(t)$ as $a_1,a_2,\dots$ so that $\settle{a_1}\le\settle{a_2}\le\cdots$.
Then every $k\le|\Lev\ell\cap\Anew(t)|$ satisfies
\begin{equation}\label{eq:settling}
 \sum_{i\le k}\left(1-\settle{a_i}\right)\le\sum_{g\in\Lev{\ell-1}\cap\Anew(t)}\left(\settle{a_k}-\settle g\right)_++E_\ell.
\end{equation}
\end{lemma}
\begin{proof}
By \ref{P:kept}, the left-hand side of \eqref{eq:settling} is $\sum_{i\le k}(\help_{a_i}^\budget+\help_{a_i}^{\ell-1})$, and the first part of this sum is at most $E_\ell$.
The agents $a_1,\dots,a_k$ receive all of their help by time $t^*=\max_{i\le k}u_{a_i}$, which satisfies $u_{j-1}<t^*\le u_j$ and $t^*\le\settle{a_k}$.
So \Cref{lem:solvency} for level $\ell-1$ at time $t^*$ gives
\[
 \sum_{i\le k}\help_{a_i}^{\ell-1}\le\sum_{a\in\Lev\ell}\help_a^{\ell-1}(t^*)\le\sum_{g\in\Lev{\ell-1}}\left(t^*-\settle g\right)_+\le\sum_{g\in\Lev{\ell-1}\cap\Anew(t)}\left(\settle{a_k}-\settle g\right)_+,
\]
where the last step holds because every $g\in\Lev{\ell-1}\setminus\Anew(t)$ satisfies $\settle g\ge t>\settle{a_k}\ge t^*$.
\end{proof}

\subsection{Growth per Agent}\label{sec:disp-growth}
We now bound the own time.
In this subsection, we show that the own time of the agents that settle before any given time grows by a factor of at most $\lambda$ per agent.

\begin{proposition}[Growth per agent]\label{prop:growth}
Every time $t$ satisfies
\[
 \sum_{a\in\Anew(t)}2^{\lev(a)}\left(1-\settle a\right)\le8(10+9\lambda)\left(\lambda^{|\Anew(t)|}-1\right)\widehat E.
\]
\end{proposition}

Throughout, we fix a time $t$, and we write $S=\Anew(t)$, as well as $S_\ell=\Lev\ell\cap S$ and $m_\ell=|S_\ell|$ for every level $\ell$.
We only use \Cref{lem:settling}, and nothing else about the run.
For a set $T\subseteq\Anew$ whose agents, listed so that $\settle{p_1}\le\settle{p_2}\le\cdots$, are $p_1,\dots,p_r$, we let
\[
 \ncomer(T)=\max_{1\le k\le r}\frac1{k+1}\sum_{i\le k}\left(1-\settle{p_i}\right),
\]
and $\ncomer(\varnothing)=0$.
Intuitively, $\ncomer(T)$ is the most time left with which agents of $T$ alone could free a \emph{newcomer} who is the sole resident of her block and is helped by nobody else: if the first $k$ agents of $T$ to settle help a newcomer from their settling times on, and she is freed with $x$ units of time left, then she has received $\sum_{i\le k}((1-\settle{p_i})-x)$ units of help, which must equal $x$, so that $x=\frac1{k+1}\sum_{i\le k}(1-\settle{p_i})$.

\begin{proof}[Proof of \Cref{prop:growth}]
We first compare consecutive levels.
Fix a level $\ell$, let $m=m_{\ell-1}$ be the number of agents at the level below, and list $S_{\ell-1}$ as $p_1,\dots,p_m$ and $S_\ell$ as $a_1,a_2,\dots$, both by settling time.
For every $k$, the agents of $S_{\ell-1}$ that settle before $a_k$ are its first $j$ agents, for some $j\le m$; so, by the definition of $\ncomer$,
\begin{align*}
 \sum_{p\in S_{\ell-1}}\left(\settle{a_k}-\settle p\right)_+
 &=\sum_{i\le j}\left(1-\settle{p_i}\right)-j\left(1-\settle{a_k}\right)\\
 &\le(j+1)\,\ncomer(S_{\ell-1})-j\left(1-\settle{a_k}\right)\\
 &\le\ncomer(S_{\ell-1})+m\left(\ncomer(S_{\ell-1})-\left(1-\settle{a_k}\right)\right)_+.
\end{align*}
\Cref{lem:settling} therefore gives, for every $k\ge1$,
\begin{equation}\label{eq:level-step}
 \sum_{i\le k}\left(1-\settle{a_i}\right)\le E_\ell+\ncomer(S_{\ell-1})+m\left(\ncomer(S_{\ell-1})-\left(1-\settle{a_k}\right)\right)_+.
\end{equation}
We now show by induction on $k$ that
\begin{equation}\label{eq:level-bound}
 \sum_{i\le k}\left(1-\settle{a_i}\right)\le(m+1)\Bigl(1-\Bigl(\frac{m}{m+1}\Bigr)^k\Bigr)\ncomer(S_{\ell-1})+E_\ell.
\end{equation}
This holds for $k=0$, as the empty sum vanishes.
For $k\ge1$, if the positive part in \eqref{eq:level-step} vanishes, then the left-hand side is at most $E_\ell+\ncomer(S_{\ell-1})$, which implies \eqref{eq:level-bound}, as the factor of $\ncomer(S_{\ell-1})$ there is at least $(m+1)(1-\frac{m}{m+1})=1$.
Otherwise, moving $m(1-\settle{a_k})$ to the left-hand side of \eqref{eq:level-step} gives
\[
 (m+1)\sum_{i\le k}\left(1-\settle{a_i}\right)-m\sum_{i<k}\left(1-\settle{a_i}\right)\le E_\ell+(m+1)\,\ncomer(S_{\ell-1}).
\]
Dividing by $m+1$ and applying \eqref{eq:level-bound} to the sum over $i<k$, we can therefore derive that
\begin{align*}
 \sum_{i\le k}\left(1-\settle{a_i}\right)
 &\le\ncomer(S_{\ell-1})+\frac{E_\ell+m\sum_{i<k}\left(1-\settle{a_i}\right)}{m+1}\\
 &\le\ncomer(S_{\ell-1})+m\Bigl(1-\Bigl(\frac{m}{m+1}\Bigr)^{k-1}\Bigr)\ncomer(S_{\ell-1})+E_\ell\\
 &=(m+1)\Bigl(1-\Bigl(\frac{m}{m+1}\Bigr)^k\Bigr)\ncomer(S_{\ell-1})+E_\ell.
\end{align*}
In the following claim, we establish a convenient simplification involving $\lambda=\sqrt{10/9}$.
\begin{claim}\label{clm:two-for-two}
For all integers $m,k\ge1$,
\[
 2(m+1)\Bigl(1-\Bigl(\frac{m}{m+1}\Bigr)^k\Bigr)\le\lambda^m(k+1),
\]
with equality only for $m=k=2$.
\end{claim}
\begin{proof}
Write $q=m/(m+1)$. Since $1/(1-q)=m+1$, we have $(m+1)(1-q^k)=\sum_{i<k}q^i$.
If $m=1$, then $q=\frac12$, and every term with $i\ge1$ is at most $\frac12$, so the left-hand side is at most $2\bigl(1+\frac{k-1}{2}\bigr)=k+1<\lambda(k+1)$.

Now let $m\ge2$. By Bernoulli's inequality, $\lambda^m=\bigl(1+\frac19\bigr)^{m/2}\ge1+\frac{m}{18}$, strictly if $m>2$.
It therefore suffices to show that
\begin{equation}\label{eq:affine}
 2(m+1)\left(1-q^k\right)\le\Bigl(1+\frac{m}{18}\Bigr)(k+1),
\end{equation}
with equality only for $k=2$ and $m\in\{2,3\}$: equality in the claim then requires equality in both steps, that is, $m=k=2$, where both indeed hold.
Multiplying by $18$ shows that \eqref{eq:affine} is equivalent to $g(k)\ge0$, where
\[
 g(k)=(m+18)(k+1)-36(m+1)\left(1-q^k\right).
\]
A direct computation gives
\begin{align*}
 g(1)&=2m,\\
 g(2)&=\frac{3(m-2)(m-3)}{m+1},\\
 g(3)&=\frac{4\bigl[(m-2)(m-3)(m-4)+2(m-4)^2+1\bigr]}{(m+1)^2}.
\end{align*}
For an integer $m\ge2$, the products $(m-2)(m-3)$ and $(m-2)(m-3)(m-4)$ are nonnegative, since each of them has a zero factor or only positive factors.
So $g(1)>0$ and $g(3)>0$, and $g(2)\ge0$, with equality exactly when $m\in\{2,3\}$.
For $k\ge1$, using $1-q=1/(m+1)$, we further have
\begin{align*}
 g(k+1)-q\,g(k)
 &=(m+18)\bigl((k+2)-q(k+1)\bigr)-36(m+1)(1-q)\\
 &=\frac{(m+18)(k+m+2)}{m+1}-36\\
 &=\frac{(m-6)(m-7)+12+(k-3)(m+18)}{m+1},
\end{align*}
which is positive for $k\ge3$, as $(m-6)(m-7)\ge0$ for every integer $m$.
Since $q>0$, induction on $k$ from $g(3)>0$ gives $g(k)>0$ for every $k\ge3$.
Hence $g(k)\ge0$ for every $k$, with equality only for $k=2$ and $m\in\{2,3\}$, which proves \eqref{eq:affine}.
\end{proof}

By \eqref{eq:level-bound} and \Cref{clm:two-for-two}, every $k$ satisfies
\[
 \sum_{i\le k}\left(1-\settle{a_i}\right)\le\tfrac12\lambda^m(k+1)\,\ncomer(S_{\ell-1})+E_\ell;
\]
for $m=0$, this is immediate, as then $\ncomer(S_{\ell-1})=0$.
Dividing by $k+1\ge2$ and taking the maximum over $k$, the definition of $\ncomer$ gives
\begin{equation}\label{eq:one-level}
 \ncomer(S_\ell)\le\tfrac12\lambda^{m_{\ell-1}}\,\ncomer(S_{\ell-1})+\tfrac12E_\ell\qquad\text{for every level }\ell.
\end{equation}

We now sum over the levels.
For every level $\ell$, let $m_{<\ell}=\sum_{\ell'<\ell}m_{\ell'}$.
Multiplying \eqref{eq:one-level} by $2^\ell$ gives
\[
 2^\ell\ncomer(S_\ell)\le\lambda^{m_{\ell-1}}\,2^{\ell-1}\ncomer(S_{\ell-1})+2^{\ell-1}E_\ell.
\]
Since $S$ is finite, $\ncomer(S_\ell)=0$ for all sufficiently small $\ell$, so chaining this over the levels gives
\[
 2^\ell\ncomer(S_\ell)\le\lambda^{m_{<\ell}}\sum_{\ell'\le\ell}2^{\ell'-1}E_{\ell'}\le\lambda^{m_{<\ell}}\,\widehat E.
\]
By the term $k=m_\ell$ in the definition of $\ncomer$, every level $\ell$ with $m_\ell\ge1$ then satisfies
\[
 \sum_{a\in S_\ell}2^\ell\left(1-\settle a\right)\le(m_\ell+1)\,2^\ell\ncomer(S_\ell)\le(m_\ell+1)\,\lambda^{m_{<\ell}}\,\widehat E\le8\lambda^{m_{<\ell}+m_\ell}\,\widehat E,
\]
as $x+1\le8\lambda^x$ for every integer $x\ge1$.\footnote{The ratio of consecutive values of $(x+1)\lambda^{-x}$ is $(x+2)/((x+1)\lambda)$, which is at least $1$ exactly when $x+1\le1/(\lambda-1)$. As $\frac{20}{19}<\lambda<\frac{19}{18}$, the largest value over the integers is attained at $x=18$, and equals $19\cdot(9/10)^9<8$.}
The exponents $m_{<\ell}+m_\ell$, over the levels $\ell$ with $m_\ell\ge1$, are distinct integers in $\{1,\dots,|S|\}$, and therefore
\[
 \sum_{a\in S}2^{\lev(a)}\left(1-\settle a\right)\le8\widehat E\sum_{i=1}^{|S|}\lambda^i=\frac{8\lambda}{\lambda-1}\left(\lambda^{|S|}-1\right)\widehat E=8(10+9\lambda)\left(\lambda^{|S|}-1\right)\widehat E,
\]
where the last step uses $\lambda/(\lambda-1)=10+9\lambda$, which holds since $\lambda^2=10/9$.
\end{proof}

\subsection{Putting Everything Together}\label{sec:disp-proof}
We can now prove \Cref{lem:displacement}.

\begin{proof}[Proof of \Cref{lem:displacement}]
Let $c=8(10+9\lambda)$, and list the agents of $\Anew$ as $a_1,\dots,a_{\Delta_j}$ so that $u_{a_1}\le\dots\le u_{a_{\Delta_j}}$.
Every $g\in\Anew(u_{a_i})$ satisfies $u_g\le\settle g<u_{a_i}$, so $|\Anew(u_{a_i})|\le i-1$, and \Cref{lem:debts} and \Cref{prop:growth} give
\[
 2^{\lev(a_i)}\debt_{a_i}\le\sum_{g\in\Anew(u_{a_i})}2^{\lev(g)}\left(1-\settle g\right)\le c\left(\lambda^{i-1}-1\right)\widehat E.
\]
Since $R(a)\le2^{\lev(a)}$ and $1-u_a=(1-\settle a)+\debt_a$ for every $a\in\Anew$, we can therefore derive that
\begin{align*}
 \sum_{a\in\Anew}R(a)\,(1-u_a)
 &\le\sum_{a\in\Anew}2^{\lev(a)}\left(1-\settle a\right)+\sum_{i=1}^{\Delta_j}2^{\lev(a_i)}\debt_{a_i}\\
 &\le c\left(\lambda^{\Delta_j}-1\right)\widehat E+c\,\widehat E\sum_{i=1}^{\Delta_j}\left(\lambda^{i-1}-1\right) \tag{\Cref{prop:growth} with $t>1$}\\
 &\le c\,\frac{\lambda}{\lambda-1}\left(\lambda^{\Delta_j}-1\right)\widehat E\\
 &=8(10+9\lambda)^2\left(\lambda^{\Delta_j}-1\right)\widehat E \tag{$\lambda/(\lambda-1)=10+9\lambda$}\\
 &\le\frac{\kappa}{2}\left(\lambda^{\Delta_j}-1\right)\budget(u_{j-1}), \tag{\Cref{clm:input}}
\end{align*}
where $\kappa=16(10+9\lambda)^2=3040+2880\lambda<6080$.
\end{proof}

\end{document}